\documentclass[epsf,12pt]{article}
\usepackage{amsmath,amssymb}
\def\bn{\bigskip\noindent}
\begin{document}
\begin{titlepage}
\renewcommand{\thefootnote}{\fnsymbol{footnote}}
 \font\csc=cmcsc10 scaled\magstep1
 {\baselineskip=14pt
 \rightline{
 \vbox{\hbox{December, 2002}
       \hbox{UT-02-61}
       \hbox{UT-Komaba/02-16} 
       \hbox{hep-th/0212020}
      }}}
\bn
\begin{center}
\Large{ 
{\bf Geometric Engineering of Seiberg-Witten Theories\\ 
with Massive Hypermultiplets}
}
\vspace{5ex}\\
\normalsize{Yukiko Konishi}

\vspace{2ex}

\normalsize{\it Department of Physics, University of Tokyo,\\
Hongo 7-3-1, Bunkyo-ku, Tokyo 113-0033, Japan}

\vspace{1ex}

\normalsize{\texttt{konishi@hep-th.phys.s.u-tokyo.ac.jp}}

\vspace{3ex}

and

\vspace{3ex}

\normalsize{Michihiro Naka}

\vspace{2ex}

\normalsize{\it Institute of Physics, University of Tokyo,\\
Komaba 3-8-1, Meguro-ku, Tokyo 153-8902, Japan}

\vspace{1ex}

\normalsize{\texttt{hiro@hep1.c.u-tokyo.ac.jp}}

\vspace{2ex}

\end{center}
\vspace{1cm}
\begin{abstract}
We analyze the geometric engineering of 
the $N=2$ $SU(2)$ gauge theories with $1\leq N_f\leq 3$
massive hypermultiplets in the vector representation. 
The set of partial differential equations satisfied by 
the periods of the Seiberg-Witten differential
is obtained from the Picard-Fuchs equations of the local B-model.
The differential equations and its solutions
are consistent with the massless case. 
We show that the Yukawa coupling of the local A-model 
gives rise to the correct instanton expansion in the gauge theory,
and propose the pattern of the distribution of the 
world-sheet instanton number from it.
As a side result, we obtain the asymptotic form of the instanton amplitude
in the gauge theories with massless hypermultiplets.
\end{abstract}
\end{titlepage}

\section{Introduction}
\hspace{5mm}
Geometric engineering of Seiberg-Witten theories
\cite{KaKlVa} is the technique for 
extracting the moduli of the Coulomb phase of the 
four-dimensional $N=2$ gauge theories 
from the moduli of the mirror symmetry model. 
It realizes 
the former
in an infinitesimally small neighborhood of 
a singularity of the latter.
This technology opened a door toward the
systematic derivation of exact results in arbitrary
$N=2$ gauge theories, for example see \cite{KaMaVa, AgGr, HaTe}. 
However, detailed analysis of the prepotentials
has been restricted 
to only one example in the $N=2$ $SU(2)$ pure Yang-Mills theory \cite{KaKlVa}.
There has been no attempt to extend such study to the gauge theories 
with hypermultiplets in various representations.

In this article we will analyze 
the geometric engineering of the $N=2$ $SU(2)$ gauge theories 
with $1 \leq N_f \leq 3$ massive hypermultiplets in the vector representation.
It is necessary for
a mirror model corresponding to the $SU(2)$ gauge theory \cite{SeWi}
to contain $A_1$-root lattice in $H_2(V,{\bf Z})$,
where $V$ is a Calabi-Yau three-fold of the A-model.
The corresponding non-compact Calabi-Yau manifold 
was found to be the canonical bundle of the Hirzebruch surface,
which has the structure of a fibration of ALE-spaces of $A_1$ type 
over ${\bf P}^1$.
It has been noticed in \cite{KaVa,KaKlVa}
that cases of the massive hypermultiplets 
should be obtained by $N_f$ point blow ups of
Hirzebruch surface ${\bf F}_2$.
The corresponding manifolds again
have the structure of a fibration over ${\bf P}^1$.
We will adopt such local mirror models in \cite{ChKlYaZa}. 
 
The geometric engineering requires that 
the moduli of the Seiberg-Witten theory 
is identified with an infinitesimal neighborhood of a singular point
within the moduli of the mirror model.
This is the limit of decoupling the gravitational effects 
by taking the string scale $M_{string}$ to $ \infty$
in the type IIA string compactification.
We will call this {\it the gauge theory limit}.
Precisely, the gauge theory limit is the limit 
$\epsilon=M_{string}^{-1}\to0$ satisfying: (1) on the A-model side,
the size of 
the base ${\bf P}^1$ must become divergent as $(4-N_f)\log\epsilon$
and the size of the exceptional curve of the ALE space must
be proportional to $\epsilon a$ where 
$a$ is the mass of the gauge field; (2)
on the B-model side, this must be the limit where
the local B-model curve degenerates to the Seiberg-Witten curve.
We will show that the gauge theory limit of the local mirror
model actually exists, as expected. 
We will analyze the mirror model  
around the gauge theory limit and obtain the following two results.

The one result in this article 
is to obtain the differential equations satisfied by 
the periods $(a, a_D)$ of 
the Seiberg-Witten differential.
This completes the previous attempts 
in the massive cases \cite{Ohta, MaSu, IsMuNuSh, Eguchi}.
Taking the gauge theory limit of the 
Picard-Fuchs operators of the local B-model,
we obtained a set of partial differential operators
for each $N_f$.
We confirmed that these actually annihilate $(a, a_D)$.
Then we solved the differential equations:
the solutions are
two functions $g_1(u,m_i)$,
$g_2(u,m_i)$ 
which are identified with $(a, a_D)$, and
the bare mass parameters $m_i \, (1\leq i\leq N_f)$.
The appearance of the mass parameters is consistent with 
the fact that the Seiberg-Witten differential 
has a linear combination of the mass parameters as its residue. 

The other result is to find out the distribution pattern of 
world-sheet instanton numbers 
for the Calabi-Yau three-fold $V$ of the local A-model.
The case of $N_f=0$ was analyzed in \cite{KaKlVa}: 
there, the asymptotic distribution of the world-sheet instanton
numbers is controlled by the instanton amplitude of the gauge theory.
Then, it is natural to ask how to extend it 
to the local A-models corresponding to 
$N_f$ hypermultiplets ($1\leq N_f\leq 3$),
and we will answer this question affirmatively.
Let us explain our results briefly. 
At the gauge theory limit $\epsilon \to 0$, 
the K\"ahler parameter $t_1$ of the base ${\bf P}^1$ 
behaves as $(4-N_f)\log\epsilon$.
The K\"ahler parameters
$t_2, \dots, t_{N_f+2}$  of the other two-cycles 
are proportional to $\epsilon$ and linear combinations of $a$ and $m_i$
$(1\leq i\leq N_f)$.
We found out that 
if we denote the two-cycle whose K\"aher parameter is $-2\epsilon a $ by 
$R_0$ and the two-cycle whose K\"ahler parameter is $-\epsilon(a+m_i)$
by $R_i$, the asymptotic form of
the world-sheet instanton number $d_{\beta}$ 
for a homology class $\beta=n_1[{\bf P}^1]+\sum_{i=0}^{N_f}k_i[R_i]$
is
\begin{equation}
\label{form1}
d_{\beta}\, \sim\, \gamma_{n_1} \, (-1)^{k_1+\cdots+k_{N_f}}
\, (2k_0)^{4n_1-3} \, \prod_{i=1}^{N_f} \, {_{n_1}C_{k_i}}
\qquad 
(0 \leq {^\forall} k_i\leq n_1),
\end{equation}
at the region $k_0\gg n_1$ for fixed $n_1 \, (\geq 1)$.
For the other values of $k_i$'s, $d_{\beta}$ is negligible.
Here $\gamma_{n_1}$ is assumed to  depend on $n_1$, and
it turned  out that 
$\gamma_{n_1}$ is related to the instanton amplitude ${\cal F}_n$ 
of the $SU(2)$ Seiberg-Witten theory without hypermultiplets as follows:
\begin{equation}
\frac
{\gamma_{n_1}}{d_{\beta=[R_0]}}
\propto \frac{2\cdot 4^{n_1}{\cal F}_{n_1}}{\Gamma(4n_1-2)}.
\end{equation}
This relation holds
for all the local A-models because of the decoupling of a hypermultiplet
in the gauge theories. 

Although not directly related to the geometric engineering,
we studied the
asymptotic form of the instanton amplitude with large instanton number
in the Seiberg-Witten theories when all mass parameters are zero. 
The idea is the same as \cite{CadeGrPa} that
such asymptotic form is governed by a singularity of the moduli space.

This article is organized as follows.
In section 2, we will give a brief review on 
the Seiberg-Witten theories, 
and derive the asymptotic form of the instanton amplitude with large 
instanton numbers.
In section \ref{local}, we will analyze mirror models
that we use in the geometric engineering.
In section \ref{geom}, we 
will carry out the geometric engineering of the 
Seiberg-Witten theories with $N_f$ hypermultiplets.
A set of partial differential equations satisfied by 
periods $(a, a_D)$ and its solutions
will be derived in subsection \ref{geo22}.
We will suggest
the pattern of the distribution of the world-sheet instanton number
in subsection \ref{gw}.
Section 5 includes a conclusion and an outlook.
Appendices contain:
A: GKZ-hypergeometric differential system,
B: Table of the world-sheet instanton numbers ($N_f=0,1$).
C: the Yukawa coupling at the gauge theory limit
$(N_f=0,1,2,3)$,
D: the Picard-Fuchs differential operator for the periods $(a,a_D)$
$(N_f=0,1,2)$.

We will use the following notations:
\noindent (1) $\theta_x:=x \partial_x$ is a logarithmic derivative.
\noindent (2) 
$a^b$ is a multi-index notation
of $a_1^{b_1}a_2^{b_2}\cdots a_m^{b_m}$ 
for given vectors $a=(a_1,\cdots,a_m)$ and $b=(b_1,\cdots,b_m)$.
\noindent (3) 
For a series in several variables
with summation $\sum_{n_i}$,
the summation is assumed over non-negative integers $n_i$
so that the arguments in all the factorials are non-negative integers.

\section{$N=2$ $SU(2)$ Gauge Theories}\label{gauge}
\hspace{5mm}
In this section, we will summarize basic facts about the
exact solution of the $N=2$ $SU(2)$ gauge theories.
Then in subsection 2.2, we will derive the distribution pattern of 
the instanton amplitude with large instanton numbers when 
all mass parameters become zero.
This result will be obtained by 
an application of the techniques in a context of mirror symmetry
\cite{CadeGrPa}. 
\subsection{Seiberg-Witten Curves and Periods}\label{swcp1}
\hspace{5mm}
The moduli space of the Coulomb branch of 
the $N=2$ $SU(2)$ gauge theory in four-dimensions 
with $0\leq N_f \leq 3$ hypermultiplets 
is determined by a holomorphic function 
${\cal F}_{gauge}(a)$ called prepotential 
\cite{SeWi}.
We define an expectation value of 
the $2\times 2$ matrix-valued complex scalar field $\phi$ 
in the $N=2$ vector multiplet and the parameter $u$
of the moduli space as follows
\begin{equation}
u:=\frac{1}{2}\langle\mbox{Tr}\,\phi^2\rangle,
\hskip 0.4 in
\langle\phi\rangle
=\left(\begin{array}{cc}
a&0\\0&-a
\end{array}\right).
\end{equation}
We denote by $m_1, \cdots, m_{N_f}$ the bare mass parameters 
of $N_f$ hypermultiplets.
In the Seiberg-Witten theory,
$a$ and $a_D:=\frac{\partial {\cal F}_{gauge}}{\partial a}$ 
are periods of the meromorphic one-form 
(Seiberg-Witten differential $\lambda_{SW}$) on an
elliptic curve (Seiberg-Witten curve) 
parameterized by $u$, $m_1,\cdots,m_{N_f}$ and 
the dynamical mass parameter $\Lambda$.

\begin{table}[htop]
\begin{center}
\begin{equation*}
\begin{array}{|c|l|l|}\hline
N_f&F(x)&H(x)\\\hline
0&x^2-u&\Lambda^4\\\hline
1&x^2-u&
\Lambda^3(x+m_1)\\\hline
2&x^2-u+\frac{\Lambda^2}{8}&
 \Lambda^2(x+m_1)(x+m_2)\\\hline
3&x^2-u+\frac{\Lambda x}{4}+\frac{\Lambda(m_1+m_2+m_3)}{8}&
\Lambda(x+m_1)(x+m_2)(x+m_3)\\\hline
\end{array}
\hskip 0.4 in
\begin{array}{|c|c|c|c|}\hline
N_f&\alpha&z&C\\\hline
0&
\frac{1}{4}&\frac{\Lambda^4}{u^2}
&1
\\\hline
1&\frac{1}{6}&
\frac{3^3\Lambda^6}{-4^4u^3}
&
\frac{3^3}{-4^4}
\\\hline
2&\frac{1}{4}&
\frac{\Lambda^4}{4^3u^2}
&\frac{1}{4^3}
\\\hline
3&\frac{1}{2}&
\frac{\Lambda^2}{4^4u}
&\frac{1}{4^4}
\\\hline
\end{array}
\end{equation*}
\end{center}
\caption{Left: $F(x)$ and $H(x)$ in (\ref{lam}). 
Right: {$\alpha,z,C$ in (\ref{pfeq}).}}\label{swcurve}
\end{table}

\begin{figure}[h]
\setlength{\unitlength}{.007\textwidth}
\begin{center}
\begin{picture}(80,25)(0,20)
\thicklines
\thinlines
\put(7,30){$e_1$}
\put(10,30){\circle*{1}}
\put(30,30){\circle*{1}}\put(31,30){$e_2$}
\put(50,30){\circle*{1}}\put(47,30){$e_3$}
\put(70,30){\circle*{1}}\put(71,30){$e_4$}
\put(10,30.2){\line(1,0){20}}
\put(10,29.8){\line(1,0){20}}\put(17.5,27){{\small cut}}
\put(50,30.2){\line(1,0){20}}
\put(50,29.8){\line(1,0){20}}\put(57.5,27){{\small cut}}
\thinlines
\put(18,40){\line(2,1){4}}
\put(18,40){\line(2,-1){4}}
\put(38,40){\line(2,1){4}}
\put(38,40){\line(2,-1){4}}
\put(20,43){$\alpha$}
\put(40,43){$\beta$}
\thicklines
 \put( 25.,30.){\circle*{0.2}}\put( 25.0365,29.3024){\circle*{0.2}}
  \put( 25.146,28.6083){\circle*{0.2}}\put( 25.3278,27.9209){\circle*{0.2}}
  \put( 25.5811,27.2436){\circle*{0.2}}\put( 25.9046,26.5798){\circle*{0.2}}
  \put( 26.2968,25.9326){\circle*{0.2}}\put( 26.7558,25.3053){\circle*{0.2}}
  \put( 27.2793,24.7008){\circle*{0.2}}\put( 27.8647,24.1221){\circle*{0.2}}
  \put( 28.5093,23.5721){\circle*{0.2}}\put( 29.2099,23.0534){\circle*{0.2}}
  \put( 29.963,22.5686){\circle*{0.2}}\put( 30.7651,22.1199){\circle*{0.2}}
  \put( 31.6121,21.7096){\circle*{0.2}}\put( 32.5,21.3397){\circle*{0.2}}
  \put( 33.4244,21.0121){\circle*{0.2}}\put( 34.3809,20.7282){\circle*{0.2}}
  \put( 35.3647,20.4894){\circle*{0.2}}\put( 36.3712,20.297){\circle*{0.2}}
  \put( 37.3953,20.1519){\circle*{0.2}}\put( 38.4321,20.0548){\circle*{0.2}}
  \put( 39.4765,20.0061){\circle*{0.2}}\put( 40.5235,20.0061){\circle*{0.2}}
  \put( 41.5679,20.0548){\circle*{0.2}}\put( 42.6047,20.1519){\circle*{0.2}}
  \put( 43.6288,20.297){\circle*{0.2}}\put( 44.6353,20.4894){\circle*{0.2}}
  \put( 45.6191,20.7282){\circle*{0.2}}\put( 46.5756,21.0121){\circle*{0.2}}
  \put( 47.5,21.3397){\circle*{0.2}}\put( 48.3879,21.7096){\circle*{0.2}}
  \put( 49.2349,22.1199){\circle*{0.2}}\put( 50.037,22.5686){\circle*{0.2}}
  \put( 50.7901,23.0534){\circle*{0.2}}\put( 51.4907,23.5721){\circle*{0.2}}
  \put( 52.1353,24.1221){\circle*{0.2}}\put( 52.7207,24.7008){\circle*{0.2}}
  \put( 53.2442,25.3053){\circle*{0.2}}\put( 53.7032,25.9326){\circle*{0.2}}
  \put( 54.0954,26.5798){\circle*{0.2}}\put( 54.4189,27.2436){\circle*{0.2}}
  \put( 54.6722,27.9209){\circle*{0.2}}\put( 54.854,28.6083){\circle*{0.2}}
  \put( 54.9635,29.3024){\circle*{0.2}}\put( 55.,30.){\circle*{0.2}}\put( 55.,30.){\circle*{0.2}}
  \put( 54.9994,30.0873){\circle*{0.2}}\put( 54.9977,30.1745){\circle*{0.2}}
  \put( 54.9949,30.2618){\circle*{0.2}}\put( 54.9909,30.349){\circle*{0.2}}
  \put( 54.9857,30.4362){\circle*{0.2}}\put( 54.9794,30.5234){\circle*{0.2}}
  \put( 54.972,30.6105){\circle*{0.2}}\put( 54.9635,30.6976){\circle*{0.2}}
  \put( 54.9538,30.7846){\circle*{0.2}}\put( 54.9429,30.8716){\circle*{0.2}}
  \put( 54.9309,30.9585){\circle*{0.2}}\put( 54.9178,31.0453){\circle*{0.2}}
  \put( 54.9036,31.132){\circle*{0.2}}\put( 54.8882,31.2187){\circle*{0.2}}
  \put( 54.8717,31.3053){\circle*{0.2}}\put( 54.854,31.3917){\circle*{0.2}}
  \put( 54.8352,31.4781){\circle*{0.2}}\put( 54.8153,31.5643){\circle*{0.2}}
  \put( 54.7943,31.6505){\circle*{0.2}}\put( 54.7721,31.7365){\circle*{0.2}}
  \put( 54.7488,31.8224){\circle*{0.2}}\put( 54.7244,31.9081){\circle*{0.2}}
  \put( 54.6989,31.9937){\circle*{0.2}}\put( 54.6722,32.0791){\circle*{0.2}}
  \put( 54.6444,32.1644){\circle*{0.2}}\put( 54.6156,32.2495){\circle*{0.2}}
  \put( 54.5855,32.3345){\circle*{0.2}}\put( 54.5544,32.4192){\circle*{0.2}}
  \put( 54.5222,32.5038){\circle*{0.2}}\put( 54.4889,32.5882){\circle*{0.2}}
  \put( 54.4545,32.6724){\circle*{0.2}}\put( 54.4189,32.7564){\circle*{0.2}}
  \put( 54.3823,32.8402){\circle*{0.2}}\put( 54.3446,32.9237){\circle*{0.2}}
  \put( 54.3058,33.0071){\circle*{0.2}}\put( 54.2658,33.0902){\circle*{0.2}}
  \put( 54.2249,33.173){\circle*{0.2}}\put( 54.1828,33.2557){\circle*{0.2}}
  \put( 54.1396,33.3381){\circle*{0.2}}\put( 54.0954,33.4202){\circle*{0.2}}
  \put( 54.0501,33.5021){\circle*{0.2}}\put( 54.0037,33.5837){\circle*{0.2}}
  \put( 53.9563,33.665){\circle*{0.2}}\put( 53.9078,33.7461){\circle*{0.2}}
  \put( 53.8582,33.8268){\circle*{0.2}}\put( 53.8076,33.9073){\circle*{0.2}}
  \put( 53.7559,33.9875){\circle*{0.2}}\put( 53.7032,34.0674){\circle*{0.2}}
  \put( 53.6494,34.1469){\circle*{0.2}}\put( 53.5946,34.2262){\circle*{0.2}}
  \put( 53.5388,34.3051){\circle*{0.2}}\put( 53.4819,34.3837){\circle*{0.2}}
  \put( 53.424,34.462){\circle*{0.2}}\put( 53.3651,34.5399){\circle*{0.2}}
  \put( 53.3052,34.6175){\circle*{0.2}}\put( 53.2442,34.6947){\circle*{0.2}}
  \put( 53.1823,34.7716){\circle*{0.2}}\put( 53.1193,34.8481){\circle*{0.2}}
  \put( 53.0553,34.9242){\circle*{0.2}}\put( 52.9904,35.){\circle*{0.2}}
  \put( 52.9244,35.0754){\circle*{0.2}}\put( 52.8575,35.1504){\circle*{0.2}}
  \put( 52.7896,35.225){\circle*{0.2}}\put( 52.7207,35.2992){\circle*{0.2}}
  \put( 52.6509,35.373){\circle*{0.2}}\put( 52.5801,35.4464){\circle*{0.2}}
  \put( 52.5083,35.5194){\circle*{0.2}}\put( 52.4356,35.5919){\circle*{0.2}}
  \put( 52.3619,35.6641){\circle*{0.2}}\put( 52.2873,35.7358){\circle*{0.2}}
  \put( 52.2117,35.807){\circle*{0.2}}\put( 52.1353,35.8779){\circle*{0.2}}
  \put( 52.0579,35.9482){\circle*{0.2}}\put( 51.9795,36.0182){\circle*{0.2}}
  \put( 51.9003,36.0876){\circle*{0.2}}\put( 51.8202,36.1566){\circle*{0.2}}
  \put( 51.7391,36.2251){\circle*{0.2}}\put( 51.6572,36.2932){\circle*{0.2}}
  \put( 51.5744,36.3608){\circle*{0.2}}\put( 51.4907,36.4279){\circle*{0.2}}
  \put( 51.4061,36.4945){\circle*{0.2}}\put( 51.3206,36.5606){\circle*{0.2}}
  \put( 51.2343,36.6262){\circle*{0.2}}\put( 51.1472,36.6913){\circle*{0.2}}
  \put( 51.0592,36.7559){\circle*{0.2}}\put( 50.9703,36.82){\circle*{0.2}}
  \put( 50.8806,36.8835){\circle*{0.2}}\put( 50.7901,36.9466){\circle*{0.2}}
  \put( 50.6988,37.0091){\circle*{0.2}}\put( 50.6066,37.0711){\circle*{0.2}}
  \put( 50.5136,37.1325){\circle*{0.2}}\put( 50.4199,37.1934){\circle*{0.2}}
  \put( 50.3253,37.2537){\circle*{0.2}}\put( 50.23,37.3135){\circle*{0.2}}
  \put( 50.1339,37.3728){\circle*{0.2}}\put( 50.037,37.4314){\circle*{0.2}}
  \put( 49.9393,37.4896){\circle*{0.2}}\put( 49.8409,37.5471){\circle*{0.2}}
  \put( 49.7417,37.6041){\circle*{0.2}}\put( 49.6418,37.6604){\circle*{0.2}}
  \put( 49.5412,37.7162){\circle*{0.2}}\put( 49.4398,37.7715){\circle*{0.2}}
  \put( 49.3377,37.8261){\circle*{0.2}}\put( 49.2349,37.8801){\circle*{0.2}}
  \put( 49.1314,37.9335){\circle*{0.2}}\put( 49.0272,37.9864){\circle*{0.2}}
  \put( 48.9223,38.0386){\circle*{0.2}}\put( 48.8168,38.0902){\circle*{0.2}}
  \put( 48.7105,38.1412){\circle*{0.2}}\put( 48.6036,38.1915){\circle*{0.2}}
  \put( 48.4961,38.2413){\circle*{0.2}}\put( 48.3879,38.2904){\circle*{0.2}}
  \put( 48.2791,38.3389){\circle*{0.2}}\put( 48.1696,38.3867){\circle*{0.2}}
  \put( 48.0595,38.4339){\circle*{0.2}}\put( 47.9488,38.4805){\circle*{0.2}}
  \put( 47.8375,38.5264){\circle*{0.2}}\put( 47.7256,38.5717){\circle*{0.2}}
  \put( 47.6131,38.6163){\circle*{0.2}}\put( 47.5,38.6603){\circle*{0.2}}
  \put( 47.3864,38.7036){\circle*{0.2}}\put( 47.2721,38.7462){\circle*{0.2}}
  \put( 47.1574,38.7882){\circle*{0.2}}\put( 47.0421,38.8295){\circle*{0.2}}
  \put( 46.9262,38.8701){\circle*{0.2}}\put( 46.8099,38.9101){\circle*{0.2}}
  \put( 46.693,38.9493){\circle*{0.2}}\put( 46.5756,38.9879){\circle*{0.2}}
  \put( 46.4577,39.0259){\circle*{0.2}}\put( 46.3393,39.0631){\circle*{0.2}}
  \put( 46.2204,39.0996){\circle*{0.2}}\put( 46.101,39.1355){\circle*{0.2}}
  \put( 45.9812,39.1706){\circle*{0.2}}\put( 45.861,39.205){\circle*{0.2}}
  \put( 45.7403,39.2388){\circle*{0.2}}\put( 45.6191,39.2718){\circle*{0.2}}
  \put( 45.4975,39.3042){\circle*{0.2}}\put( 45.3755,39.3358){\circle*{0.2}}
  \put( 45.2531,39.3667){\circle*{0.2}}\put( 45.1303,39.3969){\circle*{0.2}}
  \put( 45.0071,39.4264){\circle*{0.2}}\put( 44.8835,39.4552){\circle*{0.2}}
  \put( 44.7596,39.4832){\circle*{0.2}}\put( 44.6353,39.5106){\circle*{0.2}}
  \put( 44.5106,39.5372){\circle*{0.2}}\put( 44.3856,39.563){\circle*{0.2}}
  \put( 44.2602,39.5882){\circle*{0.2}}\put( 44.1346,39.6126){\circle*{0.2}}
  \put( 44.0086,39.6363){\circle*{0.2}}\put( 43.8823,39.6593){\circle*{0.2}}
  \put( 43.7557,39.6815){\circle*{0.2}}\put( 43.6288,39.703){\circle*{0.2}}
  \put( 43.5017,39.7237){\circle*{0.2}}\put( 43.3743,39.7437){\circle*{0.2}}
  \put( 43.2466,39.763){\circle*{0.2}}\put( 43.1187,39.7815){\circle*{0.2}}
  \put( 42.9905,39.7992){\circle*{0.2}}\put( 42.8621,39.8163){\circle*{0.2}}
  \put( 42.7335,39.8325){\circle*{0.2}}\put( 42.6047,39.8481){\circle*{0.2}}
  \put( 42.4757,39.8629){\circle*{0.2}}\put( 42.3465,39.8769){\circle*{0.2}}
  \put( 42.2171,39.8902){\circle*{0.2}}\put( 42.0876,39.9027){\circle*{0.2}}
  \put( 41.9579,39.9144){\circle*{0.2}}\put( 41.828,39.9255){\circle*{0.2}}
  \put( 41.698,39.9357){\circle*{0.2}}\put( 41.5679,39.9452){\circle*{0.2}}
  \put( 41.4377,39.954){\circle*{0.2}}\put( 41.3073,39.9619){\circle*{0.2}}
  \put( 41.1769,39.9692){\circle*{0.2}}\put( 41.0463,39.9756){\circle*{0.2}}
  \put( 40.9157,39.9813){\circle*{0.2}}\put( 40.785,39.9863){\circle*{0.2}}
  \put( 40.6543,39.9905){\circle*{0.2}}\put( 40.5235,39.9939){\circle*{0.2}}
  \put( 40.3927,39.9966){\circle*{0.2}}\put( 40.2618,39.9985){\circle*{0.2}}
  \put( 40.1309,39.9996){\circle*{0.2}}\put( 40.,40.){\circle*{0.2}}
  \put( 39.8691,39.9996){\circle*{0.2}}\put( 39.7382,39.9985){\circle*{0.2}}
  \put( 39.6073,39.9966){\circle*{0.2}}\put( 39.4765,39.9939){\circle*{0.2}}
  \put( 39.3457,39.9905){\circle*{0.2}}\put( 39.215,39.9863){\circle*{0.2}}
  \put( 39.0843,39.9813){\circle*{0.2}}\put( 38.9537,39.9756){\circle*{0.2}}
  \put( 38.8231,39.9692){\circle*{0.2}}\put( 38.6927,39.9619){\circle*{0.2}}
  \put( 38.5623,39.954){\circle*{0.2}}\put( 38.4321,39.9452){\circle*{0.2}}
  \put( 38.302,39.9357){\circle*{0.2}}\put( 38.172,39.9255){\circle*{0.2}}
  \put( 38.0421,39.9144){\circle*{0.2}}\put( 37.9124,39.9027){\circle*{0.2}}
  \put( 37.7829,39.8902){\circle*{0.2}}\put( 37.6535,39.8769){\circle*{0.2}}
  \put( 37.5243,39.8629){\circle*{0.2}}\put( 37.3953,39.8481){\circle*{0.2}}
  \put( 37.2665,39.8325){\circle*{0.2}}\put( 37.1379,39.8163){\circle*{0.2}}
  \put( 37.0095,39.7992){\circle*{0.2}}\put( 36.8813,39.7815){\circle*{0.2}}
  \put( 36.7534,39.763){\circle*{0.2}}\put( 36.6257,39.7437){\circle*{0.2}}
  \put( 36.4983,39.7237){\circle*{0.2}}\put( 36.3712,39.703){\circle*{0.2}}
  \put( 36.2443,39.6815){\circle*{0.2}}\put( 36.1177,39.6593){\circle*{0.2}}
  \put( 35.9914,39.6363){\circle*{0.2}}\put( 35.8654,39.6126){\circle*{0.2}}
  \put( 35.7398,39.5882){\circle*{0.2}}\put( 35.6144,39.563){\circle*{0.2}}
  \put( 35.4894,39.5372){\circle*{0.2}}\put( 35.3647,39.5106){\circle*{0.2}}
  \put( 35.2404,39.4832){\circle*{0.2}}\put( 35.1165,39.4552){\circle*{0.2}}
  \put( 34.9929,39.4264){\circle*{0.2}}\put( 34.8697,39.3969){\circle*{0.2}}
  \put( 34.7469,39.3667){\circle*{0.2}}\put( 34.6245,39.3358){\circle*{0.2}}
  \put( 34.5025,39.3042){\circle*{0.2}}\put( 34.3809,39.2718){\circle*{0.2}}
  \put( 34.2597,39.2388){\circle*{0.2}}\put( 34.139,39.205){\circle*{0.2}}
  \put( 34.0188,39.1706){\circle*{0.2}}\put( 33.899,39.1355){\circle*{0.2}}
  \put( 33.7796,39.0996){\circle*{0.2}}\put( 33.6607,39.0631){\circle*{0.2}}
  \put( 33.5423,39.0259){\circle*{0.2}}\put( 33.4244,38.9879){\circle*{0.2}}
  \put( 33.307,38.9493){\circle*{0.2}}\put( 33.1901,38.9101){\circle*{0.2}}
  \put( 33.0738,38.8701){\circle*{0.2}}\put( 32.9579,38.8295){\circle*{0.2}}
  \put( 32.8426,38.7882){\circle*{0.2}}\put( 32.7279,38.7462){\circle*{0.2}}
  \put( 32.6136,38.7036){\circle*{0.2}}\put( 32.5,38.6603){\circle*{0.2}}
  \put( 32.3869,38.6163){\circle*{0.2}}\put( 32.2744,38.5717){\circle*{0.2}}
  \put( 32.1625,38.5264){\circle*{0.2}}\put( 32.0512,38.4805){\circle*{0.2}}
  \put( 31.9405,38.4339){\circle*{0.2}}\put( 31.8304,38.3867){\circle*{0.2}}
  \put( 31.7209,38.3389){\circle*{0.2}}\put( 31.6121,38.2904){\circle*{0.2}}
  \put( 31.5039,38.2413){\circle*{0.2}}\put( 31.3964,38.1915){\circle*{0.2}}
  \put( 31.2895,38.1412){\circle*{0.2}}\put( 31.1832,38.0902){\circle*{0.2}}
  \put( 31.0777,38.0386){\circle*{0.2}}\put( 30.9728,37.9864){\circle*{0.2}}
  \put( 30.8686,37.9335){\circle*{0.2}}\put( 30.7651,37.8801){\circle*{0.2}}
  \put( 30.6623,37.8261){\circle*{0.2}}\put( 30.5602,37.7715){\circle*{0.2}}
  \put( 30.4588,37.7162){\circle*{0.2}}\put( 30.3582,37.6604){\circle*{0.2}}
  \put( 30.2583,37.6041){\circle*{0.2}}\put( 30.1591,37.5471){\circle*{0.2}}
  \put( 30.0607,37.4896){\circle*{0.2}}\put( 29.963,37.4314){\circle*{0.2}}
  \put( 29.8661,37.3728){\circle*{0.2}}\put( 29.77,37.3135){\circle*{0.2}}
  \put( 29.6747,37.2537){\circle*{0.2}}\put( 29.5801,37.1934){\circle*{0.2}}
  \put( 29.4864,37.1325){\circle*{0.2}}\put( 29.3934,37.0711){\circle*{0.2}}
  \put( 29.3012,37.0091){\circle*{0.2}}\put( 29.2099,36.9466){\circle*{0.2}}
  \put( 29.1194,36.8835){\circle*{0.2}}\put( 29.0297,36.82){\circle*{0.2}}
  \put( 28.9408,36.7559){\circle*{0.2}}\put( 28.8528,36.6913){\circle*{0.2}}
  \put( 28.7657,36.6262){\circle*{0.2}}\put( 28.6794,36.5606){\circle*{0.2}}
  \put( 28.5939,36.4945){\circle*{0.2}}\put( 28.5093,36.4279){\circle*{0.2}}
  \put( 28.4256,36.3608){\circle*{0.2}}\put( 28.3428,36.2932){\circle*{0.2}}
  \put( 28.2609,36.2251){\circle*{0.2}}\put( 28.1798,36.1566){\circle*{0.2}}
  \put( 28.0997,36.0876){\circle*{0.2}}\put( 28.0205,36.0182){\circle*{0.2}}
  \put( 27.9421,35.9482){\circle*{0.2}}\put( 27.8647,35.8779){\circle*{0.2}}
  \put( 27.7883,35.807){\circle*{0.2}}\put( 27.7127,35.7358){\circle*{0.2}}
  \put( 27.6381,35.6641){\circle*{0.2}}\put( 27.5644,35.5919){\circle*{0.2}}
  \put( 27.4917,35.5194){\circle*{0.2}}\put( 27.4199,35.4464){\circle*{0.2}}
  \put( 27.3491,35.373){\circle*{0.2}}\put( 27.2793,35.2992){\circle*{0.2}}
  \put( 27.2104,35.225){\circle*{0.2}}\put( 27.1425,35.1504){\circle*{0.2}}
  \put( 27.0756,35.0754){\circle*{0.2}}\put( 27.0096,35.){\circle*{0.2}}
  \put( 26.9447,34.9242){\circle*{0.2}}\put( 26.8807,34.8481){\circle*{0.2}}
  \put( 26.8177,34.7716){\circle*{0.2}}\put( 26.7558,34.6947){\circle*{0.2}}
  \put( 26.6948,34.6175){\circle*{0.2}}\put( 26.6349,34.5399){\circle*{0.2}}
  \put( 26.576,34.462){\circle*{0.2}}\put( 26.5181,34.3837){\circle*{0.2}}
  \put( 26.4612,34.3051){\circle*{0.2}}\put( 26.4054,34.2262){\circle*{0.2}}
  \put( 26.3506,34.1469){\circle*{0.2}}\put( 26.2968,34.0674){\circle*{0.2}}
  \put( 26.2441,33.9875){\circle*{0.2}}\put( 26.1924,33.9073){\circle*{0.2}}
  \put( 26.1418,33.8268){\circle*{0.2}}\put( 26.0922,33.7461){\circle*{0.2}}
  \put( 26.0437,33.665){\circle*{0.2}}\put( 25.9963,33.5837){\circle*{0.2}}
  \put( 25.9499,33.5021){\circle*{0.2}}\put( 25.9046,33.4202){\circle*{0.2}}
  \put( 25.8604,33.3381){\circle*{0.2}}\put( 25.8172,33.2557){\circle*{0.2}}
  \put( 25.7751,33.173){\circle*{0.2}}\put( 25.7342,33.0902){\circle*{0.2}}
  \put( 25.6942,33.0071){\circle*{0.2}}\put( 25.6554,32.9237){\circle*{0.2}}
  \put( 25.6177,32.8402){\circle*{0.2}}\put( 25.5811,32.7564){\circle*{0.2}}
  \put( 25.5455,32.6724){\circle*{0.2}}\put( 25.5111,32.5882){\circle*{0.2}}
  \put( 25.4778,32.5038){\circle*{0.2}}\put( 25.4456,32.4192){\circle*{0.2}}
  \put( 25.4145,32.3345){\circle*{0.2}}\put( 25.3844,32.2495){\circle*{0.2}}
  \put( 25.3556,32.1644){\circle*{0.2}}\put( 25.3278,32.0791){\circle*{0.2}}
  \put( 25.3011,31.9937){\circle*{0.2}}\put( 25.2756,31.9081){\circle*{0.2}}
  \put( 25.2512,31.8224){\circle*{0.2}}\put( 25.2279,31.7365){\circle*{0.2}}
  \put( 25.2057,31.6505){\circle*{0.2}}\put( 25.1847,31.5643){\circle*{0.2}}
  \put( 25.1648,31.4781){\circle*{0.2}}\put( 25.146,31.3917){\circle*{0.2}}
  \put( 25.1283,31.3053){\circle*{0.2}}\put( 25.1118,31.2187){\circle*{0.2}}
  \put( 25.0964,31.132){\circle*{0.2}}\put( 25.0822,31.0453){\circle*{0.2}}
  \put( 25.0691,30.9585){\circle*{0.2}}\put( 25.0571,30.8716){\circle*{0.2}}
  \put( 25.0462,30.7846){\circle*{0.2}}\put( 25.0365,30.6976){\circle*{0.2}}
  \put( 25.028,30.6105){\circle*{0.2}}\put( 25.0206,30.5234){\circle*{0.2}}
  \put( 25.0143,30.4362){\circle*{0.2}}\put( 25.0091,30.349){\circle*{0.2}}
  \put( 25.0051,30.2618){\circle*{0.2}}\put( 25.0023,30.1745){\circle*{0.2}}
  \put( 25.0006,30.0873){\circle*{0.2}}\put( 25.,30.){\circle*{0.2}} 
 \put( 35.,30.){\circle*{0.2}}\put( 34.9994,30.0873){\circle*{0.2}}
  \put( 34.9977,30.1745){\circle*{0.2}}\put( 34.9949,30.2618){\circle*{0.2}}
  \put( 34.9909,30.349){\circle*{0.2}}\put( 34.9857,30.4362){\circle*{0.2}}
  \put( 34.9794,30.5234){\circle*{0.2}}\put( 34.972,30.6105){\circle*{0.2}}
  \put( 34.9635,30.6976){\circle*{0.2}}\put( 34.9538,30.7846){\circle*{0.2}}
  \put( 34.9429,30.8716){\circle*{0.2}}\put( 34.9309,30.9585){\circle*{0.2}}
  \put( 34.9178,31.0453){\circle*{0.2}}\put( 34.9036,31.132){\circle*{0.2}}
  \put( 34.8882,31.2187){\circle*{0.2}}\put( 34.8717,31.3053){\circle*{0.2}}
  \put( 34.854,31.3917){\circle*{0.2}}\put( 34.8352,31.4781){\circle*{0.2}}
  \put( 34.8153,31.5643){\circle*{0.2}}\put( 34.7943,31.6505){\circle*{0.2}}
  \put( 34.7721,31.7365){\circle*{0.2}}\put( 34.7488,31.8224){\circle*{0.2}}
  \put( 34.7244,31.9081){\circle*{0.2}}\put( 34.6989,31.9937){\circle*{0.2}}
  \put( 34.6722,32.0791){\circle*{0.2}}\put( 34.6444,32.1644){\circle*{0.2}}
  \put( 34.6156,32.2495){\circle*{0.2}}\put( 34.5855,32.3345){\circle*{0.2}}
  \put( 34.5544,32.4192){\circle*{0.2}}\put( 34.5222,32.5038){\circle*{0.2}}
  \put( 34.4889,32.5882){\circle*{0.2}}\put( 34.4545,32.6724){\circle*{0.2}}
  \put( 34.4189,32.7564){\circle*{0.2}}\put( 34.3823,32.8402){\circle*{0.2}}
  \put( 34.3446,32.9237){\circle*{0.2}}\put( 34.3058,33.0071){\circle*{0.2}}
  \put( 34.2658,33.0902){\circle*{0.2}}\put( 34.2249,33.173){\circle*{0.2}}
  \put( 34.1828,33.2557){\circle*{0.2}}\put( 34.1396,33.3381){\circle*{0.2}}
  \put( 34.0954,33.4202){\circle*{0.2}}\put( 34.0501,33.5021){\circle*{0.2}}
  \put( 34.0037,33.5837){\circle*{0.2}}\put( 33.9563,33.665){\circle*{0.2}}
  \put( 33.9078,33.7461){\circle*{0.2}}\put( 33.8582,33.8268){\circle*{0.2}}
  \put( 33.8076,33.9073){\circle*{0.2}}\put( 33.7559,33.9875){\circle*{0.2}}
  \put( 33.7032,34.0674){\circle*{0.2}}\put( 33.6494,34.1469){\circle*{0.2}}
  \put( 33.5946,34.2262){\circle*{0.2}}\put( 33.5388,34.3051){\circle*{0.2}}
  \put( 33.4819,34.3837){\circle*{0.2}}\put( 33.424,34.462){\circle*{0.2}}
  \put( 33.3651,34.5399){\circle*{0.2}}\put( 33.3052,34.6175){\circle*{0.2}}
  \put( 33.2442,34.6947){\circle*{0.2}}\put( 33.1823,34.7716){\circle*{0.2}}
  \put( 33.1193,34.8481){\circle*{0.2}}\put( 33.0553,34.9242){\circle*{0.2}}
  \put( 32.9904,35.){\circle*{0.2}}\put( 32.9244,35.0754){\circle*{0.2}}
  \put( 32.8575,35.1504){\circle*{0.2}}\put( 32.7896,35.225){\circle*{0.2}}
  \put( 32.7207,35.2992){\circle*{0.2}}\put( 32.6509,35.373){\circle*{0.2}}
  \put( 32.5801,35.4464){\circle*{0.2}}\put( 32.5083,35.5194){\circle*{0.2}}
  \put( 32.4356,35.5919){\circle*{0.2}}\put( 32.3619,35.6641){\circle*{0.2}}
  \put( 32.2873,35.7358){\circle*{0.2}}\put( 32.2117,35.807){\circle*{0.2}}
  \put( 32.1353,35.8779){\circle*{0.2}}\put( 32.0579,35.9482){\circle*{0.2}}
  \put( 31.9795,36.0182){\circle*{0.2}}\put( 31.9003,36.0876){\circle*{0.2}}
  \put( 31.8202,36.1566){\circle*{0.2}}\put( 31.7391,36.2251){\circle*{0.2}}
  \put( 31.6572,36.2932){\circle*{0.2}}\put( 31.5744,36.3608){\circle*{0.2}}
  \put( 31.4907,36.4279){\circle*{0.2}}\put( 31.4061,36.4945){\circle*{0.2}}
  \put( 31.3206,36.5606){\circle*{0.2}}\put( 31.2343,36.6262){\circle*{0.2}}
  \put( 31.1472,36.6913){\circle*{0.2}}\put( 31.0592,36.7559){\circle*{0.2}}
  \put( 30.9703,36.82){\circle*{0.2}}\put( 30.8806,36.8835){\circle*{0.2}}
  \put( 30.7901,36.9466){\circle*{0.2}}\put( 30.6988,37.0091){\circle*{0.2}}
  \put( 30.6066,37.0711){\circle*{0.2}}\put( 30.5136,37.1325){\circle*{0.2}}
  \put( 30.4199,37.1934){\circle*{0.2}}\put( 30.3253,37.2537){\circle*{0.2}}
  \put( 30.23,37.3135){\circle*{0.2}}\put( 30.1339,37.3728){\circle*{0.2}}
  \put( 30.037,37.4314){\circle*{0.2}}\put( 29.9393,37.4896){\circle*{0.2}}
  \put( 29.8409,37.5471){\circle*{0.2}}\put( 29.7417,37.6041){\circle*{0.2}}
  \put( 29.6418,37.6604){\circle*{0.2}}\put( 29.5412,37.7162){\circle*{0.2}}
  \put( 29.4398,37.7715){\circle*{0.2}}\put( 29.3377,37.8261){\circle*{0.2}}
  \put( 29.2349,37.8801){\circle*{0.2}}\put( 29.1314,37.9335){\circle*{0.2}}
  \put( 29.0272,37.9864){\circle*{0.2}}\put( 28.9223,38.0386){\circle*{0.2}}
  \put( 28.8168,38.0902){\circle*{0.2}}\put( 28.7105,38.1412){\circle*{0.2}}
  \put( 28.6036,38.1915){\circle*{0.2}}\put( 28.4961,38.2413){\circle*{0.2}}
  \put( 28.3879,38.2904){\circle*{0.2}}\put( 28.2791,38.3389){\circle*{0.2}}
  \put( 28.1696,38.3867){\circle*{0.2}}\put( 28.0595,38.4339){\circle*{0.2}}
  \put( 27.9488,38.4805){\circle*{0.2}}\put( 27.8375,38.5264){\circle*{0.2}}
  \put( 27.7256,38.5717){\circle*{0.2}}\put( 27.6131,38.6163){\circle*{0.2}}
  \put( 27.5,38.6603){\circle*{0.2}}\put( 27.3864,38.7036){\circle*{0.2}}
  \put( 27.2721,38.7462){\circle*{0.2}}\put( 27.1574,38.7882){\circle*{0.2}}
  \put( 27.0421,38.8295){\circle*{0.2}}\put( 26.9262,38.8701){\circle*{0.2}}
  \put( 26.8099,38.9101){\circle*{0.2}}\put( 26.693,38.9493){\circle*{0.2}}
  \put( 26.5756,38.9879){\circle*{0.2}}\put( 26.4577,39.0259){\circle*{0.2}}
  \put( 26.3393,39.0631){\circle*{0.2}}\put( 26.2204,39.0996){\circle*{0.2}}
  \put( 26.101,39.1355){\circle*{0.2}}\put( 25.9812,39.1706){\circle*{0.2}}
  \put( 25.861,39.205){\circle*{0.2}}\put( 25.7403,39.2388){\circle*{0.2}}
  \put( 25.6191,39.2718){\circle*{0.2}}\put( 25.4975,39.3042){\circle*{0.2}}
  \put( 25.3755,39.3358){\circle*{0.2}}\put( 25.2531,39.3667){\circle*{0.2}}
  \put( 25.1303,39.3969){\circle*{0.2}}\put( 25.0071,39.4264){\circle*{0.2}}
  \put( 24.8835,39.4552){\circle*{0.2}}\put( 24.7596,39.4832){\circle*{0.2}}
  \put( 24.6353,39.5106){\circle*{0.2}}\put( 24.5106,39.5372){\circle*{0.2}}
  \put( 24.3856,39.563){\circle*{0.2}}\put( 24.2602,39.5882){\circle*{0.2}}
  \put( 24.1346,39.6126){\circle*{0.2}}\put( 24.0086,39.6363){\circle*{0.2}}
  \put( 23.8823,39.6593){\circle*{0.2}}\put( 23.7557,39.6815){\circle*{0.2}}
  \put( 23.6288,39.703){\circle*{0.2}}\put( 23.5017,39.7237){\circle*{0.2}}
  \put( 23.3743,39.7437){\circle*{0.2}}\put( 23.2466,39.763){\circle*{0.2}}
  \put( 23.1187,39.7815){\circle*{0.2}}\put( 22.9905,39.7992){\circle*{0.2}}
  \put( 22.8621,39.8163){\circle*{0.2}}\put( 22.7335,39.8325){\circle*{0.2}}
  \put( 22.6047,39.8481){\circle*{0.2}}\put( 22.4757,39.8629){\circle*{0.2}}
  \put( 22.3465,39.8769){\circle*{0.2}}\put( 22.2171,39.8902){\circle*{0.2}}
  \put( 22.0876,39.9027){\circle*{0.2}}\put( 21.9579,39.9144){\circle*{0.2}}
  \put( 21.828,39.9255){\circle*{0.2}}\put( 21.698,39.9357){\circle*{0.2}}
  \put( 21.5679,39.9452){\circle*{0.2}}\put( 21.4377,39.954){\circle*{0.2}}
  \put( 21.3073,39.9619){\circle*{0.2}}\put( 21.1769,39.9692){\circle*{0.2}}
  \put( 21.0463,39.9756){\circle*{0.2}}\put( 20.9157,39.9813){\circle*{0.2}}
  \put( 20.785,39.9863){\circle*{0.2}}\put( 20.6543,39.9905){\circle*{0.2}}
  \put( 20.5235,39.9939){\circle*{0.2}}\put( 20.3927,39.9966){\circle*{0.2}}
  \put( 20.2618,39.9985){\circle*{0.2}}\put( 20.1309,39.9996){\circle*{0.2}}
  \put( 20.,40.){\circle*{0.2}}\put( 19.8691,39.9996){\circle*{0.2}}
  \put( 19.7382,39.9985){\circle*{0.2}}\put( 19.6073,39.9966){\circle*{0.2}}
  \put( 19.4765,39.9939){\circle*{0.2}}\put( 19.3457,39.9905){\circle*{0.2}}
  \put( 19.215,39.9863){\circle*{0.2}}\put( 19.0843,39.9813){\circle*{0.2}}
  \put( 18.9537,39.9756){\circle*{0.2}}\put( 18.8231,39.9692){\circle*{0.2}}
  \put( 18.6927,39.9619){\circle*{0.2}}\put( 18.5623,39.954){\circle*{0.2}}
  \put( 18.4321,39.9452){\circle*{0.2}}\put( 18.302,39.9357){\circle*{0.2}}
  \put( 18.172,39.9255){\circle*{0.2}}\put( 18.0421,39.9144){\circle*{0.2}}
  \put( 17.9124,39.9027){\circle*{0.2}}\put( 17.7829,39.8902){\circle*{0.2}}
  \put( 17.6535,39.8769){\circle*{0.2}}\put( 17.5243,39.8629){\circle*{0.2}}
  \put( 17.3953,39.8481){\circle*{0.2}}\put( 17.2665,39.8325){\circle*{0.2}}
  \put( 17.1379,39.8163){\circle*{0.2}}\put( 17.0095,39.7992){\circle*{0.2}}
  \put( 16.8813,39.7815){\circle*{0.2}}\put( 16.7534,39.763){\circle*{0.2}}
  \put( 16.6257,39.7437){\circle*{0.2}}\put( 16.4983,39.7237){\circle*{0.2}}
  \put( 16.3712,39.703){\circle*{0.2}}\put( 16.2443,39.6815){\circle*{0.2}}
  \put( 16.1177,39.6593){\circle*{0.2}}\put( 15.9914,39.6363){\circle*{0.2}}
  \put( 15.8654,39.6126){\circle*{0.2}}\put( 15.7398,39.5882){\circle*{0.2}}
  \put( 15.6144,39.563){\circle*{0.2}}\put( 15.4894,39.5372){\circle*{0.2}}
  \put( 15.3647,39.5106){\circle*{0.2}}\put( 15.2404,39.4832){\circle*{0.2}}
  \put( 15.1165,39.4552){\circle*{0.2}}\put( 14.9929,39.4264){\circle*{0.2}}
  \put( 14.8697,39.3969){\circle*{0.2}}\put( 14.7469,39.3667){\circle*{0.2}}
  \put( 14.6245,39.3358){\circle*{0.2}}\put( 14.5025,39.3042){\circle*{0.2}}
  \put( 14.3809,39.2718){\circle*{0.2}}\put( 14.2597,39.2388){\circle*{0.2}}
  \put( 14.139,39.205){\circle*{0.2}}\put( 14.0188,39.1706){\circle*{0.2}}
  \put( 13.899,39.1355){\circle*{0.2}}\put( 13.7796,39.0996){\circle*{0.2}}
  \put( 13.6607,39.0631){\circle*{0.2}}\put( 13.5423,39.0259){\circle*{0.2}}
  \put( 13.4244,38.9879){\circle*{0.2}}\put( 13.307,38.9493){\circle*{0.2}}
  \put( 13.1901,38.9101){\circle*{0.2}}\put( 13.0738,38.8701){\circle*{0.2}}
  \put( 12.9579,38.8295){\circle*{0.2}}\put( 12.8426,38.7882){\circle*{0.2}}
  \put( 12.7279,38.7462){\circle*{0.2}}\put( 12.6136,38.7036){\circle*{0.2}}
  \put( 12.5,38.6603){\circle*{0.2}}\put( 12.3869,38.6163){\circle*{0.2}}
  \put( 12.2744,38.5717){\circle*{0.2}}\put( 12.1625,38.5264){\circle*{0.2}}
  \put( 12.0512,38.4805){\circle*{0.2}}\put( 11.9405,38.4339){\circle*{0.2}}
  \put( 11.8304,38.3867){\circle*{0.2}}\put( 11.7209,38.3389){\circle*{0.2}}
  \put( 11.6121,38.2904){\circle*{0.2}}\put( 11.5039,38.2413){\circle*{0.2}}
  \put( 11.3964,38.1915){\circle*{0.2}}\put( 11.2895,38.1412){\circle*{0.2}}
  \put( 11.1832,38.0902){\circle*{0.2}}\put( 11.0777,38.0386){\circle*{0.2}}
  \put( 10.9728,37.9864){\circle*{0.2}}\put( 10.8686,37.9335){\circle*{0.2}}
  \put( 10.7651,37.8801){\circle*{0.2}}\put( 10.6623,37.8261){\circle*{0.2}}
  \put( 10.5602,37.7715){\circle*{0.2}}\put( 10.4588,37.7162){\circle*{0.2}}
  \put( 10.3582,37.6604){\circle*{0.2}}\put( 10.2583,37.6041){\circle*{0.2}}
  \put( 10.1591,37.5471){\circle*{0.2}}\put( 10.0607,37.4896){\circle*{0.2}}
  \put( 9.96304,37.4314){\circle*{0.2}}\put( 9.86615,37.3728){\circle*{0.2}}
  \put( 9.77002,37.3135){\circle*{0.2}}\put( 9.67468,37.2537){\circle*{0.2}}
  \put( 9.58012,37.1934){\circle*{0.2}}\put( 9.48636,37.1325){\circle*{0.2}}
  \put( 9.3934,37.0711){\circle*{0.2}}\put( 9.30124,37.0091){\circle*{0.2}}
  \put( 9.2099,36.9466){\circle*{0.2}}\put( 9.11938,36.8835){\circle*{0.2}}
  \put( 9.02969,36.82){\circle*{0.2}}\put( 8.94084,36.7559){\circle*{0.2}}
  \put( 8.85283,36.6913){\circle*{0.2}}\put( 8.76566,36.6262){\circle*{0.2}}
  \put( 8.67936,36.5606){\circle*{0.2}}\put( 8.59391,36.4945){\circle*{0.2}}
  \put( 8.50933,36.4279){\circle*{0.2}}\put( 8.42563,36.3608){\circle*{0.2}}
  \put( 8.34281,36.2932){\circle*{0.2}}\put( 8.26088,36.2251){\circle*{0.2}}
  \put( 8.17984,36.1566){\circle*{0.2}}\put( 8.0997,36.0876){\circle*{0.2}}
  \put( 8.02047,36.0182){\circle*{0.2}}\put( 7.94215,35.9482){\circle*{0.2}}
  \put( 7.86475,35.8779){\circle*{0.2}}\put( 7.78827,35.807){\circle*{0.2}}
  \put( 7.71272,35.7358){\circle*{0.2}}\put( 7.63811,35.6641){\circle*{0.2}}
  \put( 7.56444,35.5919){\circle*{0.2}}\put( 7.49171,35.5194){\circle*{0.2}}
  \put( 7.41994,35.4464){\circle*{0.2}}\put( 7.34913,35.373){\circle*{0.2}}
  \put( 7.27928,35.2992){\circle*{0.2}}\put( 7.2104,35.225){\circle*{0.2}}
  \put( 7.14249,35.1504){\circle*{0.2}}\put( 7.07556,35.0754){\circle*{0.2}}
  \put( 7.00962,35.){\circle*{0.2}}\put( 6.94466,34.9242){\circle*{0.2}}
  \put( 6.8807,34.8481){\circle*{0.2}}\put( 6.81774,34.7716){\circle*{0.2}}
  \put( 6.75579,34.6947){\circle*{0.2}}\put( 6.69484,34.6175){\circle*{0.2}}
  \put( 6.6349,34.5399){\circle*{0.2}}\put( 6.57598,34.462){\circle*{0.2}}
  \put( 6.51809,34.3837){\circle*{0.2}}\put( 6.46122,34.3051){\circle*{0.2}}
  \put( 6.40538,34.2262){\circle*{0.2}}\put( 6.35058,34.1469){\circle*{0.2}}
  \put( 6.29682,34.0674){\circle*{0.2}}\put( 6.2441,33.9875){\circle*{0.2}}
  \put( 6.19243,33.9073){\circle*{0.2}}\put( 6.14181,33.8268){\circle*{0.2}}
  \put( 6.09224,33.7461){\circle*{0.2}}\put( 6.04374,33.665){\circle*{0.2}}
  \put( 5.99629,33.5837){\circle*{0.2}}\put( 5.94992,33.5021){\circle*{0.2}}
  \put( 5.90461,33.4202){\circle*{0.2}}\put( 5.86038,33.3381){\circle*{0.2}}
  \put( 5.81722,33.2557){\circle*{0.2}}\put( 5.77515,33.173){\circle*{0.2}}
  \put( 5.73415,33.0902){\circle*{0.2}}\put( 5.69425,33.0071){\circle*{0.2}}
  \put( 5.65543,32.9237){\circle*{0.2}}\put( 5.6177,32.8402){\circle*{0.2}}
  \put( 5.58107,32.7564){\circle*{0.2}}\put( 5.54554,32.6724){\circle*{0.2}}
  \put( 5.51111,32.5882){\circle*{0.2}}\put( 5.47779,32.5038){\circle*{0.2}}
  \put( 5.44556,32.4192){\circle*{0.2}}\put( 5.41445,32.3345){\circle*{0.2}}
  \put( 5.38445,32.2495){\circle*{0.2}}\put( 5.35556,32.1644){\circle*{0.2}}
  \put( 5.32779,32.0791){\circle*{0.2}}\put( 5.30113,31.9937){\circle*{0.2}}
  \put( 5.27559,31.9081){\circle*{0.2}}\put( 5.25118,31.8224){\circle*{0.2}}
  \put( 5.22788,31.7365){\circle*{0.2}}\put( 5.20572,31.6505){\circle*{0.2}}
  \put( 5.18467,31.5643){\circle*{0.2}}\put( 5.16476,31.4781){\circle*{0.2}}
  \put( 5.14598,31.3917){\circle*{0.2}}\put( 5.12833,31.3053){\circle*{0.2}}
  \put( 5.11181,31.2187){\circle*{0.2}}\put( 5.09642,31.132){\circle*{0.2}}
  \put( 5.08217,31.0453){\circle*{0.2}}\put( 5.06906,30.9585){\circle*{0.2}}
  \put( 5.05708,30.8716){\circle*{0.2}}\put( 5.04624,30.7846){\circle*{0.2}}
  \put( 5.03654,30.6976){\circle*{0.2}}\put( 5.02798,30.6105){\circle*{0.2}}
  \put( 5.02056,30.5234){\circle*{0.2}}\put( 5.01428,30.4362){\circle*{0.2}}
  \put( 5.00914,30.349){\circle*{0.2}}\put( 5.00514,30.2618){\circle*{0.2}}
  \put( 5.00228,30.1745){\circle*{0.2}}\put( 5.00057,30.0873){\circle*{0.2}}
  \put( 5.,30.){\circle*{0.2}}\put( 5.00057,29.9127){\circle*{0.2}}
  \put( 5.00228,29.8255){\circle*{0.2}}\put( 5.00514,29.7382){\circle*{0.2}}
  \put( 5.00914,29.651){\circle*{0.2}}\put( 5.01428,29.5638){\circle*{0.2}}
  \put( 5.02056,29.4766){\circle*{0.2}}\put( 5.02798,29.3895){\circle*{0.2}}
  \put( 5.03654,29.3024){\circle*{0.2}}\put( 5.04624,29.2154){\circle*{0.2}}
  \put( 5.05708,29.1284){\circle*{0.2}}\put( 5.06906,29.0415){\circle*{0.2}}
  \put( 5.08217,28.9547){\circle*{0.2}}\put( 5.09642,28.868){\circle*{0.2}}
  \put( 5.11181,28.7813){\circle*{0.2}}\put( 5.12833,28.6947){\circle*{0.2}}
  \put( 5.14598,28.6083){\circle*{0.2}}\put( 5.16476,28.5219){\circle*{0.2}}
  \put( 5.18467,28.4357){\circle*{0.2}}\put( 5.20572,28.3495){\circle*{0.2}}
  \put( 5.22788,28.2635){\circle*{0.2}}\put( 5.25118,28.1776){\circle*{0.2}}
  \put( 5.27559,28.0919){\circle*{0.2}}\put( 5.30113,28.0063){\circle*{0.2}}
  \put( 5.32779,27.9209){\circle*{0.2}}\put( 5.35556,27.8356){\circle*{0.2}}
  \put( 5.38445,27.7505){\circle*{0.2}}\put( 5.41445,27.6655){\circle*{0.2}}
  \put( 5.44556,27.5808){\circle*{0.2}}\put( 5.47779,27.4962){\circle*{0.2}}
  \put( 5.51111,27.4118){\circle*{0.2}}\put( 5.54554,27.3276){\circle*{0.2}}
  \put( 5.58107,27.2436){\circle*{0.2}}\put( 5.6177,27.1598){\circle*{0.2}}
  \put( 5.65543,27.0763){\circle*{0.2}}\put( 5.69425,26.9929){\circle*{0.2}}
  \put( 5.73415,26.9098){\circle*{0.2}}\put( 5.77515,26.827){\circle*{0.2}}
  \put( 5.81722,26.7443){\circle*{0.2}}\put( 5.86038,26.6619){\circle*{0.2}}
  \put( 5.90461,26.5798){\circle*{0.2}}\put( 5.94992,26.4979){\circle*{0.2}}
  \put( 5.99629,26.4163){\circle*{0.2}}\put( 6.04374,26.335){\circle*{0.2}}
  \put( 6.09224,26.2539){\circle*{0.2}}\put( 6.14181,26.1732){\circle*{0.2}}
  \put( 6.19243,26.0927){\circle*{0.2}}\put( 6.2441,26.0125){\circle*{0.2}}
  \put( 6.29682,25.9326){\circle*{0.2}}\put( 6.35058,25.8531){\circle*{0.2}}
  \put( 6.40538,25.7738){\circle*{0.2}}\put( 6.46122,25.6949){\circle*{0.2}}
  \put( 6.51809,25.6163){\circle*{0.2}}\put( 6.57598,25.538){\circle*{0.2}}
  \put( 6.6349,25.4601){\circle*{0.2}}\put( 6.69484,25.3825){\circle*{0.2}}
  \put( 6.75579,25.3053){\circle*{0.2}}\put( 6.81774,25.2284){\circle*{0.2}}
  \put( 6.8807,25.1519){\circle*{0.2}}\put( 6.94466,25.0758){\circle*{0.2}}
  \put( 7.00962,25.){\circle*{0.2}}\put( 7.07556,24.9246){\circle*{0.2}}
  \put( 7.14249,24.8496){\circle*{0.2}}\put( 7.2104,24.775){\circle*{0.2}}
  \put( 7.27928,24.7008){\circle*{0.2}}\put( 7.34913,24.627){\circle*{0.2}}
  \put( 7.41994,24.5536){\circle*{0.2}}\put( 7.49171,24.4806){\circle*{0.2}}
  \put( 7.56444,24.4081){\circle*{0.2}}\put( 7.63811,24.3359){\circle*{0.2}}
  \put( 7.71272,24.2642){\circle*{0.2}}\put( 7.78827,24.193){\circle*{0.2}}
  \put( 7.86475,24.1221){\circle*{0.2}}\put( 7.94215,24.0518){\circle*{0.2}}
  \put( 8.02047,23.9818){\circle*{0.2}}\put( 8.0997,23.9124){\circle*{0.2}}
  \put( 8.17984,23.8434){\circle*{0.2}}\put( 8.26088,23.7749){\circle*{0.2}}
  \put( 8.34281,23.7068){\circle*{0.2}}\put( 8.42563,23.6392){\circle*{0.2}}
  \put( 8.50933,23.5721){\circle*{0.2}}\put( 8.59391,23.5055){\circle*{0.2}}
  \put( 8.67936,23.4394){\circle*{0.2}}\put( 8.76566,23.3738){\circle*{0.2}}
  \put( 8.85283,23.3087){\circle*{0.2}}\put( 8.94084,23.2441){\circle*{0.2}}
  \put( 9.02969,23.18){\circle*{0.2}}\put( 9.11938,23.1165){\circle*{0.2}}
  \put( 9.2099,23.0534){\circle*{0.2}}\put( 9.30124,22.9909){\circle*{0.2}}
  \put( 9.3934,22.9289){\circle*{0.2}}\put( 9.48636,22.8675){\circle*{0.2}}
  \put( 9.58012,22.8066){\circle*{0.2}}\put( 9.67468,22.7463){\circle*{0.2}}
  \put( 9.77002,22.6865){\circle*{0.2}}\put( 9.86615,22.6272){\circle*{0.2}}
  \put( 9.96304,22.5686){\circle*{0.2}}\put( 10.0607,22.5104){\circle*{0.2}}
  \put( 10.1591,22.4529){\circle*{0.2}}\put( 10.2583,22.3959){\circle*{0.2}}
  \put( 10.3582,22.3396){\circle*{0.2}}\put( 10.4588,22.2838){\circle*{0.2}}
  \put( 10.5602,22.2285){\circle*{0.2}}\put( 10.6623,22.1739){\circle*{0.2}}
  \put( 10.7651,22.1199){\circle*{0.2}}\put( 10.8686,22.0665){\circle*{0.2}}
  \put( 10.9728,22.0136){\circle*{0.2}}\put( 11.0777,21.9614){\circle*{0.2}}
  \put( 11.1832,21.9098){\circle*{0.2}}\put( 11.2895,21.8588){\circle*{0.2}}
  \put( 11.3964,21.8085){\circle*{0.2}}\put( 11.5039,21.7587){\circle*{0.2}}
  \put( 11.6121,21.7096){\circle*{0.2}}\put( 11.7209,21.6611){\circle*{0.2}}
  \put( 11.8304,21.6133){\circle*{0.2}}\put( 11.9405,21.5661){\circle*{0.2}}
  \put( 12.0512,21.5195){\circle*{0.2}}\put( 12.1625,21.4736){\circle*{0.2}}
  \put( 12.2744,21.4283){\circle*{0.2}}\put( 12.3869,21.3837){\circle*{0.2}}
  \put( 12.5,21.3397){\circle*{0.2}}\put( 12.6136,21.2964){\circle*{0.2}}
  \put( 12.7279,21.2538){\circle*{0.2}}\put( 12.8426,21.2118){\circle*{0.2}}
  \put( 12.9579,21.1705){\circle*{0.2}}\put( 13.0738,21.1299){\circle*{0.2}}
  \put( 13.1901,21.0899){\circle*{0.2}}\put( 13.307,21.0507){\circle*{0.2}}
  \put( 13.4244,21.0121){\circle*{0.2}}\put( 13.5423,20.9741){\circle*{0.2}}
  \put( 13.6607,20.9369){\circle*{0.2}}\put( 13.7796,20.9004){\circle*{0.2}}
  \put( 13.899,20.8645){\circle*{0.2}}\put( 14.0188,20.8294){\circle*{0.2}}
  \put( 14.139,20.795){\circle*{0.2}}\put( 14.2597,20.7612){\circle*{0.2}}
  \put( 14.3809,20.7282){\circle*{0.2}}\put( 14.5025,20.6958){\circle*{0.2}}
  \put( 14.6245,20.6642){\circle*{0.2}}\put( 14.7469,20.6333){\circle*{0.2}}
  \put( 14.8697,20.6031){\circle*{0.2}}\put( 14.9929,20.5736){\circle*{0.2}}
  \put( 15.1165,20.5448){\circle*{0.2}}\put( 15.2404,20.5168){\circle*{0.2}}
  \put( 15.3647,20.4894){\circle*{0.2}}\put( 15.4894,20.4628){\circle*{0.2}}
  \put( 15.6144,20.437){\circle*{0.2}}\put( 15.7398,20.4118){\circle*{0.2}}
  \put( 15.8654,20.3874){\circle*{0.2}}\put( 15.9914,20.3637){\circle*{0.2}}
  \put( 16.1177,20.3407){\circle*{0.2}}\put( 16.2443,20.3185){\circle*{0.2}}
  \put( 16.3712,20.297){\circle*{0.2}}\put( 16.4983,20.2763){\circle*{0.2}}
  \put( 16.6257,20.2563){\circle*{0.2}}\put( 16.7534,20.237){\circle*{0.2}}
  \put( 16.8813,20.2185){\circle*{0.2}}\put( 17.0095,20.2008){\circle*{0.2}}
  \put( 17.1379,20.1837){\circle*{0.2}}\put( 17.2665,20.1675){\circle*{0.2}}
  \put( 17.3953,20.1519){\circle*{0.2}}\put( 17.5243,20.1371){\circle*{0.2}}
  \put( 17.6535,20.1231){\circle*{0.2}}\put( 17.7829,20.1098){\circle*{0.2}}
  \put( 17.9124,20.0973){\circle*{0.2}}\put( 18.0421,20.0856){\circle*{0.2}}
  \put( 18.172,20.0745){\circle*{0.2}}\put( 18.302,20.0643){\circle*{0.2}}
  \put( 18.4321,20.0548){\circle*{0.2}}\put( 18.5623,20.046){\circle*{0.2}}
  \put( 18.6927,20.0381){\circle*{0.2}}\put( 18.8231,20.0308){\circle*{0.2}}
  \put( 18.9537,20.0244){\circle*{0.2}}\put( 19.0843,20.0187){\circle*{0.2}}
  \put( 19.215,20.0137){\circle*{0.2}}\put( 19.3457,20.0095){\circle*{0.2}}
  \put( 19.4765,20.0061){\circle*{0.2}}\put( 19.6073,20.0034){\circle*{0.2}}
  \put( 19.7382,20.0015){\circle*{0.2}}\put( 19.8691,20.0004){\circle*{0.2}}
  \put( 20.,20.){\circle*{0.2}}\put( 20.1309,20.0004){\circle*{0.2}}
  \put( 20.2618,20.0015){\circle*{0.2}}\put( 20.3927,20.0034){\circle*{0.2}}
  \put( 20.5235,20.0061){\circle*{0.2}}\put( 20.6543,20.0095){\circle*{0.2}}
  \put( 20.785,20.0137){\circle*{0.2}}\put( 20.9157,20.0187){\circle*{0.2}}
  \put( 21.0463,20.0244){\circle*{0.2}}\put( 21.1769,20.0308){\circle*{0.2}}
  \put( 21.3073,20.0381){\circle*{0.2}}\put( 21.4377,20.046){\circle*{0.2}}
  \put( 21.5679,20.0548){\circle*{0.2}}\put( 21.698,20.0643){\circle*{0.2}}
  \put( 21.828,20.0745){\circle*{0.2}}\put( 21.9579,20.0856){\circle*{0.2}}
  \put( 22.0876,20.0973){\circle*{0.2}}\put( 22.2171,20.1098){\circle*{0.2}}
  \put( 22.3465,20.1231){\circle*{0.2}}\put( 22.4757,20.1371){\circle*{0.2}}
  \put( 22.6047,20.1519){\circle*{0.2}}\put( 22.7335,20.1675){\circle*{0.2}}
  \put( 22.8621,20.1837){\circle*{0.2}}\put( 22.9905,20.2008){\circle*{0.2}}
  \put( 23.1187,20.2185){\circle*{0.2}}\put( 23.2466,20.237){\circle*{0.2}}
  \put( 23.3743,20.2563){\circle*{0.2}}\put( 23.5017,20.2763){\circle*{0.2}}
  \put( 23.6288,20.297){\circle*{0.2}}\put( 23.7557,20.3185){\circle*{0.2}}
  \put( 23.8823,20.3407){\circle*{0.2}}\put( 24.0086,20.3637){\circle*{0.2}}
  \put( 24.1346,20.3874){\circle*{0.2}}\put( 24.2602,20.4118){\circle*{0.2}}
  \put( 24.3856,20.437){\circle*{0.2}}\put( 24.5106,20.4628){\circle*{0.2}}
  \put( 24.6353,20.4894){\circle*{0.2}}\put( 24.7596,20.5168){\circle*{0.2}}
  \put( 24.8835,20.5448){\circle*{0.2}}\put( 25.0071,20.5736){\circle*{0.2}}
  \put( 25.1303,20.6031){\circle*{0.2}}\put( 25.2531,20.6333){\circle*{0.2}}
  \put( 25.3755,20.6642){\circle*{0.2}}\put( 25.4975,20.6958){\circle*{0.2}}
  \put( 25.6191,20.7282){\circle*{0.2}}\put( 25.7403,20.7612){\circle*{0.2}}
  \put( 25.861,20.795){\circle*{0.2}}\put( 25.9812,20.8294){\circle*{0.2}}
  \put( 26.101,20.8645){\circle*{0.2}}\put( 26.2204,20.9004){\circle*{0.2}}
  \put( 26.3393,20.9369){\circle*{0.2}}\put( 26.4577,20.9741){\circle*{0.2}}
  \put( 26.5756,21.0121){\circle*{0.2}}\put( 26.693,21.0507){\circle*{0.2}}
  \put( 26.8099,21.0899){\circle*{0.2}}\put( 26.9262,21.1299){\circle*{0.2}}
  \put( 27.0421,21.1705){\circle*{0.2}}\put( 27.1574,21.2118){\circle*{0.2}}
  \put( 27.2721,21.2538){\circle*{0.2}}\put( 27.3864,21.2964){\circle*{0.2}}
  \put( 27.5,21.3397){\circle*{0.2}}\put( 27.6131,21.3837){\circle*{0.2}}
  \put( 27.7256,21.4283){\circle*{0.2}}\put( 27.8375,21.4736){\circle*{0.2}}
  \put( 27.9488,21.5195){\circle*{0.2}}\put( 28.0595,21.5661){\circle*{0.2}}
  \put( 28.1696,21.6133){\circle*{0.2}}\put( 28.2791,21.6611){\circle*{0.2}}
  \put( 28.3879,21.7096){\circle*{0.2}}\put( 28.4961,21.7587){\circle*{0.2}}
  \put( 28.6036,21.8085){\circle*{0.2}}\put( 28.7105,21.8588){\circle*{0.2}}
  \put( 28.8168,21.9098){\circle*{0.2}}\put( 28.9223,21.9614){\circle*{0.2}}
  \put( 29.0272,22.0136){\circle*{0.2}}\put( 29.1314,22.0665){\circle*{0.2}}
  \put( 29.2349,22.1199){\circle*{0.2}}\put( 29.3377,22.1739){\circle*{0.2}}
  \put( 29.4398,22.2285){\circle*{0.2}}\put( 29.5412,22.2838){\circle*{0.2}}
  \put( 29.6418,22.3396){\circle*{0.2}}\put( 29.7417,22.3959){\circle*{0.2}}
  \put( 29.8409,22.4529){\circle*{0.2}}\put( 29.9393,22.5104){\circle*{0.2}}
  \put( 30.037,22.5686){\circle*{0.2}}\put( 30.1339,22.6272){\circle*{0.2}}
  \put( 30.23,22.6865){\circle*{0.2}}\put( 30.3253,22.7463){\circle*{0.2}}
  \put( 30.4199,22.8066){\circle*{0.2}}\put( 30.5136,22.8675){\circle*{0.2}}
  \put( 30.6066,22.9289){\circle*{0.2}}\put( 30.6988,22.9909){\circle*{0.2}}
  \put( 30.7901,23.0534){\circle*{0.2}}\put( 30.8806,23.1165){\circle*{0.2}}
  \put( 30.9703,23.18){\circle*{0.2}}\put( 31.0592,23.2441){\circle*{0.2}}
  \put( 31.1472,23.3087){\circle*{0.2}}\put( 31.2343,23.3738){\circle*{0.2}}
  \put( 31.3206,23.4394){\circle*{0.2}}\put( 31.4061,23.5055){\circle*{0.2}}
  \put( 31.4907,23.5721){\circle*{0.2}}\put( 31.5744,23.6392){\circle*{0.2}}
  \put( 31.6572,23.7068){\circle*{0.2}}\put( 31.7391,23.7749){\circle*{0.2}}
  \put( 31.8202,23.8434){\circle*{0.2}}\put( 31.9003,23.9124){\circle*{0.2}}
  \put( 31.9795,23.9818){\circle*{0.2}}\put( 32.0579,24.0518){\circle*{0.2}}
  \put( 32.1353,24.1221){\circle*{0.2}}\put( 32.2117,24.193){\circle*{0.2}}
  \put( 32.2873,24.2642){\circle*{0.2}}\put( 32.3619,24.3359){\circle*{0.2}}
  \put( 32.4356,24.4081){\circle*{0.2}}\put( 32.5083,24.4806){\circle*{0.2}}
  \put( 32.5801,24.5536){\circle*{0.2}}\put( 32.6509,24.627){\circle*{0.2}}
  \put( 32.7207,24.7008){\circle*{0.2}}\put( 32.7896,24.775){\circle*{0.2}}
  \put( 32.8575,24.8496){\circle*{0.2}}\put( 32.9244,24.9246){\circle*{0.2}}
  \put( 32.9904,25.){\circle*{0.2}}\put( 33.0553,25.0758){\circle*{0.2}}
  \put( 33.1193,25.1519){\circle*{0.2}}\put( 33.1823,25.2284){\circle*{0.2}}
  \put( 33.2442,25.3053){\circle*{0.2}}\put( 33.3052,25.3825){\circle*{0.2}}
  \put( 33.3651,25.4601){\circle*{0.2}}\put( 33.424,25.538){\circle*{0.2}}
  \put( 33.4819,25.6163){\circle*{0.2}}\put( 33.5388,25.6949){\circle*{0.2}}
  \put( 33.5946,25.7738){\circle*{0.2}}\put( 33.6494,25.8531){\circle*{0.2}}
  \put( 33.7032,25.9326){\circle*{0.2}}\put( 33.7559,26.0125){\circle*{0.2}}
  \put( 33.8076,26.0927){\circle*{0.2}}\put( 33.8582,26.1732){\circle*{0.2}}
  \put( 33.9078,26.2539){\circle*{0.2}}\put( 33.9563,26.335){\circle*{0.2}}
  \put( 34.0037,26.4163){\circle*{0.2}}\put( 34.0501,26.4979){\circle*{0.2}}
  \put( 34.0954,26.5798){\circle*{0.2}}\put( 34.1396,26.6619){\circle*{0.2}}
  \put( 34.1828,26.7443){\circle*{0.2}}\put( 34.2249,26.827){\circle*{0.2}}
  \put( 34.2658,26.9098){\circle*{0.2}}\put( 34.3058,26.9929){\circle*{0.2}}
  \put( 34.3446,27.0763){\circle*{0.2}}\put( 34.3823,27.1598){\circle*{0.2}}
  \put( 34.4189,27.2436){\circle*{0.2}}\put( 34.4545,27.3276){\circle*{0.2}}
  \put( 34.4889,27.4118){\circle*{0.2}}\put( 34.5222,27.4962){\circle*{0.2}}
  \put( 34.5544,27.5808){\circle*{0.2}}\put( 34.5855,27.6655){\circle*{0.2}}
  \put( 34.6156,27.7505){\circle*{0.2}}\put( 34.6444,27.8356){\circle*{0.2}}
  \put( 34.6722,27.9209){\circle*{0.2}}\put( 34.6989,28.0063){\circle*{0.2}}
  \put( 34.7244,28.0919){\circle*{0.2}}\put( 34.7488,28.1776){\circle*{0.2}}
  \put( 34.7721,28.2635){\circle*{0.2}}\put( 34.7943,28.3495){\circle*{0.2}}
  \put( 34.8153,28.4357){\circle*{0.2}}\put( 34.8352,28.5219){\circle*{0.2}}
  \put( 34.854,28.6083){\circle*{0.2}}\put( 34.8717,28.6947){\circle*{0.2}}
  \put( 34.8882,28.7813){\circle*{0.2}}\put( 34.9036,28.868){\circle*{0.2}}
  \put( 34.9178,28.9547){\circle*{0.2}}\put( 34.9309,29.0415){\circle*{0.2}}
  \put( 34.9429,29.1284){\circle*{0.2}}\put( 34.9538,29.2154){\circle*{0.2}}
  \put( 34.9635,29.3024){\circle*{0.2}}\put( 34.972,29.3895){\circle*{0.2}}
  \put( 34.9794,29.4766){\circle*{0.2}}\put( 34.9857,29.5638){\circle*{0.2}}
  \put( 34.9909,29.651){\circle*{0.2}}\put( 34.9949,29.7382){\circle*{0.2}}
  \put( 34.9977,29.8255){\circle*{0.2}}\put( 34.9994,29.9127){\circle*{0.2}}
  \put( 35.,30.){\circle*{0.2}}
\end{picture}
\caption{ $\alpha$-
and $\beta$- cycles.
}\label{cycles1}
\end{center}
\end{figure}
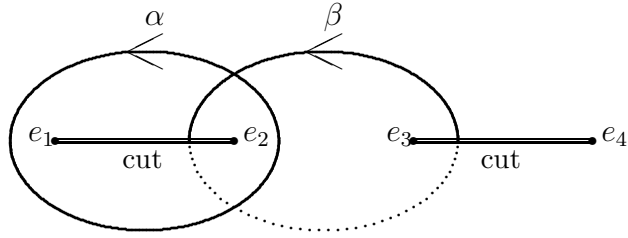

The Seiberg-Witten curve and the Seiberg-Witten differential are 
written as follows
\cite{HaOz}
\begin{equation}
\label{lam}
y^2=F(x)^2-H(x),
\hskip 0.4 in
\lambda_{SW}=\frac{1}{2\pi i}\frac{xdx}{y}
\Big[-F'(x)+\frac{F(x)H'(x)}{2H(x)}\Big],
\end{equation}
where $'$ denotes the differentiation with respect to $x$.
The functions $F(x), \, H(x)$ are shown in Table \ref{swcurve}. 
The Seiberg-Witten differential
$\lambda_{SW}$ is a meromorphic one-form  determined so that it satisfies
$\partial_u \lambda_{SW}\propto\frac{dx}{y}$. 
Then
$a$ (resp. $a_D$) is represented as a period integral 
of $\lambda_{SW}$ along the $\alpha$- (resp. $\beta$-) cycle
\begin{equation}
\label{period}
a=\oint_{\alpha}\lambda_{SW},\qquad
a_D=\oint_{\beta}\lambda_{SW}.
\end{equation}
Here we specify the $\alpha$- and $\beta$-cycles.
We have four branching points of the curve at
$x=e_1,e_2,e_3,e_4$ with $e_i$ as follows
(for $N_f=0$, $e_1=-\sqrt{u+\Lambda^4}$, $e_2=-\sqrt{u-\Lambda^4}$,
$e_3=\sqrt{u-\Lambda^4}$, $e_4=\sqrt{u+\Lambda^4}$)
\begin{equation}
{
\begin{bmatrix}e_1\\e_2\end{bmatrix}}
=-\sqrt{u}\mp\frac{1}{2}\Big({\frac{H(-\sqrt{u})}{u}}\Big)^{\frac{1}{2}}
+{\cal O}(\Lambda^{4-N_f}),
\quad
\begin{bmatrix}
e_3\\e_4
\end{bmatrix}
=\sqrt{u}\mp\frac{1}{2}\Big({\frac{H(\sqrt{u})}{u}}\Big)^{\frac{1}{2}}
+{\cal O}(\Lambda^{4-N_f}),
\end{equation}
which give rise to the cuts to run from $e_1$ to $e_2$ and $e_3$ to $e_4$.
Then $\alpha$- (resp. $\beta$-) cycle is chosen to be 
a loop going around a pair of the points $e_1$ and $e_2$ 
(resp. $e_2$ and $e_3$)
counterclockwise as shown in Figure \ref{cycles1}.

{\it Massless case}

First consider the case where all mass parameters are zero.
We can tell the 
behavior of the periods  at the weak coupling region $|u|\to \infty$
to the lowest order in $\Lambda$,
\begin{equation}\label{aad}
a\sim\sqrt{u},
\qquad
a_D
\sim \frac{i(4-N_f)}{2\pi }\sqrt{u}
\Big[
\log \frac{u}{\Lambda^2}-2+\mbox{const.} \, a
\Big],
\end{equation}
by evaluating the period integrals (\ref{period}).
Meanwhile
$a$ and $a_D$ are solutions of 
the Picard-Fuchs differential equation \cite{ItYa, KlLeTh, KlLeThYa}
\begin{equation}\label{pfeq}
\Big[
(\theta_{z}+\alpha)^2-
z\theta_{z}(\theta_{z}+2\alpha)\Big]
\Big(
\begin{array}{l}
~a\\a_D
\end{array}
\Big)
=0.
\end{equation}
The values of the constant $\alpha$ and the variable $z$ 
are given in  Table \ref{swcurve}.
This is an ordinary differential equation of second order
with only regular singular points at $z=0,1,\infty$.
It has therefore two solutions. Around $z=0$, they are
\begin{eqnarray}
&&w_1(z)= w(z;0)=z^{-\alpha}\sum_{n=0}^{\infty}
\frac{(\alpha)_n(-\alpha)_n}{(n!)^2}z^n,
\quad \mbox{with } \quad w(z;\rho)=
\sum_{n=0}^{\infty}
\frac{(\alpha+\rho)_n(-\alpha+\rho)_n}{((1+\rho)_n)^2}z^{n+\rho-\alpha},
\nonumber\\
&&w_2(z)=\partial_{\rho} w(z;\rho)|_{\rho=0}
=w_1\log z+z^{-\alpha}\sum_{n=1}^{\infty}
\frac{(\alpha)_n(-\alpha)_n}{(n!)^2}z^n
\\&&~~~~~~~~~~~~~~~~~~~~~~~~~~~~~~~\times
\Big[
\psi(\alpha+n)+\psi(-\alpha+n)-2\psi(n+1)-\psi(\alpha)-\psi(-\alpha)+
2\psi(1)
\Big],\nonumber
\label{temp4}
\end{eqnarray}
Comparing the solutions $w_1(z)$, $w_2(z)$  and 
$a$, $a_D$ in (\ref{aad}), we obtain
$
a=C^{\alpha}\Lambda w_1(z)$, $
a_D=\frac{i(4-N_f)}{2\pi}C^{\alpha}\Lambda 
\Big[-2\alpha
w_2(z)+
\mbox{ const. }
w_1(z)
\Big]$
where $C$ is defined by
$z=C(\frac{u}{\Lambda^2})^{\frac{1}{2\alpha}}$. 

Substituting the inverted series of $a$ into $a_D$ and
then integrating it by $a$, we obtain the prepotential. 
Since $u$ is expanded in terms of $(\frac{\Lambda}{a})^{\frac{n}{\alpha}}$,
so is the instanton correction part of the prepotential.
Therefore the form of the prepotential 
must be
\begin{equation}\label{instantonexpansion}
{\cal F}_{gauge}=\frac{i(4-N_f)}{4\pi}a^2
\Big[
\log\frac{a^2}{\Lambda^2}-\sum_{n=1}^{\infty}{\cal F}_n
\Big(\frac{\Lambda}{a}\Big)^{\frac{n}{\alpha}}
\Big].
\end{equation}
The instanton amplitude ${\cal F}_n$ is shown in Table
\ref{instantonnumber4}. 
\begin{table}[htop]
\begin{equation*}
\begin{array}{|r|l|}\hline
n&{\cal F}_n\\\hline
1&\frac{1}{16}~~~~~~~~~~~~~~~~~~~N_f=0\\
2&\frac{5}{2048}\\
3&\frac{3}{8192}\\
4&\frac{1469}{16777216}\\
5&\frac{4471}{167772160}\\
6&\frac{40397}{4294967296}\\
7&\frac{441325}{120259084288}\\
8&\frac{866589165}{562949953421312}\\
\hline
1&-\frac{1}{512}~~~~~~~~~~~~~~~~~ N_f=1\\
2&\frac{51}{4194304}\\
3&-\frac{385}{1610612736}\\
4&\frac{1016295}{140737488355328}\\
5&-\frac{2466711}{9007199254740992}\\
6&\frac{1765076797}{147573952589676412928}\\
7&-\frac{38024186031}{66113130760175032991744}\\
8&\frac{2349612953305695}{79228162514264337593543950336}\\
\hline    
\end{array}
\begin{array}{|r|l|}\hline
n&{\cal F}_n\\\hline
1&\frac{1}{1024}~~~~~~~~~~~~~~~~~~~~~ N_f=2\\
2&\frac{5}{8388608}\\
3&\frac{3}{2147483648}\\
4&\frac{1469}{281474976710656}\\
5&\frac{4471}{180143985094819840}\\
6&\frac{40397}{295147905179352825856}\\
7&\frac{441325}{528905046081400263933952}\\
8&\frac{866589165}{158456325028528675187087900672}\\
\hline
1&0~~~~~~~~~~~~~~~~~~~~~~~~ N_f=3\\
2&\frac{1}{4194304}\\
3&0\\
4&\frac{5}{140737488355328}\\
5&0\\
6&\frac{3}{147573952589676412928}\\
7&0\\
8&\frac{1469}{79228162514264337593543950336}\\
\hline
\end{array}
\end{equation*}
\caption{Instanton amplitude ${\cal F}_n$ with small instanton number $n$
$(1\leq n\leq 8)$.}
\label{instantonnumber4}
\end{table}

\vspace*{5mm}

{\it Massive case }

When  values of the mass parameters are  generic,
the analysis of the instanton correction 
is not as simple as that in the massless case.
Already there exist much effort to obtain the prepotential in the 
massive case \cite{Ohta, MaSu, IsMuNuSh}.
Rather than review these results,
we will only remark on two points 
that will be of use later; 
on  differential equations satisfied by  $(a,a_D)$ 
and on general structure of the prepotential.

The extension of the Picard-Fuchs equation (\ref{pfeq}) to the massive case 
$[p(u,m_i)\partial_u^2+
q(u,m_i)\partial_u+r(u,m_i)](a,a_D)=0$
has been known \cite{Ohta, MaSu, IsMuNuSh, Eguchi}.
This should be regarded as an ordinary differential equation 
with one variable $u$ keeping the
mass parameters $m_i$ $(1\leq i\leq N_f)$ constant. 
This equation, in principle, allows us to obtain $a$ and $a_D$,
and actually the instanton correction to some low order in $\Lambda$
has been calculated. 
However, the complicated expression of 
$p(u,m_i),q(u,m_i),r(u.m_i)$ makes it difficult 
to solve the equation completely.

As it turns out, we could obtain exact solutions
from a different approach. 
In section \ref{geom}, we will show that
the geometric engineering of the gauge theory 
provides a set of partial differential equations
that $a$ and $a_D$ must satisfy. 
There, the mass parameters $m_i$ as well will be treated as variables.  
We will solve the equations and obtain expressions
of $a$ and $a_D$ for generic values of the mass parameters.
Moreover, we will obtain
an ordinary differential operator with respect to $u$
and it
is consistent with the known results
and the massless Picard-Fuchs equation (\ref{pfeq}).

Next, we turn to general structure of the 
prepotential. 
The parameters in the case of $N_f$ and that of $N_f-1$  
are related by the following limit corresponding to 
the decoupling of a hypermultiplet:
\begin{equation}
\label{decouple}
m_{N_f}\rightarrow \infty,\hskip 0.2 in 
m_{N_f}\Lambda^{4-N_f}\rightarrow \Lambda^{4-(N_f-1)}.
\end{equation}
In the second relation, $\Lambda$ in the left-hand side 
is that of the gauge theory with $N_f$ hypermultiplets and 
one in the right-hand side is that of the gauge theory with 
$N_f-1$ hypermultiplets.
This relation imposes a very strong constraint
on the possible form of $u$ as a inverse series of $a$ and 
on that of $\partial_a^3{\cal F}_{gauge}$. 

First, $u$ must take the form
\begin{equation}
\label{formofa}
\begin{split}
u&=
a^2\, \Big[
1+\sum_{n=1}^{\infty}\, b_n
\, \cdot\, 
\Big(\frac{\Lambda}{a}\Big)^{(4-N_f)n}
\Big], 
\\
b_n& : \quad 
\left(
\begin{array}{l}
\mbox{polynomial of degree $n$ in each $\frac{m_i}{a}$,}\\
\mbox{symmetric with respect to 
$\frac{m_1}{a},\cdots,\frac{m_{N_f}}{a}$.}
\end{array}\right).
\end{split}
\end{equation} 
The reason for the form of $b_n$ is clear for $N_f=0$ 
because $\frac{u}{a^2}$ is a series in $(\frac{\Lambda}{a})^4$,
see (\ref{temp4}).
For $N_f\geq 1$, this form is determined so that 
the instanton correction comes in the power of $\Lambda^{4-N_f}$
and the terms of 
${\Lambda}^{4-N_f}$ descend to those of $\Lambda^{5-N_f}$
at the limit (\ref{decouple}) of the decoupling of a hypermultiplet.

Second,
$\partial_a^3{\cal F}_{gauge}$ must take the following form
\begin{equation}
\label{formofyukawa}
\begin{split}
\partial_a^3{\cal F}_{gauge}&\propto
-{{8}\over{a}}+\sum_{i=1}^{N_f}
{
\Big(
{{1}\over{a+m_i}}
+
{{1}\over{a-m_i}}
\Big)
}
-{{8}\over{a}}\sum_{n=1}^{\infty}P_n\cdot
\Big({{\Lambda}\over{a}}\Big)^{(4-N_f)n}
,\\
P_n&=B_n\, \prod_{i=1}^{N_f} \,
\Big({{m_i}\over{a}}\Big)^n+\cdots \quad : \quad 
\left(
\begin{array}{l}
\mbox{polynomial of degree $n$ in each $\frac{m_i}{a}$,}\\
\mbox{symmetric with respect to 
$\frac{m_1}{a},\cdots,\frac{m_{N_f}}{a}$.}
\end{array}\right).
\end{split}
\end{equation}
Note that the terms $\frac{1}{a},\frac{1}{a\pm m_i}$ 
correspond to the one-loop corrected terms 
while the
higher order terms in 
$\Lambda$ correspond to contributions of the instanton correction.
Here $B_n$ is a number common to all the cases with $N_f=0,1,2,3$. 
And it is written in terms of the instanton amplitude 
${\cal F}_n$ of the gauge theory without hypermultiplets ($N_f=0$)
in (\ref{instantonexpansion}) as follows
\begin{equation}
B_n=\frac{4n(4n-1)(4n-2){\cal F}_n}{4}.
\end{equation}
The reason for this form
is the following.
Consider first the case of $N_f=0$.
From the Picard-Fuchs equation 
(\ref{pfeq}), 
we can obtain the relation
$\partial_a^3 {\cal F}_{gauge}\propto(\frac{du}{da})^3\frac{1}{u^2-\Lambda^4}$
(derived from the argument in next paragraph).
Considering that
$\frac{u}{a^2}$ 
is a series in $(\frac{\Lambda}{a})^4$, 
$\partial_a^3 {\cal F}_{gauge}$ must be  proportional to
$\frac{1}{a}\times
\left[\mbox{ series in }(\frac{\Lambda}{a})^{4}\right]$.
Thus (\ref{formofyukawa}) is proved for the case with $N_f=0$.
We denote by $B_n$ the coefficient of $\frac{1}{a}(\frac{\Lambda}{a})^{4n}$
divided by that of $\frac{1}{a}$.
Next consider the case with $N_f\geq 1$.
The one-loop correction gives rise to not only the term
$a^2\log \frac{a}{\Lambda}$
but also the terms 
$(a\pm m_i)^2\log \frac{a\pm m_i}{\Lambda}$
because a massless particle also appears at 
$a=\pm m_i$.
This comes from the terms in the Lagrangian,
$m_i\widetilde{Q}_iQ_i+\widetilde{Q}_i\Phi Q_i$
with the $N=1$ chiral superfield $\Phi$ in the $N=2$ vector multiplet
and two $N=1$ chiral superfields $Q_i, \widetilde{Q}_i$ $(1\leq i \leq N_f)$
in the $N=2$ hypermultiplets.
The ratio of coefficients of $\frac{1}{a}$ and $\frac{1}{a\pm m_i}$
are determined as follows:
if we assume the ratio to be $1$ to $r$, then 
the prepotential at the massless limit is proportional to 
$
(1+2r)
a^2 \log a+\cdots
$. 
Comparison with  (\ref{instantonexpansion}) gives $r=-\frac{1}{8}$.
As to the instanton correction term, 
it must take the form above by the consistency with the decoupling limit
of a hypermultiplet.

We note the identity
\begin{equation}
\label{gaugeyukawa}
\partial_a^3{\cal F}_{gauge}=
\Big(\frac{da}{du}\Big)^{-3}\Big(
\frac{d^2a_D}{du^2}
\frac{da}{du}-\frac{d^2a}{du^2}\frac{da_D}{du}
\Big).
\end{equation}
The second factor in the right hand side is the Wronskian of 
$\partial_u a$ and $\partial_u a_D$.
If we knew the 
differential operator that annihilates $(a,a_D)$, 
we would have the differential operator of the form
\begin{equation}
\label{equationpqr}
{\cal P}={P}(u,m_i)\partial_u^3+{Q}(u,m_i)\partial_u^2+
{R}(u,m_i)\partial_u,
\end{equation}
to provide the Wronskian as follows
\begin{equation}
\frac{d^2a_D}{du^2}\frac{da}{du}-\frac{d^2a}{du^2}\frac{da_D}{du}
\propto e^{-\int du \frac{Q}{P}}.
\end{equation}
This follows from the differential equation
${\partial_u a \cal P}a_D-\partial_u a_D{\cal P}a=
(P\partial_u+Q)(\partial_ua \partial_u^2a_D-
\partial_u a_D \partial_u^2a)=0$.
We will actually derive the differential operator in section \ref{geom}.
Then we will obtain expressions of   
 $u$ and $\partial_a^3{\cal F}_{gauge}$ for generic values of the
mass parameters to some low orders in $\Lambda$, and
find that they actually satisfy (\ref{formofa}) and (\ref{formofyukawa}).

\subsection{Asymptotic Form of Instanton Amplitude}
\hspace{5mm}
Now
let us return to the massless case.
We could see 
from Table \ref{instantonnumber4}
that
the instanton amplitude ${\cal F}_n$
decreases rapidly as $n$ increases.
Such distribution of the instanton amplitude 
turns out to be  governed by a
singularity of the moduli space at $z=1$.
We will derive the asymptotic form of the instanton amplitude 
${\cal F}_n$ by the analysis of $\partial_a^3{\cal F}_{gauge}$
around $z=1$.

Before proceeding, we must remark that 
our analysis was inspired by and is almost the same as
that of the number of rational curves of large degree
in the quintic hypersurface in ${\bf P}^4$ \cite{CadeGrPa}.
The apparent difference that our analysis is on  
 the instanton amplitude of the gauge theory,
and the one in \cite{CadeGrPa} is on the world-sheet instanton 
number of the quintic,
does not matter here. It is  because 
quantities used in the analysis are determined only by a
generalized hypergeometric differential equation,
where similarity between the two system lies.
 
Note that the third derivative of the prepotential can be written as follows
\begin{equation}\label{triple}
\partial_a^3{\cal F}_{gauge}=\frac{C_2}{C_1^2\Lambda}
\Big(\frac{dw_1}{dz}\Big)^{-3}
\frac{\alpha^2}{(1-z)z^{2\alpha+3}},
\end{equation}
where $C_1$,(resp. $C_2$) is the coefficient of $w_1(z)$ (resp. $w_2(z)$)
in $a$ (resp. $a_D$) divided by $\Lambda$.
The last factor in the right-hand side has
come from the Wronskian of $\partial_u a$ and $\partial_u a_D$. 
Together with 
the behavior of 
$w_1(z)$ around $z=1$
\begin{equation}
w_1(z)\sim 
\frac{\sin\pi\alpha}{\pi\alpha}\big[
1+\alpha^2 z'\log z'+\cdots
\big],\quad
z':=\frac{z-1}{z},
\end{equation}
we can see that (\ref{triple})
diverges at $z=1$. Therefore
the radius of convergence of the instanton
expansion (\ref{instantonexpansion}) is determined by the 
value of $a$ at $z=1$.
The instanton expansion converges
on the domain $|\frac{\Lambda}{a(z)}|<|\frac{\Lambda}{a(1)}|$
and, by the theorem of Hadamard,
the asymptotic form of ${\cal F}_n$ should satisfy
$\lim_{n\rightarrow \infty} {^n\sqrt{{\cal F}_n}}=
|\frac{\Lambda}{a(1)}|^{\frac{1}{\alpha}}$.

To obtain a little  more elaborate asymptotic form of 
${\cal F}_n$ for large 
$n$, let us  adopt the ansatz similar to the one in \cite{CadeGrPa}
\begin{equation}
{(\beta n-2)(\beta n-1)(\beta n)}
{\cal F}_n\sim
Bn^{\lambda}(\log n)^{\mu}\left|\frac{a(1)}{\Lambda}\right|^{\beta n}
\qquad
(n\gg 1),
\end{equation}
with three constants  to be determined, $\lambda, \mu, B$.
$\beta$ is $\frac{1}{\alpha}=4,6,4$ for $N_f=0,1,2$.
For $N_f=3$, the instanton expansion is actually an
expansion by $(\frac{\Lambda}{a})^4$ rather than $(\frac{\Lambda}{a})^2$,
and we should redefine ${\cal F}_{2n}$ as ${\cal F}_n$ and set $\beta=4$.
Substituting this into (\ref{instantonexpansion}), differentiating by 
$a$ three times and evaluating it around $z'=0$,
we obtain
\begin{equation}
\partial_a^3{\cal F}_{gauge}\sim\frac{i(4-N_f)}{4\pi  a}
\frac{B\Gamma(\lambda+1)(-\log z')^{\mu}}
{(\alpha^2\beta)^{\lambda+1}(z'\log z')^{\lambda+1}},
\qquad z'\sim 0.
\end{equation}
In the process we have 
replaced the summation in $n$ with an integration
and
changed the variable $n$ into $t:=n \beta\log\frac{a(z)}{|a(1)|}$.
On the other hand, near $z'=0$, the expression (\ref{triple})
becomes 
$\frac{C_2}{C_1^2\Lambda}
\Big(\frac{\pi\alpha}{\sin\pi\alpha}\Big)^3
\frac{1}{\alpha^4}\frac{-1}{z'(\log z')^3}$.
Comparing these, we could obtain
$\lambda=0$, $\mu=-2$ and $B=\frac{2\beta}{\alpha}
(\frac{\pi\alpha}{\sin\pi\alpha})^2$.

Therefore we conclude that 
the asymptotic form of the instanton amplitude is as follows
\begin{equation}\label{rn}
{\cal F}_n\sim\frac{4}{\alpha\beta^2}
\Big(\frac{\pi\alpha}{\sin \pi\alpha}\Big)^2
\Big(\frac{|C^{\alpha}| \sin\pi\alpha}{\pi\alpha}\Big)^{\beta n}
\frac{1}{n^3(\log n)^2}=:r_n
\hskip 0.2 in
(n\gg 1).
\end{equation}
Note that 
$r_n$ declines as $n$ becomes large 
because in our cases, $|\frac{a(1)}{\Lambda}|=
\frac{|C^{\alpha}|\sin\pi\alpha}{\pi\alpha}$
is smaller than 1. (
$0.900316,0.656385,0.31831,0.0397887$
for $N_f=0,1,2,3$.)
Note also that
the convergence would be very slow because of the factor $(\log n)^{-2}$
in $r_n$.

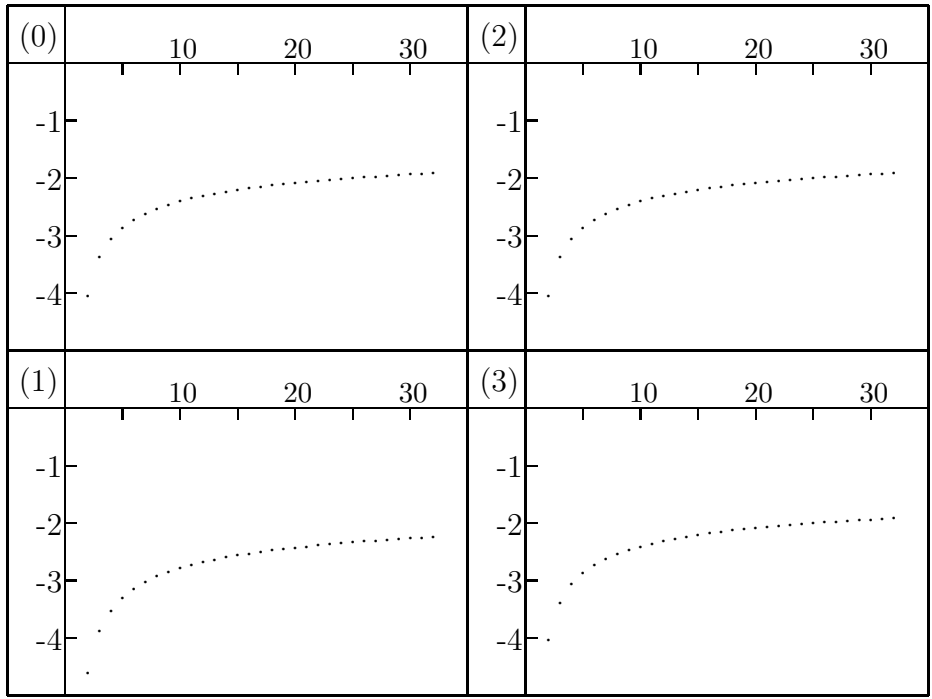
\begin{figure}[htop]
\setlength{\unitlength}{.009\textwidth}
\begin{center}
\begin{picture}(80,60)(0,0)
\thicklines
\put(0,0){\line(0,1){60}}
\put(40,0){\line(0,1){60}}
\put(0,0){\line(1,0){80}}
\put(0,30){\line(1,0){80}}
\put(80,60){\line(0,-1){60}}
\put(80,60){\line(-1,0){80}}
\thinlines

\put(5,0){\line(0,1){60}}
\put(45,0){\line(0,1){60}}
\put(0,25){\line(1,0){80}}
\put(0,55){\line(1,0){80}}
\put(5,0){\line(1,0){1}}
\put(5,5){\line(1,0){1}}\put(2.5,4){-4}
\put(5,10){\line(1,0){1}}\put(2.5,9){-3}
\put(5,15){\line(1,0){1}}\put(2.5,14){-2}
\put(5,20){\line(1,0){1}}\put(2.5,19){-1}
\put(5,25){\line(1,0){1}}
\put(5,35){\line(1,0){1}}\put(2.5,34){-4}
\put(5,40){\line(1,0){1}}\put(2.5,39){-3}
\put(5,45){\line(1,0){1}}\put(2.5,44){-2}
\put(5,50){\line(1,0){1}}\put(2.5,49){-1}
\put(5,55){\line(1,0){1}}
\put(45,0){\line(1,0){1}}
\put(45,5){\line(1,0){1}}\put(42.5,4){-4}
\put(45,10){\line(1,0){1}}\put(42.5,9){-3}
\put(45,15){\line(1,0){1}}\put(42.5,14){-2}
\put(45,20){\line(1,0){1}}\put(42.5,19){-1}
\put(45,25){\line(1,0){1}}
\put(45,35){\line(1,0){1}}\put(42.5,34){-4}
\put(45,40){\line(1,0){1}}\put(42.5,39){-3}
\put(45,45){\line(1,0){1}}\put(42.5,44){-2}
\put(45,50){\line(1,0){1}}\put(42.5,49){-1}
\put(45,55){\line(1,0){1}}
\put(10,25){\line(0,-1){1}}
\put(15,25){\line(0,-1){1}}\put(14,25.5){\small{10}}
\put(20,25){\line(0,-1){1}}
\put(25,25){\line(0,-1){1}}\put(24,25.5){\small{20}}
\put(30,25){\line(0,-1){1}}
\put(35,25){\line(0,-1){1}}\put(34,25.5){\small{30}}
\put(10,55){\line(0,-1){1}}
\put(15,55){\line(0,-1){1}}\put(14,55.5){\small{10}}
\put(20,55){\line(0,-1){1}}
\put(25,55){\line(0,-1){1}}\put(24,55.5){\small{20}}
\put(30,55){\line(0,-1){1}}
\put(35,55){\line(0,-1){1}}\put(34,55.5){\small{30}}
\put(50,25){\line(0,-1){1}}
\put(55,25){\line(0,-1){1}}\put(54,25.5){\small{10}}
\put(60,25){\line(0,-1){1}}
\put(65,25){\line(0,-1){1}}\put(64,25.5){\small{20}}
\put(70,25){\line(0,-1){1}}
\put(75,25){\line(0,-1){1}}\put(74,25.5){\small{30}}
\put(50,55){\line(0,-1){1}}
\put(55,55){\line(0,-1){1}}\put(54,55.5){\small{10}}
\put(60,55){\line(0,-1){1}}
\put(65,55){\line(0,-1){1}}\put(64,55.5){\small{20}}
\put(70,55){\line(0,-1){1}}
\put(75,55){\line(0,-1){1}}\put(74,55.5){\small{30}}

\put(1,56.5){(0)}
\put(7, 34.8064){\circle*{0.2}}
\put(8, 38.1086){\circle*{0.2}}
\put(9, 39.6954){\circle*{0.2}}
\put(10, 40.6875){\circle*{0.2}}
\put(11, 41.3884){\circle*{0.2}}
\put(12, 41.9202){\circle*{0.2}}
\put(13, 42.3433){\circle*{0.2}}
\put(14, 42.6912){\circle*{0.2}}
\put(15, 42.9846){\circle*{0.2}}
\put(16, 43.2368){\circle*{0.2}}
\put(17,43.4569){\circle*{0.2}}
\put(18, 43.6515){\circle*{0.2}}
\put(19, 43.8254){\circle*{0.2}}
\put(20, 43.982){\circle*{0.2}}
\put(21, 44.1243){\circle*{0.2}}
\put(22,44.2544){\circle*{0.2}}
\put( 23, 44.374){\circle*{0.2}}
\put(24, 44.4845){\circle*{0.2}}
\put(25, 44.587){\circle*{0.2}}
\put(26, 44.6826){\circle*{0.2}}
\put(27, 44.7719){\circle*{0.2}}
\put(28, 44.8558){\circle*{0.2}}
\put(29, 44.9347){\circle*{0.2}}
\put(30, 45.0092){\circle*{0.2}}
\put(31, 45.0796){\circle*{0.2}}
\put(32,45.1464){\circle*{0.2}}
\put(33, 45.2098){\circle*{0.2}}
\put(34, 45.2702){\circle*{0.2}}
\put(35, 45.3278){\circle*{0.2}}
\put(36, 45.3829){\circle*{0.2}}
\put(37, 45.4355){\circle*{0.2}}
\put(1,26.5){(1)}
\put(7, 1.97163){\circle*{0.2}}
\put(8, 5.64332){\circle*{0.2}}
\put(9, 7.41661){\circle*{0.2}}
\put(10, 8.5257){\circle*{0.2}}
\put(11, 9.3089){\circle*{0.2}}
\put(12, 9.90289){\circle*{0.2}}
\put(13, 10.3751){\circle*{0.2}}
\put(14, 10.7633){\circle*{0.2}}
\put(15, 11.0905){\circle*{0.2}}
\put(16, 11.3717){\circle*{0.2}}
\put(17,11.6171){\circle*{0.2}}
\put(18, 11.834){\circle*{0.2}}
\put(19, 12.0278){\circle*{0.2}}
\put(20, 12.2025){\circle*{0.2}}
\put(21, 12.361){\circle*{0.2}}
\put(22,12.506){\circle*{0.2}}
\put(23, 12.6393){\circle*{0.2}}
\put(24, 12.7624){\circle*{0.2}}
\put(25, 12.8767){\circle*{0.2}}
\put(26, 12.9833){\circle*{0.2}}
\put(27,13.0829){\circle*{0.2}}
\put(28, 13.1764){\circle*{0.2}}
\put(29, 13.2644){\circle*{0.2}}
\put(30, 13.3474){\circle*{0.2}}
\put(31, 13.426){\circle*{0.2}}
\put(32,13.5005){\circle*{0.2}}
\put(33, 13.5713){\circle*{0.2}}
\put(34, 13.6386){\circle*{0.2}}
\put(35, 13.7029){\circle*{0.2}}
\put(36, 13.7643){\circle*{0.2}}
\put(37,13.8231){\circle*{0.2}}
\put(41,56.5){(2)}
\put(47, 34.8064){\circle*{0.2}}
\put(48, 38.1086){\circle*{0.2}}
\put(49, 39.6954){\circle*{0.2}}
\put(50, 40.6875){\circle*{0.2}}
\put(51,41.3884){\circle*{0.2}}
\put(52, 41.9202){\circle*{0.2}}
\put(53, 42.3433){\circle*{0.2}}
\put(54, 42.6912){\circle*{0.2}}
\put(55, 42.9846){\circle*{0.2}}
\put(56,43.2368){\circle*{0.2}}
\put(57, 43.4569){\circle*{0.2}}
\put(58, 43.6515){\circle*{0.2}}
\put(59, 43.8254){\circle*{0.2}}
\put(60, 43.982){\circle*{0.2}}
\put(61,44.1243){\circle*{0.2}}
\put(62, 44.2544){\circle*{0.2}}
\put(63, 44.374){\circle*{0.2}}
\put(64, 44.4845){\circle*{0.2}}
\put(65, 44.587){\circle*{0.2}}
\put(66,44.6826){\circle*{0.2}}
\put(67, 44.7719){\circle*{0.2}}
\put(68, 44.8558){\circle*{0.2}}
\put(69, 44.9347){\circle*{0.2}}
\put(70, 45.0092){\circle*{0.2}}
\put(71,45.0796){\circle*{0.2}}
\put(72, 45.1464){\circle*{0.2}}
\put(73, 45.2098){\circle*{0.2}}
\put(74, 45.2702){\circle*{0.2}}
\put(75,45.3278){\circle*{0.2}}
\put(76, 45.3829){\circle*{0.2}}
\put(77, 45.4355){\circle*{0.2}}
\put(41,26.5){(3)}
\put(47, 4.80645){\circle*{0.2}}
\put(48, 8.10861){\circle*{0.2}}
\put(49, 9.69542){\circle*{0.2}}
\put(50, 10.6875){\circle*{0.2}}
\put(51, 11.3884){\circle*{0.2}}
\put(52, 11.9202){\circle*{0.2}}
\put(53, 12.3433){\circle*{0.2}}
\put(54, 12.6912){\circle*{0.2}}
\put(55, 12.9846){\circle*{0.2}}
\put(56,13.2368){\circle*{0.2}}
\put(57, 13.4569){\circle*{0.2}}
\put(58, 13.6515){\circle*{0.2}}
\put(59, 13.8254){\circle*{0.2}}
\put(60, 13.982){\circle*{0.2}}
\put(61,14.1243){\circle*{0.2}}
\put(62, 14.2544){\circle*{0.2}}
\put(63, 14.374){\circle*{0.2}}
\put(64, 14.4845){\circle*{0.2}}
\put(65, 14.587){\circle*{0.2}}
\put(66, 14.6826){\circle*{0.2}}
\put(67, 14.7719){\circle*{0.2}}
\put(68, 14.8558){\circle*{0.2}}
\put(69, 14.9347){\circle*{0.2}}
\put(70, 15.0092){\circle*{0.2}}
\put(71,15.0796){\circle*{0.2}}
\put(72, 15.1464){\circle*{0.2}}
\put(73, 15.2098){\circle*{0.2}}
\put(74, 15.2702){\circle*{0.2}}
\put(75, 15.3278){\circle*{0.2}}
\put(76, 15.3829){\circle*{0.2}}
\put(77, 15.4355){\circle*{0.2}}
\end{picture}
\end{center}
\caption{
$\log_e \frac{{\cal F}_n}{r_n}$. 
$N_f=0$ (top left), $1$ (bottom left),
$2$ (top right) and $3$ (bottom right). 
}
\label{logratio}
\end{figure}

We plot
the logarithm of ratio $\log_e\frac{{\cal F}_n}{r_n}$
up to $2\leq n\leq 32$ in Figure \ref{logratio}.
Clearly the data is not enough to show the 
convergence to 1, but 
it is natural believe that the computation up to higher value of $n$
would show the correctness of the asymptotic form 
(\ref{rn}).

\vspace{5mm}

\section{Models of Mirror Symmetry}\label{local}
\hspace{5mm}
In this section we will construct four examples of the 
mirror symmetry. These will appear as the mirror models
that reproduce the Seiberg-Witten theories in the previous section
via geometric engineering in the next section.
First we will construct them
within the framework of local mirror symmetry \cite{ChKlYaZa}
in subsection \ref{localmirror}.
Then in subsection \ref{yukawacplg},
we will derive their Yukawa coupling 
using the framework of mirror symmetry \cite{Batyrev, HoKlThYa, CoKa}.
We note that a compactification of the non-compact 
Calabi-Yau three-fold of the local A-model is 
the Calabi-Yau three-fold  of the A-model.
We will utilize both frameworks because
the former is
convenient to identify the Seiberg-Witten theory,
and the latter is necessary to see a structure of
the Yukawa coupling.
A definition of the Yukawa coupling of the B-model is not 
yet clear in the framework of local mirror symmetry. 

\subsection{Local Mirror Construction}
\label{localmirror}
\hspace*{5mm}
The local mirror symmetry \cite{ChKlYaZa} is the 
duality between two moduli spaces.
The one is the complexified K\"ahler moduli of the 
canonical bundle $V$ of a two-dimensional toric variety
${\bf P}_{base}$. On the K\"ahler moduli,
there is a holomorphic function called prepotential which 
can be written in terms of world-sheet instanton numbers.
The other is the moduli space of 
monomial deformations of the curve in another two-dimensional
toric variety. 
There, period integrals of a meromorphic one-form on the curve. 

A local mirror model is constructed from a two-dimensional polytope.
We will study four local mirror models 
associated with the reflexive polytopes $\triangle_{local}$ 
shown in Figure \ref{polytope}.
The word ``Model $i$'' $(0\leq i\leq 3)$ under each polytope means
that we will call the local mirror model constructed from it by Model $i$.
Integral points in each polytope 
are denoted by $\nu_1,\cdots,\nu_r$ $(r=k+3)$, see Table \ref{onecones}.
\begin{figure}[htop]
\begin{center}
\setlength{\unitlength}{.006\textwidth}
\begin{picture}(180,45)(35,-23)
\thicklines
\put(70,10){\line(1,-1){10}}
\put(70,10){\line(-1,-1){20}}
\put(80,0){\line(-3,-1){30}}
\thinlines
\put(70,0){\line(1,0){10}}
\put(70,0){\line(0,1){10}}
\put(70,0){\line(-1,0){10}}
\put(70,0){\line(-2,-1){20}}
\put(70,1){\small 5}
\put(80,0){\small 1}
\put(70,11){\small 2}
\put(57,0){\small 3}
\put(50,-14){\small 4}
\put(60,-20){\small Model 0}
\thicklines
\put(110,0){\line(-1,1){10}}
\put(100,10){\line(-1,0){10}}
\put(90,10){\line(0,-1){20}}
\put(90,-10){\line(2,1){20}}
\thinlines
\put(100,0){\line(1,0){10}}
\put(100,0){\line(-1,1){10}}
\put(100,0){\line(0,1){10}}
\put(100,0){\line(-1,0){10}}
\put(100,0){\line(-1,-1){10}}

\put(100,1){\small 6}
\put(110,0){\small 1}
\put(100,11){\small 2}
\put(87,11){\small 3}
\put(87,0){\small 4}
\put(87,-14){\small 5}

\put(90,-20){\small Model 1}

\thicklines
\put(140,0){\line(0,1){10}}
\put(140,10){\line(-1,0){20}}
\put(120,10){\line(0,-1){20}}
\put(120,-10){\line(2,1){20}}

\thinlines
\put(130,0){\line(1,0){10}}
\put(130,0){\line(1,1){10}}
\put(130,0){\line(-1,1){10}}
\put(130,0){\line(0,1){10}}
\put(130,0){\line(-1,0){10}}
\put(130,0){\line(-1,-1){10}}

\put(130,1){\small 7}
\put(140,0){\small 1}
\put(140,11){\small 2}
\put(130,11){\small 3}
\put(117,11){\small 4}
\put(117,0){\small 5}
\put(117,-14){\small 6}
\put(120,-20){\small Model 2}

\thicklines
\put(170,0){\line(0,1){10}}
\put(170,10){\line(-1,0){20}}
\put(150,10){\line(0,-1){20}}
\put(150,-10){\line(1,0){10}}
\put(160,-10){\line(1,1){10}}
\thinlines
\put(160,0){\line(1,0){10}}
\put(160,0){\line(1,1){10}}
\put(160,0){\line(-1,1){10}}
\put(160,0){\line(0,1){10}}
\put(160,0){\line(-1,0){10}}
\put(160,0){\line(-1,-1){10}}
\put(160,0){\line(0,-1){10}}

\put(160,1){\small 8}
\put(170,0){\small 1}
\put(170,11){\small 2}
\put(160,11){\small 3}
\put(147,11){\small 4}
\put(147,0){\small 5}
\put(147,-14){\small 6}
\put(160,-14){\small 7}

\put(150,-20){\small Model 3}
\end{picture}
\end{center}
\caption{Polytope $\triangle_{local}$.
}
\label{polytope}
\end{figure}


\begin{table}[htop]
\begin{center}
\begin{tabular}{|c|l|l|}\hline
Model &$k$&$(\nu_1,\cdots,\nu_{k+3})$\\\hline
0&2&$(1,0),(0,1),(-1,0),(-2,-1),(0,0)$\\
1&3&$(1,0),(0,1),(-1,1),(-1,0),(-1,-1),(0,0)$\\
2&4&$(1,0),(1,1),(0,1),(-1,1),(-1,0),(-1,-1),(0,0)$\\
3&5&$(0,1),(1,1),(0,1),(-1,1),(-1,0),(-1,-1),(0,-1),(0,0)$\\\hline
\end{tabular}
\end{center}
\caption{Integral points in $\triangle_{local}$. 
}
\label{onecones}
\end{table}

Let us define the lattice $L$
\begin{equation}
\label{defofl}
\begin{split}
&\tilde{\nu}_i:=
\left(\begin{array}{c}1\\\nu_i\end{array}\right)\in{\bf Z}^3,
\quad
A:=\Big(\tilde{\nu}_1,\cdots,\tilde{\nu}_{k+3}\Big)
\mbox{ : $3\times ({k+3})$ matrix},
\\
&L:=\Big\{l\in{\bf Z}^{k+3}\,:\, 
A\cdot l\,=\vec{0}
 \Big\},
\\
&l^{(1)},\cdots l^{(k)}\mbox{ : basis of $L$}.
\end{split}
\end{equation}
Our choice of the basis $l^{(i)}$ is shown in Table \ref{basisofL}.
\begin{table}[htop]
\begin{equation*}
\begin{array}{|c|l|c|l|}\hline
\mbox{Model} &{l}^{(i)},\quad (1\leq i \leq k)
&
\mbox{Model} &{l}^{(i)},\quad (1\leq i \leq k)
\\\hline
0&\begin{bmatrix}
0&1&-2&1&0\\1&0&1&0&-2\\
\end{bmatrix}
&
2&\begin{bmatrix}
0&0&0&1&-2&1&0\\
0&0&1&-1&1&0&-1\\
0&1&-2&1&0&0&0\\
1&-1&1&0&0&0&-1\\
\end{bmatrix}
\\\hline
1&\begin{bmatrix}
0&0&1&-2&1&0\\
0&1&-1&1&0&-1\\
1&-1&1&0&0&-1\\
\end{bmatrix}
&
3&\begin{bmatrix}
0&0&0&1&-2&1&0&0\\ 
-1&0&1&-1&1&-1&1&0\\
0&1&-2&1&0&0&0&0\\
1&-1&1&0&0&0&0&-1\\
1&0&0&0&0&1&-1&-1\\
\end{bmatrix}
\\\hline
\end{array}
\end{equation*}
\caption{The basis $l^{(1)},\cdots, l^{(k)}$. 
$i$-th row of the matrix is $l^{(i)}$.}
\label{basisofL}
\end{table}

\vspace{5mm}

{\it Local A-model}

Let us regard a polytope $\triangle_{local}$  
as a two-dimensional complete fan $\Sigma_{local}$ whose 1-cones
have $\nu_{i}$ $(1\leq i\leq k+2)$ as its generators;
cones in $\Sigma_{local}$ are 
\begin{equation}
\begin{split}
&2\mbox{-cones}: [\nu_i,\nu_{i+1}] \quad(1\leq i\leq k+1),\quad
[\nu_1,\nu_{k+2}],\\
&1\mbox{-cones}: [\nu_i] \quad(1\leq i\leq k+2),
\qquad 0\mbox{-cone}: \{0\}.
\end{split}
\end{equation}
Here $[v_1,\cdots,v_j]$ means the cone spanned by 
vectors $v_1,\cdots,v_j\in{\bf Z}^2$, i.e. 
the set of points in ${\bf R}^2$ 
which are a linear combination of $v_1,\cdots,v_j$ with 
non-negative real coefficients.
Then, we can construct ${\bf P}_{base}$ of the canonical bundle $V$ 
as  the toric variety ${\bf P}_{\Sigma_{local}}$:
${\bf P}_{base}$ is the Hirzebruch surface ${\bf F}_2$
for Model $0$ and its $N_f$ point blow ups for Model $N_f$ $(1\leq N_f\leq 3)$.

We note that $H_2({\bf P}_{base},{\bf Z})=H_2(V,{\bf Z})$ 
can be identified with the lattice $L$.
Let us denote by $D_i$ a divisor corresponding to 1-cone $\nu_i$
($1\leq i\leq k+2$).
Between the divisors, there are two linear equivalence relations
$\sum_{i=1}^{k+2}\nu_iD_i=\vec{0}$. 
Then, if we define a vector 
$l=(C\cdot D_1,\cdots,C\cdot D_{k+2},-C\cdot (D_1+\cdots+D_{k+2}))$
from a two-cycle $C$ in $V$, 
it satisfies $A\cdot l=0$. 
Thus we can identify the space $L=\mbox{ker}_{\bf Z}A$ with 
$H_2(V,{\bf Z})$.
We will identify
the basis $\{l^{(1)},\cdots,l^{(k)}\}$ of $L$
with the basis of $H_2(V;{\bf Z})$,
and also use $l^{(i)}$  to  denote the corresponding two-cycle.

We study the complexified K\"ahler moduli space of $V$
which is the space of complexified K\"ahler classes.
The complexified K\"ahler class is  
a cohomology class of the sum $\omega$ 
of a two-form $B$ and a K\"ahler form $J$ of $V$ multiplied by $\sqrt{-1}$:
$\omega=B+iJ$.
We define the K\"ahler parameter of the two-cycle $\ell^{(i)}$
\begin{equation}
t_i:=\oint_{l^{(i)}}2\pi i\omega \qquad (1\leq i\leq k).
\end{equation}
Then the object of the study is the prepotential.
Its general form is
\begin{equation}
{\cal F}_{mirror}=\sum_{1\leq i, j, p\leq k}
{{J_i \cdot J_j \cdot J_p}\over{6}}
t_i t_j t_p+
\sum_{1 \leq i \leq k}
{{c_2 \cdot J_i}\over{24}} t_i
-i{{\zeta(3)}\over{2(2\pi)^3}} c_3+
\sum_{n_1, \cdots, n_k} d_{\vec{n}} \, \mbox{Li}_3
( e^{n_1t_1+\cdots+n_k t_k} ),
\label{defofmirrorprepotential}
\end{equation}
where $d_{\vec{n}=(n_1,\dots,n_k)}$ is 
the world-sheet instanton number.
$J_i\in H^2(V,{\bf R})$ is the dual of the two-cycle $l^{(i)}$ 
by the pairing of $H^2(V,{\bf R})$ and $H_2(V,{\bf R})$.
$\mbox{ Li}_3(x)=\sum_{k=1}^{\infty}\frac{x^k}{k^3}$,
$c_2$ is the second Chern class of $V$
and $c_3$ is the third Chern number of $V$.
The sum in the last term is over $(n_1,\dots,n_k)$ such that 
$\sum_i n_i l^{(i)}$ is the homology class of curves in $V$.
For Model 0,1 and 2, 
the summation is over ${^\forall} n_i \geq 0$.
For model 3, the summation should be over the domain where 
$n_1, n_3, n_4, n_5 \geq 0$ and for given $(n_1, n_3, n_4, n_5)$,
over $n_2$ that satisfies $n_2 \leq n_5$ or $n_2 \leq 2 n_3$  
or $n_2 \leq n_3, n_4$.
This is because $l^{(2)}$ does not correspond to a curve in $V$ 
(see also Table \ref{generatorsoftoricideal}).
The Yukawa coupling is defined to be
the third derivative of the prepotential by the K\"ahler parameters
\begin{equation}
F_{t_i t_j t_p}:=\partial_{t_i} \partial_{t_j} \partial_{t_p} {\cal F}_{mirror}
=J_i \cdot J_j \cdot J_p+
\sum_{n_1, \cdots, n_k} d_{\vec{n}}
{{e^{n_1t_1+\cdots+n_k t_k}}\over{1-e^{n_1t_1+\cdots n_kt_k}}}.
\end{equation}

\vspace*{5mm}
{\it Local B-model}

The curve of the local B-model is a 
hypersurface in the toric variety ${\bf P}_{\triangle_{local}}$
\begin{equation}
P_{local}(X_1,X_2)=a_1 X^{\nu_1}+\cdots+a_{k+2}X^{\nu_{k+2}}+a_{k+3}=0,
\end{equation}
with $k+3$ parameters $a_1,\cdots, a_{k+3}$.
$X_1,X_2$ are local coordinates of ${\bf P}_{\triangle_{local}}$.
The dimensions of moduli space of the curve is 
$k$ because we must subtract from the number of parameters $k+3$ the 
dimensions of the toric automorphism $X_i\frac{\partial}{\partial X_i}$
$(1\leq i\leq 2)$ and one degree of freedom of multiplying $P_{local}$
by a constant. 
Then, we have the following local coordinates of the moduli space
using the basis of $L$
\begin{equation}
z_i=a^{l^{(i)}}\qquad (1\leq i\leq k).
\end{equation}

\begin{table}[htop]
\begin{center}
\begin{tabular}{|c|l|l|}\hline
Model&$l$\\\hline
0&$~l^{(1)},l^{(2)}$
\\\hline
1&$~l^{(1)},l^{(2)},l^{(3)},l^{(1)}+l^{(2)},l^{(2)}+l^{(3)}$
  \\\hline
2&$\begin{array}{l}
 l^{(1)},
 l^{(2)},
 l^{(3)},
 l^{(4)},
 l^{(1)}+l^{(2)},
 l^{(2)}+l^{(3)},
 l^{(3)}+l^{(4)},\\
 l^{(2)}+l^{(3)}+l^{(4)},
 l^{(1)}+2l^{(2)}+l^{(3)}
 \end{array}$
\\\hline
3&
  $\begin{array}{l}
  l^{(1)},
l^{(3)},      
l^{(4)},
l^{(5)},
l^{(2)}+l^{(5)},
l^{(3)}+l^{(4)},
l^{(1)}+l^{(2)}+l^{(5)},\\
l^{(2)}+l^{(3)}+l^{(5)},
l^{(2)}+l^{(3)}+l^{(4)},
l^{(2)}+l^{(3)}+l^{(4)}+l^{(5)},\\
l^{(1)}+l^{(2)}+l^{(3)}+l^{(4)},
l^{(1)}+2l^{(2)}+l^{(3)}+2l^{(5)},\\ 
l^{(1)}+2l^{(2)}+l^{(3)}+l^{(5)},
 l^{(1)}+2l^{(2)}+l^{(3)}+l^{(4)}+l^{(5)}
\end{array}$\\\hline
\end{tabular}
\end{center}
\caption{The minimal generators of the toric ideal. 
A vector $l$ corresponds to a differential operator 
$
\Pi_{i;l_i>0}\partial_{a_i}^{l_i}-\Pi_{i;l_i<0}\partial_{a_i}^{-l_i}
$ in the toric ideal.}
\label{generatorsoftoricideal}
\end{table}

On the moduli space, we consider period integrals of a meromorphic
one-form
\begin{equation}
\oint_{P_{local}=0}\log P_{local}
\,\frac{dX_1}{X_1}\wedge\frac{dX_2}{X_2}.
\end{equation}
Then, the Picard-Fuchs equations that annihilate  this period 
integrals are provided by the
GKZ-hypergeometric differential system 
$H_A(\beta)$ with $A$ in (\ref{defofl}) and $\beta=(0,0,0)$
\cite{ChKlYaZa}.
$H_A(\beta)$
consists of  two parts.
The one is the part made of the three differential operators
\begin{equation}
\theta_{a_1}+\cdots+\theta_{a_{k+3}},\qquad
\sum_{i=1}^{k+3}(\nu_i)_j\theta_{a_i}\quad(j=1,2).
\label{indeponlyonz}
\end{equation}
The other is the toric ideal part generated by 
\begin{equation}
\prod_{i;l_i>0}\partial_{a_i}^{l_i}-\prod_{i;l_i<0}\partial_{a_i}^{-l_i}
\qquad {}^{\forall}l\in L .
\end{equation}
This toric ideal is actually generated by finite number of 
generators (see Appendix A).
We show the generators in Table \ref{generatorsoftoricideal}.
The first part (\ref{indeponlyonz}) means that the period integrals 
depend only on $z_i$'s and
a differential operator 
$\theta_{a_j}$ acts on the period integrals as 
$(l^{(i)})_j\theta_{z_i}$.
Therefore, if we write differential operators in the toric 
ideal using $\theta_{a_j}$'s and $z_i$'s 
and further replace $\theta_{a_j}$ with
$(l^{(i)})_j\theta_{z_i}$,
we arrive at Picard-Fuchs operators.
For example, in the case of Model $1$,
the Picard-Fuchs operators are 
\begin{equation}
\begin{split}
&{\cal L}_1=
\theta_1(\theta_1-\theta_2+\theta_3)-
z_1(-2\theta_1+\theta_2)(-2\theta_1+\theta_2-1),
\\
&{\cal L}_2=(\theta_2-\theta_3)(-2\theta_1+\theta_2)-z_2
(\theta_1-\theta_2+\theta_3)(-\theta_2-\theta_3),
\\
&{\cal L}_3=\theta_3(-\theta_1+\theta_2-\theta_3)-
z_3(\theta_2-\theta_3)(-\theta_2-\theta_3),
\\
&{\cal L}_4=(\theta_2-\theta_3)\theta_1
-z_1z_2(-2\theta_1+\theta_2)(-\theta_2-\theta_3),
\\
&{\cal L}_5=\theta_3(-2\theta_1+\theta_2)
-z_2z_3(-\theta_2-\theta_3)(-\theta_2-\theta_3-1).
\end{split}
\end{equation}
Now we can obtain the period integrals as solutions of the Picard-Fuchs
equations. 
Let us define
\begin{equation}
f(z;\rho)=\sum_{\vec{n}=(n_1,\dots,n_k)}a_{\vec{n}}z^{\vec{n}+\rho},\qquad
a_{\vec{n}}=\prod_{i=1}^{k+3}
\frac{\Gamma (1+\sum_{j=1}^k\rho_j\left(l^{(j)}\right)_i)}
{\Gamma (1+\sum_{j=1}^k(n_j+\rho_j)\left(l^{(j)}\right)_i)},
\end{equation}
where $\rho=(\rho_1,\cdots,\rho_k)$ is a 
set of parameters.
Then the solutions are 
\begin{align}
f(z;\vec{0})&=1,\notag\\
\label{solution}
\partial_{\rho_i}f(z;\rho)|_{\rho=\vec{0}}&=\log z_i+\cdots \hskip 0.2 in 
 (1\leq i\leq k),\\
\sum_{1\leq i\leq j\leq k}c_{i,j}
\partial_{\rho_i}\partial_{\rho_j}f(z;\rho)|_{\rho=\vec{0}}&=
\sum_{1\leq i \leq j \leq k}
 c_{i,j}\log z_i\log z_j+\cdots.\notag
\end{align}
Coefficients $(c_{i,j})$ in the last double logarithmic 
solution are shown in Table \ref{tableofc}.
These coefficients can be obtained from the formula in \cite{ChKlYaZa} 
or by directly substituting the solution into the Picard-Fuchs equations.
There are enough equations ($\frac{k(k+1)}{2}-1$ of them)
for the $\frac{k(k+1)}{2}$ coefficients $c_{ij}$ $(1\leq i\leq j \leq k)$.
\begin{table}[htop]
\begin{center}
\begin{tabular}{|c|l|}\hline
Model&$(c_{i,j})$\\\hline
0&$(0,1,1)$\\\hline
1&$(0,2,2,2,4,1)$\\\hline
2&$(0,2,2,2,2,4,4,1,2,0)$\\\hline
3&$(0,4,4,4,4,4,8,8,8,3,4,8,0,8,2)$\\\hline
\end{tabular}
\end{center}
\caption{The coefficients $(c_{i,j})$ in the double 
logarithmic solution.
They are arranged in the lexicographic order
$(c_{1,1},c_{1,2},\cdots,c_{2,2}\cdots,c_{k,k})$.
}
\label{tableofc}
\end{table}

The local mirror symmetry states that 
the single logarithmic solutions in (\ref{solution})
constitute the mirror map to the K\"ahler parameters of the local A-model
as follows
\begin{equation}
\label{mirrormap}
t_i=\partial_{\rho_i}f(z;\rho)|_{\rho=\vec{0}}=\log z_i+\cdots
\qquad (1\leq i \leq k).
\end{equation}
And the double logarithmic solution in (\ref{solution}) is
translated into the following derivative of the prepotential 
(\ref{defofmirrorprepotential})
under the mirror map (\ref{mirrormap})
\begin{equation}
\left\{-2\partial_{t_2},\,
-(\partial_{t_2}+\partial_{t_3}),\,
-(\partial_{t_2}+\partial_{t_4}), \,
-(\partial_{t_4}+\partial_{t_5})\right\}\, {\cal F}_{mirror},
\end{equation}
for Model 0,1,2 and 3.
This allows us to obtain the world-sheet instanton numbers of the
canonical bundle $V$ of the local A-model
(see Table in Appendix B for the results of $N_f=0, 1$).

\subsection{Yukawa Coupling}\label{yukawacplg}
\hspace{5mm}
In this subsection, we will derive the Yukawa coupling.
Since the Yukawa coupling of the B-model is not 
yet incorporated into the framework  of local mirror symmetry,
we must resort to the  mirror symmetry.
We will construct a mirror model corresponding to
each local mirror model by
reversing the process that leads to the local mirror symmetry 
\cite{ChKlYaZa}.
We will only give minimal explanation.
The reader may consult \cite{Batyrev, HoKlThYa, ChKlYaZa} for more detail.

We construct a following four-dimensional polytope $\triangle^*$
which contain $\triangle_{local}$ as a two-dimensional face
\begin{equation}
\begin{split}
\triangle^*&=\mbox{conv}[\overline{\nu}_1,\cdots,\overline{\nu}_r,
\overline{\nu}_{r+1},\overline{\nu}_{r+2}
],
\\
\overline{\nu}_i&=(\nu_i,2,3)\quad(1\leq i\leq r),\quad
\overline{\nu}_{r+1}=(0,0,-1,0),\quad
\overline{\nu}_{r+2}=(0,0,0,-1).
\end{split}
\end{equation}
$\mbox{conv}[p_1,\cdots,p_s]$ means the convex hull of 
points $p_1,\cdots, p_s$, i.e. the set of all points in ${\bf R}^4$
that can be written as $c_1 p_1+\cdots +c_s p_s$
with ${^\forall c_j}\geq 0, c_1+\cdots+c_s=1$.
This is a reflexive integral polytope, and
has the dual integral polytope $\triangle$. 
We will denote by $\Sigma$ (resp. $\Sigma^*$ ) 
a fan determined by a maximal triangulation of 
$\triangle^*$ (resp. $\triangle$).

\vspace{5mm}

{\it A-model}

A Calabi-Yau three-fold  $\widetilde{V}$ in the A-model 
is realized as a hypersurface in the toric variety 
${\bf P}_{\Sigma}$ associated to the fan $\Sigma$.
This is a compactification of the canonical bundle $V$ of 
${\bf P}_{base}$ in the local A-model:
$\widetilde{V}$ has a structure of  a fibration of elliptic curves over 
${\bf P}_{base}$. 
Quantities in the complexified K\"ahler moduli space of $\widetilde{V}$ 
are defined in the same way as the local A-model.
Let us denote by $t_0$ the K\"ahler parameter
corresponding to the elliptic fiber.
This is an integral of a complexified K\"ahler form over the elliptic fiber.
Then, the K\"ahler moduli space of $V$ of the local A-model 
appears at the limit $t_0\to-\infty$.

\vspace*{5mm}

{\it B-model}

The moduli space of the B-model is 
that of complex deformations of another Calabi-Yau three-fold 
$\widetilde{V}^*$.
It is a hypersurface determined by $P^*=0$ in the toric variety 
${\bf P}_{\Sigma^*}$
\begin{equation}
P^*=1+z_0X_3^2X_4^3P_{local}+\frac{1}{X_3}+\frac{1}{X_4}.
\end{equation}
$X_1,\cdots,X_4$ are local coordinates of ${\bf P}_{\Sigma^*}$
and $z_0$ is a one more parameter other than $a_1,\cdots,a_{k+3}$.
The limit $t_0\to -\infty$ in the A-model corresponds to 
the limit $z_0\to 0$.
On this moduli space, 
we consider period integrals of a holomorphic three-form $\Omega$
over three-cycles in $\widetilde{V}^*$. 
The period integrals
are solutions of Picard-Fuchs equations.
The Picard-Fuchs operators of the B-model are provided with the
Picard-Fuchs operators of the local B-model ${\cal L}_i$
\begin{equation}
\begin{split}
&\widetilde{{\cal L}}_0=(\theta_{z_0}+\theta_{a_{k+3}})\theta_{z_0}-12z_0
(6\theta_{z_0}+1)(6\theta_{z_0}+5),\\
&\widetilde{{\cal L}_i}={\cal L}_i  \quad
\mbox{ with  $\theta_{a_{k+3}}$ replaced by 
$\theta_{z_0}+\theta_{a_{k+3}}$}.
\end{split}
\label{mirrorpicardfuchs}
\end{equation}
Then, we have a mirror map by solutions of 
the Picard-Fuchs equations.

We are ready to introduce the Yukawa coupling of the mirror B-model.
We can always fix a symplectic basis of $H_3(\widetilde{V}^*,{\bf Z})$,
with $\alpha_I^*,\beta^{I*}$ $I=1,\dots h^{2,1}+1$ $(h^{2,1}=k+1)$
satisfying $\alpha_I^*\cdot\beta^{J*}=\delta_{I}^J$,
$\alpha_I^*\cdot\alpha_J^*=\beta^{I*}\cdot\beta^{J*}=0$,
and define the period integrals
\begin{equation}
A^I:=\int_{\alpha_I^*}\Omega,
\quad
B_I:=\int_{{\beta^I}^*}\Omega.
\end{equation}
Then, the Yukawa coupling of the B-model is defined as
\begin{equation}
F_{z_iz_jz_p}:=
\sum_{I=1}^{k+2}
\Big[
A^I(z)\partial_{z_i}\partial_{z_j}\partial_{z_p}
B_I(z)-
B_I(z)\partial_{z_i}\partial_{z_j}\partial_{z_p}
A^I(z)
\Big].
\end{equation}
This is a contravariant tensor of rank three because of the Griffith
transversality.
The Yukawa coupling is determined
from the Picard-Fuchs operators (\ref{mirrorpicardfuchs}).
We note that it does not depend on a choice of the symplectic basis.

We can obtain the Yukawa coupling of the local B-model
at the limit $z_0\to 0$.
The results are too long to list here.
In the next section, we will transform the variables of the B-model
into $(z_0,\cdots,z_k)$ to the variables 
$(z_0,\epsilon,u,m_1,\cdots,m_{k-2})$.
Here $(u,m_1,\cdots,m_{k-2})$ are parameters of the Seiberg-Witten theory
and $\epsilon$ is another parameter to be $\epsilon \to 0$.
We will first transform $F_{z_iz_jz_p}$ $(0\leq i,j,p\leq k)$
into $F_{z_0z_0z_0},\cdots, F_{uuu},\cdots$
and then obtain the Yukawa coupling of the A-model
$F_{t_it_jt_p}$ as a transform of these by the mirror map.
Remarkably, we find out that
we can neglect
$F_{z_0**}$ and even $F_{\epsilon **}$ (here $*=z_0,\epsilon,u$, or $m_i$)
to obtain the Yukawa coupling of the local A-model
$F_{t_it_jt_p}$ with $2\leq i,j, p\leq k$
at the gauge theory limit $z_0\to 0$ and $\epsilon\to 0$.

\section{Geometric Engineering of Seiberg-Witten Theories}
\label{geom}
\hspace{5mm}
In this section, we will carry out 
the geometric engineering of the Seiberg-Witten theories.
We will pick up the Model $N_f$ 
as the local mirror model corresponding to the gauge theory
with $N_f$-hypermultiplets \cite{KaKlVa, KaVa}.
In subsection \ref{geo21},
We will first 
identify the moduli coordinates so that
the curve of the local B-model reproduces the
Seiberg-Witten curve at the gauge theory limit $\epsilon\to 0$.
Then we will study the behavior of the 
period integrals under the limit $\epsilon\to 0$.
We will also obtain a set of differential equations that
annihilates the period integrals $(a,a_D)$ and check the equivalence 
of the prepotentials at the gauge theory limit.
These will be addressed in subsection \ref{geo22}. 
Finally, in subsection \ref{gw} 
we will suggest that the asymptotic distribution of 
the world-sheet instanton numbers of the local A-model
is controlled by the instanton amplitude of the gauge theory.
This is an extension of the argument about Model $0$ in \cite{KaKlVa}.
 
\subsection{Identification of Moduli Coordinates}\label{geo21}
\hspace{5mm}
We give the identification of the moduli coordinates  $(z_1,z_2,\cdots,z_k)$
$(k=N_f+2)$
of a local B-model with parameters $(\epsilon, u, m_1, \cdots, m_{N_f})$
at the gauge theory limit $\epsilon\to 0$ in Table \ref{ztoep}.
This transforms the curve of the local B-model into the Seiberg-Witten curve 
(\ref{lam}) at the limit $\epsilon\to 0$.
This indicates that
the moduli space of the Seiberg-Witten theory is 
realized as an infinitesimal neighborhood of a singularity in the 
moduli space of the local mirror model \cite{KaKlVa}.
We can check that the discriminant of the local B-model reduces 
that of the Seiberg-Witten theory.
Then, we expect that period integrals of the local B-model 
separate into irreducible spaces under the monodromy transformation,
and one of them should be identified with the space of  
periods $(a,a_D)$ of the Seiberg-Witten differential. 
We note that the space includes the mass parameters of the gauge theory.
We will make this expectation explicit in next subsection.

\vspace{5mm}

\begin{table}[htop]
\begin{center}
\begin{tabular}{|c|l|}\hline
$N_f$&$(z_i)$~~~~$(i=1, \dots, N_f+2)$\\\hline
0&$(z_1,z_2)=({{\epsilon^4\Lambda^4}\over{4}},
 {{1-u\epsilon^2}\over{4}})$ 
\\\hline 
1&
$(z_1,z_2z_3,z_3)=(
{{\epsilon^3\Lambda^3}\over{4}},
{{1-u\epsilon^2}\over{4}},
{{1+m_1\epsilon}\over{2}})$
\\\hline
2&$
(z_1,z_2z_3z_4,z_3,z_4)=
({{\epsilon^2\Lambda^2}\over{4}},
{{1-v\epsilon^2}\over{4}},
{{1-n^2\epsilon^2}\over{4}},
1+m\epsilon,)
$\\
&$(v,m,n)=(u-{{\Lambda^2}\over{8}},
{{m_1+m_2}\over{2}},
{{m_1-m_2}\over{2}})
$\\\hline
3&
$(z_1,z_2z_3z_4z_5,z_3,z_4,z_5)=
({{\epsilon\Lambda}\over{4}},
{{1-w \epsilon^2}\over{4}},
{{1-n^2\epsilon^2}\over{4}},
1+m\epsilon,
{{1+p\epsilon}\over{2}})
$\\
&$(w,m,n,p)=(u+{{\Lambda^2}\over{64}}-{{\Lambda(m_1+m_2+m_3)}\over{8}},
{{m_1+m_2}\over{2}}-{{\Lambda}\over{8}},
{{m_1-m_2}\over{2}},
m_3-{{\Lambda}\over{8}})
$\\\hline
\end{tabular}
\end{center}
\caption{The change of the moduli coordinates of the local B-model.}
\label{ztoep}
\end{table}

\subsection{Set of Partial Differential Equations}\label{geo22}
\hspace*{5mm}
In this subsection, we will relate the prepotential of the local A-model to
that of the Seiberg-Witten theory.
For that, we will derive the behavior of 
the K\"ahler parameters $t_1,\cdots,t_k$ $(k=N_f+2)$ 
at the gauge theory limit $\epsilon \to 0$.
This should be determined from the behavior of the period integrals
of the local B-model at the limit $\epsilon\to 0$.
Thus, we will first study solutions of set 
of partial differential equations
obtained from the Picard-Fuchs equations of the local B-model 
at the limit $\epsilon \to 0$. 
For illustrative purposes, we will explain the results 
in the case $N_f=1$ as an example.

We first assume the form of a solution of the Picard-Fuchs equations 
at the limit $\epsilon \to 0$ as
$\epsilon^{\rho}\sum_{n=0}^{\infty}f_n(u,m_1)\epsilon^n$ 
and substitute it into the equations.
Then, the part of the lowest order with respect to $\epsilon$ becomes 
as follows ($m=m_1$)
\begin{equation}\label{temp5}
\begin{split}
&(\rho-1)\partial_u f_0(u,m)=0,\\
&(\rho-1)\partial_{m}f_0(u,m)=0,\\
&\Big[
\theta_m(\theta_m-1)+{{2m^2}\over{3u}}\theta_u(2\theta_u+4\theta_m-1)
\Big]f_0(u,m)=0,\\
&\Big[\theta_m(2\theta_u+\theta_m-1)+{{3m\Lambda^3}\over{4u^2}}
\theta_u(\theta_u-1)\Big]f_0(u,m)=0,
\\
&\Big[(2\theta_u+\theta_m-1)(2\theta_u+4\theta_m-1)+(1-\rho^2)
-{{9\Lambda^3}\over{8mu}}\theta_u\theta_m\Big]f_0(u,m)=0.
\end{split}
\end{equation}
The first two equation means that $f_0(u,m)$ must be a constant or 
otherwise $\rho=1$. 
Then, solving the last three equations with $\rho=1$,
we find solutions
\begin{equation}
g_1(u,m):=g(u,m;\rho)|_{\rho=0},\qquad 
g_2(u,m):=\partial_{\rho}g(u,m;\rho)|_{\rho=0},\qquad
m,
\end{equation}
where $g(u,m;\rho)$ is the following function with an 
auxiliary parameter $\rho$
\begin{equation}\!\!\!\!\!
g(u,m;\rho)=\sum_{n_1,n_2}
{{(-{{1}\over{2}}+3\rho)_{n_1+n_2}(-1)^{n_1+n_2}}\over
{(1+2\rho)_{n_1}(1+\rho)_{n_1-n_2}(-n_1+2n_2)!}}
\Bigl({{\Lambda^3}\over{16mu}}\Bigr)^{n_1-\frac{1}{3}+2\rho}
\Bigl({{4m^2}\over{u}}\Bigr)^{n_2-\frac{1}{6}+\rho}.
\end{equation}
We remind that
the Picard-Fuchs equations 
have five independent 
solutions; $1,t_1,t_2,t_3$ and one double logarithmic solution.
The behavior of $t_1=\log\frac{4z_1}{(1+\sqrt{1-4z_1})^2}$ 
at $\epsilon\to 0$ is $\log \frac{\epsilon^4\Lambda^4}{4}$.
Thus, adding $1$ and $t_1$ to the three solutions 
of order ${\cal O}(\epsilon)$ above,
we obtain the following solutions around $\epsilon=0$
\begin{equation} 
1,\qquad
\log \epsilon,\qquad
\epsilon g_1(u,m),\qquad
\epsilon g_2(u,m), \qquad
\epsilon m.
\end{equation}
It still remains to identify $t_2,t_3$ and the double logarithmic solution
with these solutions. We can perform analytic continuation from the expansion 
around $z=\vec{0}$ to the expansion around 
$(z_1,z_2z_3,z_3)=(0,\frac{1}{4},\frac{1}{2})$
with the help of the transformation formula of the Gauss'
hypergeometric function \cite{GrRy}. For $t_2,t_3$, the result is
\begin{equation}
t_2=-\epsilon \Big(\frac{\Lambda}{2}g_1(u,m)+m\Big)+{\cal O}(\epsilon^2),\qquad
t_3=-\epsilon \Big(\frac{\Lambda}{2}g_1(u,m)-m\Big)+{\cal O}(\epsilon^2).
\end{equation}

We should check 
that differential operators  
just appeared from the Picard-Fuchs equations
annihilate the periods $(a,a_D)$ of the Seiberg-Witten differential 
$\lambda_{SW}$ in (\ref{lam}).
Let us denote the differential operators in the 
last three equations of (\ref{temp5}) with $\rho=1$
by ${\cal D}_1, {\cal D}_2, {\cal D}_3$
\begin{equation}
\begin{split}
&{\cal D}_1:=
\theta_m(\theta_m-1)+{{2m^2}\over{3u}}\theta_u(2\theta_u+4\theta_m-1),
\\
&{\cal D}_2:=\theta_m(2\theta_u+\theta_m-1)+{{3m\Lambda^3}\over{4u^2}}
\theta_u(\theta_u-1),\\
&{\cal D}_3:=(2\theta_u+\theta_m-1)(2\theta_u+4\theta_m-1)
-{{9\Lambda^3}\over{8mu}}\theta_u\theta_m.
\end{split}
\end{equation}
It is easy to see that the differential operator ${\cal D}_i$ $(1\leq i\leq 3)$
satisfies
\begin{equation}
{\cal D}_i^{\prime}\frac{dx}{y}=\mbox{ non-singular function},
\quad {\cal D}_i^{\prime} \mbox{ is defined by }
{\cal D}_i^{\prime}\partial_u=\partial_u {\cal D}_i.
\end{equation}
Here, $y=\sqrt{(x^2-u)^2-\Lambda^3(x+m)}$. 
Thus, ${\cal D}_i$ annihilates $(a,a_D)$.
Then, $(a,a_D)$ must be linear combinations of $g_1(u,m), g_2(u,m)$.
We note that we 
can safely think of $(a,a_D)$ as linear combinations of $g_1,g_2$
because adding $m$ to $(a,a_D)$ does not change the massless limit.
By the comparison of the terms of the lowest order in $\Lambda$ 
at the limit $m\to 0$, we obtain
\begin{equation}
a=\frac{\Lambda}{2}g_1(u,m).
\end{equation}
After all, we may conclude that 
the space of period integrals of the Seiberg-Witten theory
(including the mass parameter $m$) is the subspace 
of the period integrals of the local B-model
that is closed under the monodromy transformation
around $\epsilon=0$.
We note that we can state 
the behavior of the K\"ahler parameter $t_2,t_3$ in terms of 
$a$
\begin{equation}
t_2=-\epsilon(a+m)+{\cal O}(\epsilon^2),\qquad
t_3=-\epsilon(a-m)+{\cal O}(\epsilon^2).
\end{equation}

The analysis of the cases with $N_f=2,3$ 
proceeds analogously. 
We obtained the behavior of the period integrals of the local B-model 
around $\epsilon=0$ as follows
\begin{equation}
1,\qquad
\log\frac{(\epsilon\Lambda)^{4-N_f}}{4},
\qquad
\epsilon g_1(u,m_i),\qquad
\epsilon g_2(u,m_i),\qquad
\epsilon m_i\quad (1\leq i\leq N_f),
\end{equation}
where
$g_1(u,m_i):=g(u,m_i;\rho)|_{\rho={0}}$ and 
$g_2(u,m_i):=\partial_{\rho}g(u,m_i;\rho)|_{\rho=0}$.
It is straightforward to see that $g_1,g_2$ and $m_i$ 
are annihilated by the differential operators ${\cal D}_j$
derived from the Picard-Fuchs equations of the local B-model.
We show independent ${\cal D}_j$'s in Table \ref{setofpartial},
and present $g(u,m_i;\rho)$ in Table \ref{solution1}.
Then, we can check that these ${\cal D}_j$'s annihilate $(a,a_D)$.
Thus, $(a,a_D)$ are linear combinations of $g_1,g_2$ 
and for $a$, the expression is
\begin{equation}
a=\big\{
\Lambda,\frac{\Lambda}{2},\frac{i\Lambda}{2},-i\Lambda
\big\}g_1(u,m_i)\qquad
\mbox{for }N_f=0,1,2,3.
\end{equation}
It is instructive to write down
the parameter $u$ as an inverse series in $a$ in Table \ref{inverta}.
This is consistent with (\ref{formofa}) of the gauge theory.
We can also derive the behaviors of 
the K\"ahler parameters at the limit $\epsilon\to 0$. 
The results are summarized in Table \ref{titoa}.
We note that
only one K\"ahler parameter $t_1$ diverges logarithmically and the other 
parameters approach to zero \cite{KaKlVa}. 
\begin{table}[htop]
\begin{center}
\begin{tabular}{|c|l|}\hline
$N_f$&differential operators
\\\hline
0&$
{\cal D}_1=(\theta_u-{{1}\over{2}})^2-{{\Lambda^4}\over{u^2}}
\theta_u(\theta_u-1)$
\\\hline
1&$
{\cal D}_1=
\theta_{m_1}(\theta_{m_1}-1)+{{2m_{1}^2}\over{3u}}
\theta_u(2\theta_u+4\theta_{m_1}-1)$\\
&${\cal D}_2=\theta_{m_1}(2\theta_u+\theta_{m_1}-1)
+{{3m_1\Lambda^3}\over{4u^2}}
\theta_u(\theta_u-1)$\\
&${\cal D}_3=(2\theta_u+\theta_{m_1}-1)(2\theta_u+4\theta_{m_1}-1)
-{{9\Lambda^3}\over{8m_1u}}\theta_u\theta_{m_1}$
\\\hline
2&
${\cal D}_1=(2\theta_v+\theta_m+\theta_n-1)(2\theta_v+2\theta_m+2\theta_n-1)
+{{\Lambda^2}\over{2v}}\theta_v(2\theta_v+\theta_m+2\theta_n-2)$\\
&${\cal D}_2=\theta_n(2\theta_v+\theta_n+\theta_m-1)-{{n^2\Lambda^2}\over{v^2}}
\theta_v(\theta_v-1)$\\
&${\cal D}_3=\theta_m(\theta_m-1)-
{{m^2}\over{n^2}}\theta_n(2\theta_v+\theta_m+2\theta_n-2)$\\
&${\cal D}_4=\theta_n(2\theta_v+\theta_m+2\theta_n-2)+
{{2n^2}\over{v}}\theta_v(2\theta_v+2\theta_m+2\theta_n-1)$\\
&${\cal D}_5=\theta_m\theta_n+{{2m^2}\over{v}}\theta_v\theta_n+
{{2n^2}\over{v}}\theta_v\theta_m$\\
&${\cal D}_6=\theta_m(2\theta_v+\theta_m+\theta_n-1)
+{{\Lambda^2}\over{2v}}\theta_m\theta_v
+{{m^2\Lambda^2}\over{v^2}}\theta_v(\theta_v-1)$
\\\hline
3&
${\cal D}_1=2\partial_n\partial_p+\Lambda n \partial_w^2$\\
&${\cal D}_2=\partial_n(\partial_m+\partial_p)+
2(m\partial_n+n\partial_m)\partial_w$\\
&${\cal D}_3=\partial_n(4\theta_w+2\theta_p+3\theta_n+2\theta_m-3)-
n\partial_m^2$\\
&${\cal D}_4=\partial_m(\partial_m+\partial_p)+2\partial_w
(4\theta_w+2\theta_p+3\theta_n+3\theta_m-2)$\\
&${\cal D}_5=\partial_p(\partial_m+\partial_p)+2\partial_w
(2\theta_w+2\theta_p+\theta_m+\theta_n-1)$\\
&${\cal D}_6=4\partial_p(m\partial_n+n\partial_m)-
\Lambda n\partial_w(\partial_p+\partial_m)$\\
&${\cal D}_7=\partial_p(4\theta_w+2\theta_p+3\theta_n+3\theta_m-2)-
\partial_m(2\theta_w+2\theta_p+\theta_n+\theta_m-1)$\\
&${\cal D}_8=\partial_n(2\theta_w+2\theta_p+\theta_n+\theta_m-1)-
(m\partial_n+n\partial_m)\partial_p$\\
&${\cal D}_9=
2(m\partial_n+n\partial_m)(2\theta_w+2\theta_p+\theta_n+\theta_m-1)-
\Lambda n \partial_w(3\theta_w+2\theta_p+2\theta_m+2\theta_n-2)$
\\ 
&${\cal D}_{10}=4(2\theta_w+2\theta_p+\theta_n+\theta_m-1)
(4\theta_w+2\theta_p+3\theta_n+3\theta_m-2)$\\
&$~~~~~~~~+2\Lambda m\partial_w(3\theta_w+2\theta_p+2\theta_n+2\theta_m-2) 
+\Lambda(\partial_p+\partial_m)(3\theta_w+2\theta_p+2\theta_n+2\theta_m-3)$\\
\hline
\end{tabular}
\end{center}
\caption{Set of partial differential operators. 
Those differential operators ${\cal D}_i$ annihilate 
the lowest order part $f_0(u;m_i)$ of the solution 
$\sum_{n=0}^{\infty}f_n(u;m_i)\epsilon^{n+1}$ for the Picard-Fuchs
equations of the local B-model.
This is also the set of 
differential operators annihilating the periods $(a,a_D)$ of 
the Seiberg-Witten differential $\lambda_{SW}$
 of the $N=2$ $SU(2)$ gauge theory.
}
\label{setofpartial}
\end{table}
\begin{table}[htop]
\begin{center}
\begin{tabular}{|c|l|}\hline
$N_f$ & $g(u,m_i;\rho)$\\\hline
0&$
\begin{array}{l}
\sum_{n}\frac{(-\frac{1}{4}+\rho)_n(\frac{1}{4}+\rho)_n}{(1+\rho)^2}
(\frac{\Lambda}{u})^{n+\rho-\frac{1}{2}}
\end{array}
$\\\hline
1&$
\begin{array}{l}
\sum_{n_1,n_2}
{{(-{{1}\over{2}}+3\rho)_{n_1+n_2}(-1)^{n_1+n_2}}\over
{(1+2\rho)_{n_1}(1+\rho)_{n_1-n_2}!(-n_1+2n_2)!}}
\bigl({{\Lambda^3}\over{16m_1u}}\bigr)^{n_1-\frac{1}{3}+2\rho}
\bigl({{4m_1^2}\over{u}}\bigr)^{n_2-\frac{1}{6}+\rho}
\end{array}
$\\\hline
2&$
\begin{array}{l}
\sum_{n_1n_2,n_3}
\frac{\Gamma(\rho+1)}{\Gamma(\rho+1+n_1)}
\frac{\Gamma(-\frac{1}{2}+\rho+n_1+n_2)}{\Gamma(-\frac{1}{2}+\rho)}
\frac{\Gamma(\frac{1}{2}+\rho+n_1-n_2+n_3)}{\Gamma(\frac{1}{2}+\rho)}
\\~~~~~\times
\frac{\Gamma(1+\rho)}{\Gamma(1+\rho+n_1-n_2)}
\frac{(-1)^{n_3}}{(n_2-n_3)!(2n_3)!}
\big(\frac{-\Lambda^2}{4v}\big)^{n_1+\rho-\frac{1}{2}}
\big(\frac{n^2}{v}\big)^{n_2}
\big(\frac{4m^2}{b^2}\big)^{n_3}\\
\end{array}
$\\&
$\begin{array}{l}
=\sum_{n_1,l,p}
\frac{\Gamma(1+\rho)}{\Gamma(1+\rho+n_1-2l)}
\frac{\Gamma(1+\rho)}{\Gamma(1+\rho+n_1-2p)}
\frac{(-\frac{1}{2}+\rho)_{n_1+p+l}}{(2l)!(2p)!}\\
~~~~~\times
\frac{\Gamma(\frac{1}{2}+n_1-l-p+\rho)}{\Gamma(\frac{1}{2}+\rho)}
\big(\frac{\Lambda^2}{4v}\big)^{n_1+\rho-\frac{1}{2}}
\big(\frac{-m_1^2}{v}\big)^p
\big(\frac{-m_2^2}{v}\big)^l
\\~~
+\frac{m_1m_2}{v}
\sum_{n_1,l,p}
\frac{\Gamma(1+\rho)}{\Gamma(\rho+n_1-2l)}
\frac{\Gamma(1+\rho)}{\Gamma(\rho+n_1-2p)}
\frac{(-\frac{1}{2}+\rho)_{n_1+p+l+1}}{(2l+1)!(2p+1)!}
\\~~~~~\times
\frac{\Gamma(-\frac{1}{2}+n_1-l-p+\rho)}{\Gamma(\frac{1}{2}+\rho)}
\big(\frac{\Lambda^2}{4v}\big)^{n_1}
\big(\frac{-m_1^2}{v}\big)^p
\big(\frac{-m_2^2}{v}\big)^l
\end{array}
$\\\hline
3&$
\begin{array}{l}
\big(\frac{-w}{\Lambda^2}\big)^{1/2-\rho}
\sum_{n_i}\frac{1}{(-2n_1+n_2-n_4)!n_1!(-n_2+2n_3)}\\
~~~~~\times
\frac{\Gamma(1+2\rho)\Gamma(1+4\rho)\Gamma(-\frac{1}{2}+\rho+n_3)
\Gamma(\frac{1}{2}+2\rho+n_4-n_1)\Gamma(\frac{1}{2}+3\rho+2n_4-n_3)}
{\Gamma(1+2\rho+n_2-2n_3+n_4)\Gamma(1+4\rho-n_2+3n_4)
\Gamma(-\frac{1}{2}+\rho)\Gamma(\frac{1}{2}+2\rho)\Gamma(\frac{1}{2}+3\rho)}
\\~~~~~\times
(-1)^{n_1+n_4}
\big(\frac{n^2}{4m^2}\big)^{n_1}
\big(\frac{m}{p}\big)^{n_2}
\big(\frac{-p^2}{w}\big)^{n_3}
\big(\frac{-\Lambda}{m}\big)^{n_4}
\end{array}$
\\
&$
\begin{array}{l}
=
\big(\frac{-w}{\Lambda^2}\big)^{\frac{1}{2}-\rho}
\sum_{0\leq p,q,k,p+q+k\leq 2n_3}
\frac{(-\frac{1}{2}+\rho)_{n_3}\Gamma(\frac{1}{2}+3\rho+3n_3-2k-2q-2p)}
{p!k!q!\Gamma(\frac{1}{2}+3\rho)}
\\~~~~~\times
\frac{\Gamma(1+2\rho+2n_3-k-p-q)\Gamma(1+2\rho)^2}
{\Gamma(1+2\rho+2n_3-2p-k-q)\Gamma(1+2\rho+2n_3-p-2k-q)}
\\~~~~~\times
\frac{1}
{\Gamma(1+2\rho+2n_3-p-k-2q)}
\big(\frac{-\Lambda^2}{4^2w}\big)^{n_3}
\big(\frac{4p_1}{-\Lambda}\big)^p
\big(\frac{4p_2}{-\Lambda}\big)^q
\big(\frac{4p_3}{-\Lambda}\big)^k
  +{\cal O}(\rho^2)
\end{array}
$\\\hline
\end{tabular}
\end{center}
\caption{$g(u,m_i; \rho)$. $g(u,m_i;0)$ and 
$\partial_{\rho}g(u,m_i;\rho)|_{\rho=0}$
and $m_i$ $(i=1,\dots, N_f)$ 
are solutions of the set of the partial differential equations in 
Table \ref{setofpartial}. 
For $N_f=3$, $p_i=m_i-\frac{\Lambda}{8}$ in the last line.}
\label{solution1}
\end{table}
\begin{table}[htop]
\begin{center}
\begin{equation*}
\begin{array}{|c|l|}\hline
N_f&u\\\hline
0&
\begin{array}{l}
a^2 + \frac{{\Lambda}^4}{8 a^2} + \frac{5 {\Lambda}^8}{512 
      a^6} + \frac{9 {\Lambda}^{12}}{4096 
      a^{10}} + \frac{1469 {\Lambda}^{16}}{2097152 
      a^{14}} + \frac{4471 {\Lambda}^{20}}{16777216 a^{18}}+\cdots
\end{array}
\\\hline
1&
\begin{array}{l}
a^2 + \frac{{\Lambda}^3m_1}{8 
      a^2} + \frac{\Lambda^6(-3a^2+5{{m_1}}^2)}{512 
      a^6} + \frac{\Lambda^9m_1( -7a^2 + 9 {{m_1}}^2 )}{4096 a^{10}}\\
+ \frac{\Lambda^{12}(153 a^4 - 1430 a^2 {{m_1}}^2 +1469 {{m_1}}^4 )}{2097152 
      a^{14}} 
+ \frac{\Lambda^{15}m_1( 1131 a^4 - 5250 
            a^2 {{m_1}}^2 + 4471 {{m_1}}^4 )}{16777216 a^{18}} +\cdots
\end{array}
\\\hline
2&\begin{array}{l}
a^2 + \frac{\Lambda^2 {m_1} {m_2}}{8a^2} \
+ \frac{\Lambda^4( a^4 + 5 {{m_1}}^2 {{m_2}}^2 - 3 
            a^2 ( {{m_1}}^2 + {{m_2}}^2 )  )}{512 a^6} \\
+ \frac{\Lambda^6{m_1} {m_2} ( 5 
            a^4 + 9 {{m_1}}^2 {{m_2}}^2 - 7 
            a^2 ( {{m_1}}^2 + {{m_2}}^2 )  ) }{4096 a^{10}} +\cdots
\end{array}
\\\hline
3&\begin{array}{l}
a^2 + \frac{\Lambda{m_1} {m_2} {m_3} }{8 a^2} 
      \\
+ \frac{\Lambda^2(  a^6 + 5 {{m_1}}^2 {{m_2}}^2 {{m_3}}^2 + 
              a^4 ( {{m_1}}^2 + {{m_2}}^2 + {{m_3}}^2 )  - 3 
            a^2 ( {{m_2}}^2 {{m_3}}^2 + {{m_1}}^2 ( {{m_2}}^2 \
+ {{m_3}}^2 )  )  ) }{512 
      a^6} + \cdots
\end{array}
\\\hline
\end{array}
\end{equation*}
\end{center}
\caption{$u$ 
as an infinite power series in $a$.}
\label{inverta}
\end{table}
\begin{table}[htop]
\begin{center}
\begin{tabular}{|c|l|}\hline
$N_f$ &$t_1, t_2, \cdots, t_k\quad (k=N_f+2)$
\\\hline
0&$\log{{\Lambda^4\epsilon^4}\over{4}},-2\epsilon a
$
\\\hline
1&$\log{{\Lambda^3\epsilon^3}\over{4}},
-\epsilon(m_1+a),-\epsilon(-m_1+a)$\\\hline
2
 &$\log{{\Lambda^2 \epsilon^2}\over{4}},-\epsilon(m_2+a),
 -\epsilon(m_1-m_2),-\epsilon(-m_1+a)$\\\hline
3&
 $\log{{\epsilon \Lambda}\over{4}},-\epsilon(m_2+m_3),-\epsilon(m_1-m_2),
 -\epsilon(a-m_1),-\epsilon(a-m_3)$\\\hline
\end{tabular}
\end{center}
\caption{Behavior of the  K\"ahler parameters
at the gauge theory limit $\epsilon\rightarrow 0$.}
\label{titoa}
\end{table}

Now we equate the prepotential of the local A-model 
${\cal F}_{mirror}$ at the limit $\epsilon \to 0$ 
with 
that of the Seiberg-Witten theory ${\cal F}_{gauge}$.
It is of use to redefine K\"ahler parameters $t_2,\cdots,t_k$ as
\begin{equation}
s_0=-a\epsilon+{\cal O}(\epsilon^2),\qquad
s_i=-m_i\epsilon+{\cal O}(\epsilon^2)\quad (1\leq i\leq N_f).
\end{equation}
The explicit transformation is shown in Table \ref{sN}.
With these parameters, the double logarithmic solution
should coincide with $\partial_{s_0}{\cal F}_{mirror}$.
Then we expect that the relation
\begin{equation}
\partial_{s_0}{\cal F}_{mirror}\sim
\mbox{ const. }+\epsilon a_{D} +{\cal O}(\epsilon^2)
\end{equation}
holds at $\epsilon\to 0$.
It would be straightforward but tedious to check this relation 
by an analytic continuation of the
double logarithmic solution to $\epsilon=0$.
Alternatively, it turned out that we will only need the following behavior
of the Yukawa coupling $\partial_{s_0}{\cal F}_{mirror}$ at 
the limit $\epsilon\to 0$ in next subsection
\begin{equation}\label{condition2}
\epsilon \partial_{s_0}^3{\cal F}_{mirror}\propto
\partial_a^3{\cal F}_{gauge}.
\end{equation}
From the B-model, we obtained the Yukawa coupling 
to some low order in $\Lambda$ in Table \ref{yukawa}.
It is easy to see that the Yukawa coupling has the structure
of $\partial^3{\cal F}_{gauge}$ in (\ref{formofyukawa}).  
\begin{table}[htop]
\begin{equation*}
\begin{array}{|c|l|l|}\hline
\mbox{Model} &(N_0,N_1,\cdots,N_{N_f})&(s_0,\cdots,s_{N_f})\\\hline
0&(2n_2)    &(\frac{t_2}{2})\\\hline
1&(n_2+n_3,n_2-n_3) &(\frac{t_2+t_3}{2},\frac{t_2-t_3}{2})\\\hline
2&(n_2+n_4,n_3-n_4,n_2-n_3) 
 &(\frac{t_2+t_3+t_4}{2},\frac{t_3-t_4}{2},\frac{t_2-t_3}{2})\\\hline
3&(n_4+n_5,n_3-n_4,n_2-n_3,n_2-n_5) 
 &(\frac{t_2+t_3+t_4+t_5}{2},\frac{t_3-t_4}{2},
   \frac{t_2-t_3}{2},\frac{t_2-t_5}{2})\\\hline
\end{array}
\end{equation*}
\caption{$s_0, s_1, \cdots, s_{N_f}$ and $N_0, N_1, \cdots, N_{N_f}$.}
\label{sN}
\end{table}
\begin{table}[htop]
\begin{center}
\begin{tabular}{|c|l|}\hline
$N_f$ &$\epsilon\partial_{s_0}^3{\cal F}_{mirror}$\\\hline
0&
$\begin{array}{l}
-{{8}\over{a}}-{{3\Lambda^4}\over{a^5}}-{{105\Lambda^8}\over{64a^9}}
-{{495\Lambda^{12}}\over{512a^{13}}}-{{154245\Lambda^{16}}
\over{25894a^{17}}}+\cdots
\end{array}$
\\\hline
 1&
 $\begin{array}{l}
-{{8}\over{a}}+{{1}\over{a+m_1}}+{{1}\over{a-m_1}}
-{{3\Lambda^3 m_1}\over{a^5}}+
{{15\Lambda^6(3a^2-14m_1)}\over{128a^9}}\\
+
{{15\Lambda^9(14a^2m_1-33m_1^2)}\over{512a^{13}}}
-{{15\Lambda^{12}(1683a^4-26026a^2m_1^2+41132m_1^4)}\over{1048576a^{17}}}
+\cdots
\end{array}$
\\\hline
2&$\begin{array}{l}
-\frac{8}{a} + \frac{1}{a + {m_1}} + \frac{1}{a - {m_1}} 
 + \frac{1}{a + {m_2}} + \frac{1}{a - {m_2}} 
- \frac{3 \
{\Lambda}^2 {m_1} {m_2}}{a^5}\\
- \frac{3 {\Lambda}^4 ( 
          a^4 + 70 {{m_1}}^2 {{m_2}}^2 - 15 
          a^2 ( {{m_1}}^2 + {{m_2}}^2 )  ) }{128 a^9}
- \frac{5 {\Lambda}^6 {m_1} {m_2} ( 14 
          a^4 + 99 {{m_1}}^2 {{m_2}}^2 - 42 
          a^2 ( {{m_1}}^2 + {{m_2}}^2 )  ) }{512 
      a^{13}}+\cdots 
\end{array}$
\\\hline
3&$
\begin{array}{l}
-\frac{8}{a} +\frac{1}{a + {m_1}} + \frac{1}{a - {m_1}} 
+ \frac{1}{a +{m_2}} 
+ \frac{1}{a - {m_2}} + \frac{1}{a + {m_3}} + \frac{1}{a - {m_3}}
\\
- \frac{3 \Lambda {m_1} {m_2} 
{m_3}}{a^5} - \frac{3 \
{\Lambda}^2 ( 
          70 {{m_1}}^2 {{m_2}}^2 {{m_3}}^2 + 
            a^4 ( {{m_1}}^2 + {{m_2}}^2 + {{m_3}}^2 )  - 15 
          a^2 ( {{m_2}}^2 {{m_3}}^2 + {{m_1}}^2 ( {{m_2}}^2 + \
{{m_3}}^2 )  )  ) }{128 a^9}+\cdots
\end{array}$
\\\hline
\end{tabular}
\end{center}
\caption{The Yukawa coupling $\partial_{s_0}^3{\cal F}_{mirror}$
multiplied by $\epsilon$.}
\label{yukawa}
\end{table}

The rest of this subsection
is devoted to a derivation of 
the behavior of Yukawa coupling (\ref{condition2}).
We explain in the case of $N_f=1$ as an example.
We redefine the K\"ahler parameters $(m_1=m)$
\begin{equation}
s_0:=\frac{t_2+t_3}{2}\sim -a\epsilon,\quad
s_1:=\frac{t_2-t_3}{2}\sim-m\epsilon.
\end{equation}
The Yukawa coupling at the limit $\epsilon\to 0$ is 
evaluated as
\begin{equation}
\begin{split}
\partial_{s_0}^3{\cal F}_{mirror}&=
\Big(\frac{\partial u}{\partial {s_0}}\Big)^3F_{uuu}+
\Big(\frac{\partial u}{\partial {s_0}}\Big)^2
\Big(\frac{\partial m}{\partial s_0}\Big)F_{uum}+
\Big(\frac{\partial u}{\partial s_0}\Big)
\Big(\frac{\partial m}{\partial s_0}\Big)^2F_{umm}+
\Big(\frac{\partial m}{\partial s_0}\Big)^3F_{mmm}
\\
&\sim
-\frac{1}{\epsilon^3}\Big(\frac{\partial u}{\partial a}\Big)^3F_{uuu}.
\end{split}
\end{equation}
The transition from the first line to the second 
follows because all of $F_{uuu},F_{uum},F_{umm},F_{mmm}$ 
have the same order in $\epsilon$ (${\cal O}(\epsilon^2)$).
Then, we calculated the Yukawa coupling $F_{uuu}$ as
\begin{equation}
F_{uuu}=
{{\epsilon^2}}\frac{
64(-3u+4m^2)}{
-256u^2(u-m^2)+32\Lambda^3m(9u-8m^2)-27\Lambda^6}+{\cal O}(\epsilon^3).
\end{equation}
This has been obtained from the Yukawa coupling of the B-model 
$F_{z_iz_jz_p}$ $(0\leq i,j,p\leq 3)$
by the transformation as the contravariant tensor of rank three.
The form shown above is the lowest order term with respect to $z_0$.
On the other hand, we can obtain the expression 
$\partial_a^3{\cal F}_{gauge}$ by (\ref{gaugeyukawa}).
Through the geometric engineering,
we do know the differential operator ${\cal P}$ in (\ref{equationpqr})
from the ${\cal D}_1,{\cal D}_2$ and ${\cal D}_3$. 
$P,Q,R$ in (\ref{equationpqr}) turned out to be
\begin{equation}
\begin{split}
P&=(4m^2-3u)(-256m^2u^2+256u^3+\Lambda^3m(256m^2-288u)+27\Lambda^6),\\
Q&=-2048m^4u+3840m^2u^2-1536u^3-384\Lambda^3m^3+81\Lambda^6,\\
R&=-8(32m^4-72m^2u+24u^2+9\Lambda^3m). 
\end{split}
\end{equation}
Performing an indefinite integration $\int du\,\frac{Q}{P}$,
we can check the relation $F_{uuu}\propto e^{-\int du \frac{Q}{P}}$
up to a factor independent of $u$.
Thus, we have confirmed the relation (\ref{condition2}).
It is straightforward to check (\ref{condition2}) for all 
cases $N_f=0,1,2,3$.
We calculated the Yukawa coupling $F_{uuu}$ in Table in Appendix C,
and obtained the differential operator ${\cal P}$ in Table 
in Appendix D.
We note that the expression of ${\cal P}$ is consistent with 
the Picard-Fuchs equation of the massless case (\ref{pfeq}).

\subsection{Distribution Pattern of 
World-sheet Instanton Numbers}\label{gw}
\hspace*{5mm}
To begin with, let us recall the definition of the Yukawa coupling 
of the local A-model.
It encodes the world-sheet instanton numbers as follows
\begin{equation}
\label{amodelexpression}
\begin{split}
\partial_{s_0}^3{\cal F}_{mirror}&:=\mbox{term of the triple intersection}\\
&~~~+
\sum_{n_1,\cdots,n_{N_f+2}} \, d_{n_1,\cdots,n_k} \,  
N_0^3 \, \frac{e^{n_1t_1+N_0s_0+\sum_{i=1}^{N_f} N_is_i}}
{1-e^{n_1t_1+N_0s_0+\sum_{i=1}^{N_f}N_is_i}}.
\end{split}
\end{equation}
Here, we have introduced $N_0, N_1, \cdots, N_{N_f}$ as coefficients of 
$s_0,\cdots,s_{N_f}$ in $n_2t_2+\cdots+n_{N_f+2}t_{N_f+2}$ 
(see Table \ref{sN}).
On the other hand, we have seen in subsection \ref{swcp1} that 
the third derivative of ${\cal F}_{gauge}$ by $a$ has 
the expansion
\begin{equation}
\label{formofyukawa2}
\begin{split}
\partial_a^3{\cal F}_{gauge}&\propto
-{{8}\over{a}}+\sum_{i=1}^{N_f}
{
\Big(
{{1}\over{a+m_i}}
+
{{1}\over{a-m_i}}
\Big)
}
-{{8}\over{a}}\sum_{n=1}^{\infty}P_n\cdot
\Big({{\Lambda}\over{a}}\Big)^{(4-N_f)n}
,\\
P_n & := \quad
\left(
\begin{array}{l}
\mbox{polynomial of degree $n$ in each $\frac{m_i}{a}$,}\\
\mbox{symmetric with respect to 
$\frac{m_1}{a},\cdots,\frac{m_{N_f}}{a}$}
\end{array}\right),\\
B_n&:=\Big[\mbox{ coefficient of }
\prod_{i=1}^{N_f}\Big(\frac{m_i}{a}\Big)^n \mbox{ in }
P_n\Big]
=\frac{4n(4n-1)(4n-2){\cal F}_n}{4},
\\
{\cal F}_n &\,:=
\Big[ \mbox{ the instanton amplitude of the gauge theory ($N_f=0$)}\Big].
\end{split}
\end{equation}
We have also explained that
${\partial}_{s_0}^3 {\cal F}_{mirror}$'s 
derived from the B-model in Table \ref{yukawa} 
actually have this structure in previous subsection.
Then, combining these matters, it is natural to suspect that 
the world-sheet instanton numbers in (\ref{amodelexpression})
would be translated into the instanton amplitudes of the gauge theory.
Actually, by comparing two instanton expansions in
(\ref{amodelexpression}) and (\ref{formofyukawa2}).
we will obtain the asymptotic distribution of 
the world-sheet instanton number 
${d_{\vec{n}=(n_1,\dots,n_{N_f+2})}}$ at the limit $\epsilon\to 0$
controlled by the instanton amplitude ${\cal F}_n$.
We note that 
the strategy in this section
is essentially the same as one in subsection 2.2 or \cite{CadeGrPa}.

Before proceeding the results, we would like to give several remarks 
on the  expression (\ref{amodelexpression}) 
at the limit $\epsilon\rightarrow 0$.
First, the constant term of the triple intersection can be  neglected
since the Yukawa coupling diverges as ${\cal O}(\epsilon^{-1})$
at the limit $\epsilon\to 0$.
Secondly,
the contributions from the terms 
of the world-sheet instanton numbers with 
$n_1=0$ and the those terms with $n_1\geq 1$ are 
different in that we can neglect 
the factor of the multiple cover contribution 
$1/(1-e^{\sum_i n_it_i})$ when $n_1\geq 1$, but we can not when $n_1=0$.
Thus we will have to treat each contribution separately.
Thirdly, for $n_1\geq 1$, the expansion 
(\ref{amodelexpression}) contains the power series in $\Lambda^{4-N_f}$
because of the factor $e^{n_1t_1}$ in the numerator.
Thus we will compare the contributions of $\Lambda^{(4-N_f)n_1}$
in (\ref{amodelexpression}) and (\ref{formofyukawa2}).
Finally, to read off and guess the 
behavior of the world-sheet instanton numbers,
it will be helpful to consult Tables of them of low degree in Appendix B.

We begin with the analysis of Model $0$ of \cite{KaKlVa}
in a slightly different manner so that 
we can continue to the case of Model $1$ smoothly.
Let us start from the terms with $n_1=0$. 
We can see from Table in Appendix B,
that values of $d_{n_1,n_2}$ is nonzero only at $n_2=1$
at least for a small value of $n_2$.
We note that we ignore the contribution of $d_{0,0}$ here because 
this could not be determined neither from the 
Yukawa coupling
nor from the double logarithmic solution. 
With the value $(n_1,n_2)=(0,1)$, the corresponding
term in (\ref{amodelexpression})
is $\frac{1}{1-e^{t_2}}\sim \frac{1}{2\epsilon a}$ at $\epsilon\to 0$.
On the other hand, we know that the term in (\ref{formofyukawa2}) 
corresponding to the contribution with $n_1=0$ is $\frac{8}{\epsilon a}$ only.
Thus, we naturally expect that $d_{0,n_2}=0$ for all $n_2\geq 2$.
Meanwhile, we turn to the terms with $n_1\geq 1$ . 
The corresponding terms in (\ref{formofyukawa2}) is
${e^{n_1t_1}}{(\epsilon s_0)^{-(4n_1+1)}}$.
This expression must coincide with
$\sum_{n_2} d_{n_1,n_2}e^{n_1t_1+2s_0n_2}$ in (\ref{amodelexpression})
up to a constant factor. 
Now we consider the case
where the contribution from the terms with large values of $n_2$ is
dominant, and the summation could be replaced with an integration.
This specialization 
is sensible because the K\"ahler parameter
 $t_2$ goes to zero at the $\epsilon\to 0$.
Then, recalling the formula $\int dx \, x^{n}\, e^{-b x}=\Gamma(n+1) b^{-n-1}$,
it is natural to adopt the following ansatz 
on $d_{n_1,n_2}$ for a given value of $n_1$
\begin{equation}
d_{0,n_2}=c_1\delta_{n_2,1}\qquad (n_2\neq 0),\hskip 0.5 in 
d_{n_1,n_2}\sim\gamma_{n_1}(2n_2)^{\alpha_{n_1}}\qquad 
(n_1\geq 1, \, n_2\gg n_1).
\end{equation}
Here, $c_1$ is a constant 
and $\gamma_{n_1}, \alpha_{n_1}$ are constants depending on $n_1$.
With this ansatz, we can estimate 
the Yukawa coupling (\ref{amodelexpression}) at the limit 
$\epsilon\to 0$ by the integration of $n_2$
\begin{equation}
\begin{split}
\partial_{s_0}^3{\cal F}_{mirror}
&\sim  {{4c_1}\over{a\epsilon}}+
\gamma_{n_1}{{\Gamma(\alpha_{n_1}+4)\cdot 
(\epsilon^4\Lambda^4)^{n_1}}\over{2(a\epsilon)^{\alpha_{n_1}+4}4^{n_1}}}.\\
\end{split}
\end{equation}
We have shown that this must be of order ${\cal O}(\epsilon^{-1})$
in previous subsection.
Hence, we obtain $\alpha_{n_1}=4n_1-3$. 
Then, the Yukawa coupling becomes
the series in $(\frac{\Lambda}{a})^4$:
$\partial_{s_0}^3{\cal F}_{mirror}\sim \frac{4c_1}{a\epsilon}
[1+
\sum_{n_1=1}^{\infty}\frac{\gamma_{n_1}\Gamma(4n_1+1)}{8c_1\cdot 4^{n_1}}
(\frac{\Lambda}{a})^{4n_1}]
$. 
By comparing this series with
the expansion (\ref{formofyukawa2}),  
we obtain the relation between $\gamma_{n}$ and ${\cal F}_{n}$.
In summary, we arrive at 
the following distribution of the world-sheet instanton numbers 
\begin{equation}\label{f222}
\begin{split}
d_{0,n_2}&= c_1\delta_{n_2,1}\quad\quad~~ (n_2\neq 0),\\
d_{n_1,n_2}&\sim\gamma_{n_1}(2n_2)^{4n_1-3}
\hskip 0.2 in (n_1\geq 1, \, n_2\gg n_1),
\qquad
\frac{\gamma_{n_1}}{c_1}
=\frac{2\cdot 4^{n_1}{\cal F}_{n_1}}{\Gamma(4n_1-2)},
\end{split}
\end{equation}
where ${\cal F}_{n}$ is the instanton amplitude
of the gauge theory with $N_f=0$.

Next we explain how the result of Model $0$ is extended to  Model $1$. 
To grasp a matter, let us consult Table of the world-sheet instanton
numbers in Appendix B again.
First, for the numbers $d_{n_1,n_2,n_3}$ with $n_1=0$,
we can see that nonzero values 
exist only at $(n_2,n_3)=(1,0),(0,1)$ and $(1,1)$. 
These values of $(n_1,n_2,n_3)$ give rise to the contributions
$\frac{1}{1-e^{-\epsilon(a\pm m)}}\sim\frac{1}{\epsilon (a\pm m)}$
and $\frac{1}{1-e^{-2a\epsilon}}\sim \frac{1}{2\epsilon a}$
in (\ref{amodelexpression}) $(m_1=m)$.
On the other hand, the terms corresponding to $n_1=0$ 
in (\ref{formofyukawa2}) give rise to just enough contributions. 
Thus,
we can conclude that $d_{0,n_2,n_3}=0$ 
unless $(n_2,n_3)=(1,0),(0,1)$ or $(1,1)$.
Then, all we have to do will be to determine the numbers 
$d_{0,1,0},d_{0,0,1}$ and $d_{0,1,1}$ with $n_1=0$. 
Secondly, we proceed to the numbers $d_{n_1,n_2,n_3}$ with $n_1\geq 1$.
From the Table, it is likely that there is a rule that
$d_{n_1,n_2,n_3}=0$ unless $0 \leq n_2-n_3 \leq n_1$.
Furthermore, the distribution of $|d_{n_1,n_2,n_3}|$ 
looks like a binomial distribution
centered around $n_2-n_3=\frac{n_1}{2}$
and the sign of $d_{n_1,n_2,n_3}$ seems to be $(-1)^{n_2-n_3}$.
Then, we arrive at the following ansatz on $d_{n_1,n_2,n_3}$
\begin{equation}
\label{ansatz1}
\begin{split}
d_{0,n_2,n_3}&=
c_{10}\delta_{n_2,1}\delta_{n_3,0}
+c_{01}\delta_{n_2,0}\delta_{n_3,1}+
c_{11}\delta_{n_2,1}\delta_{n_3,1}
\qquad((n_2,n_3)\not=(0,0)),\\
d_{n_1,n_2,n_3}&\sim
\gamma_{n_1}\, (-1)^{n_2-n_3}\, (n_2+n_3)^{\beta_{n_1}}
\, {_{n_1} C_{n_2-n_3}}
\hskip 0.4 in (n_1\geq 1, \, n_2+n_3 \gg n_1),
\end{split}
\end{equation}
where $c_{ij}$'s are constants, 
and $\beta_{n_1}$, $\gamma_{n_1}$ are numbers depending on  $ n_1$.
Substituting this into (\ref{amodelexpression}) and changing the 
variables from $(n_2, n_3)$ to $N_0=n_2+n_3$ and $N_1=n_2-n_3$, 
the Yukawa coupling is evaluated in the region $N_0\gg n_1$ 
as follows
\begin{eqnarray}
\partial_{s_0}^3{\cal F}_{mirror}&&\sim {{1}\over{\epsilon}}\Big[
{{4c_{11}}\over{a}}+{{c_{10}}\over{a+m}}+{{c_{01}}\over{a-m}}
\Big]
\nonumber\\
&&~~~
+\frac{1}{2}\sum_{n_1=1}^{\infty}\, \gamma_{n_1}\, 
\Big({{\epsilon^3\Lambda^3}\over{4}}\Big)^{n_1}
\, \int d{N_0} \, N_0^{\beta_{n_1}+3} \, e^{-\epsilon a N_0}
\, \sum_{N_1=0}^{n_1} \, (-1)^{N_1} \, {_{n_1} C_{N_1}} \, 
e^{-\epsilon m N_1},
\nonumber\\
&&\sim
{{1}\over{\epsilon}}\Big[
{{4c_{11}}\over{a}}+{{c_{10}}\over{a+m}}+{{c_{01}}\over{a-m}}
\Big]+\frac{1}{2}\sum_{n_1=1}^{\infty}
\, {\gamma_{n_1}}\, \Big({{\epsilon^3\Lambda^3}\over{4}}\Big)^{n_1}
\, {{\Gamma(\beta_{n_1}+4)}\over{(a\epsilon)^{\beta_{n_1}+4}}}
\, (m\epsilon)^{n_1}.
\end{eqnarray}
The factor $\frac{1}{2}$ before the summation 
has  entered by the Jacobian of the change of  variables. 
By imposing that the Yukawa coupling is ${\cal O}(\epsilon^{-1})$ and 
the ratio of $\frac{1}{a}$ to $\frac{1}{a\pm m}$ is $8$ to $-1$,
we obtain
\begin{equation}
\label{constnf1}
c_{10}=c_{01}=-\frac{c_{11}}{2},\quad
\beta_{n_1}=4n_1-3.
\end{equation}
After that, the Yukawa coupling is expressed as
\begin{equation}
\partial_{s_0}^3{\cal F}_{mirror}\sim
-\frac{c_{11}}{2a\epsilon}\Big[
-\frac{8}{a}+\frac{1}{a+m}+\frac{1}{a-m}-
\sum_{n_1=1}^{\infty}\gamma_{n_1}
\frac{\Gamma(4n_1+1)}{4^{n_1} c_1}
\Big(\frac{m}{a}\Big)^{n_1}
\Big(\frac{\Lambda}{a}\Big)^{3n_1}
\Big].
\label{temp3}
\end{equation}
Comparing this with (\ref{formofyukawa2}),
we can relate $\gamma_{n_1}$ to ${\cal F}_{n_1}$
\begin{equation}
\label{gammatof}
\frac{\gamma_{n_1}}{c_{11}}
=\frac{2\cdot 4^{n_1}{\cal F}_{n_1}}{\Gamma(4n_1-2)}.
\end{equation}
Therefore, we arrived at the asymptotic behavior of $d_{n_1,n_2,n_3}$
(\ref{ansatz1}) with (\ref{constnf1}) and (\ref{gammatof}). 
We note that only the terms of highest degree in $m$ in 
(\ref{formofyukawa2}) have
appeared in (\ref{temp3}).
To reproduce the remaining contributions, one may be tempted to 
adopt alternative ansatz;
$d_{n_1,n_2,n_3}\sim
\sum_{\mu=0}^{\mbox{\tiny Min}[N_1,n_1]}\gamma_{n_1}^{\mu}
N_0^{\beta_{n_1}^{\mu}}
{_{\mu}}C_{N_1}$
sorted out according to each value of $\mu$. 
Then one would find that $\beta_{n_1}^{\mu}=3n_1+\mu-3$ and 
$\gamma_{n_1}^{\mu}$ can be written in terms of 
the coefficients of $\frac{1}{a\epsilon}
(\frac{m}{a})^{\mu}(\frac{\Lambda}{a})^{3\mu}$.
However, given that $N_0\gg n_1$, it is clear that
the most dominant term in the ansatz is the term with $\mu=n_1$.  
Thus, we do not extract the subleading contributions 
in the strategy here.

The extension to Model $2$ and Model $3$ is now straightforward.
We propose the distribution of the
world-sheet instanton numbers at the gauge theory limit $\epsilon\to 0$
altogether with the notation in Table \ref{sN} as follows:
for $n_1\geq 1$,
\begin{equation}
d_{n_1,n_2,\cdots ,n_{N_f+2}}
\left\{
\begin{array}{ll}
\sim \gamma_{n_1}\, (-1)^{N_1+\cdots+N_{N_f}}\, 
N_0^{4n_1-3} \, \prod_{i=1}^{N_f} \,  {_{n_1}}C_{N_i}
&~~~(0 \leq {^\forall}N_i\leq n_1)\\
=0 &~~~\mbox{otherwise}\\
\end{array}
\right\},
\end{equation}
where this is effective for the region $N_0\gg n_1$;  for $n_1=0$,
\begin{equation}
\begin{split}
&d_{0,\vec{n}^{\prime}}=c\Big[\delta_{(\vec{n}^{\prime}),(1,\cdots,1)}
-\frac{1}{2}\sum_{\alpha\in I}\delta_{\vec{n}^{\prime},\alpha}\Big]
\qquad\qquad (\vec{n}'=(n_2,\cdots,n_{N_f})\not=\vec{0}),
\\
I=&\left\{
\begin{array}{cl}
\mbox{Model 0}&\mbox{empty set}\\           
\mbox{Model 1}&\{(1,0),(0,1)\}\\
\mbox{Model 2}&\{(1,0,0),(1,1,0),(0,1,1),(0,0,1)\}\\
\mbox{Model 3}&\{(0,0,0,1),(0,0,1,0),(0,1,1,0),(1,0,0,1),
(1,1,1,0),(1,1,0,1)\}
\end{array}\right\},
\end{split}
\end{equation}
where the constants $c$ and $\gamma_{n_1}$ are correlated as
\begin{equation}
\frac{\gamma_{n_1}}{c}
=\frac{2\cdot 4^{n_1}{\cal F}_{n_1}}{\Gamma(4n_1-2)}.
\end{equation}
Here, $\delta_{\vec{q},\vec{q^{\prime}}}:=
\prod_{i=2}^{N_f+2}\delta_{q_i,q^{\prime}_i}$ and
${\cal F}_n$ is the instanton amplitude of the gauge theory with $N_f=0$.

It is instructive to state the result with the basis of 
$H_2(V,{\bf Z})$ in Table \ref{newbasis}.
At the gauge theory limit $\epsilon \to 0$,
${\bf P}^1_b, R_0$ and $R_i$ $(i=1 \leq i \leq N_f)$ are
the curves whose K\"ahler parameter behave as
$(4-N_f)\log \epsilon, -2\epsilon a$ and $-\epsilon(a+m_i)$.
Then, the intersection matrices among the curves
$({\bf P}^1_b,R_0,R_1,\cdots,R_{N_f})$ 
in $V$ of Model $N_f$ ($N_f=0,1,2$ and  $3$) are 
\begin{align}
\begin{bmatrix}-2&1\\1&0\end{bmatrix},\quad
\begin{bmatrix}
-2&1&0\\1&0&0\\0&0&-1
\end{bmatrix},\quad
\begin{bmatrix}
-2&1&0&0\\1&0&0&0\\0&0&-1&0\\0&0&0&-1
\end{bmatrix},\quad
\begin{bmatrix}
-2&1&0&0&0\\1&0&0&0&0\\0&0&-1&0&0\\0&0&0&-1&0\\0&0&0&0&-1
\end{bmatrix}.
\end{align}
These matrices make clear the structure of the ${\bf P}_{base}$
of Model $N_f$ as $N_f$ point blow ups of ${\bf F}_2$.
Now we state the distribution of the world-sheet instanton numbers $d_{\beta}$ 
with homology class $\beta=n_1[{\bf P}^1_b]+\sum_{i=0}^{N_f}k_i[R_i]$:
for $n_1\geq 1$
\begin{equation}
d_{\beta}\sim \gamma_{n_1}(-1)^{k_1+\dots+k_{N_f}}
(2k_0)^{4n_1-3}
\prod_{i=1}^{N_f}{_{n_1}}C_{k_i}\qquad 
(2k_0\gg n_1, \, 0\leq {}^\forall k_i\leq n_1),
\end{equation}
and for other $\beta$'s, $d_{\beta}$ is negligible;
for $n_1=0$,
$d_{\beta}\not=0$ only at
$\beta=[R_0],[R_i]$ or $[R_0-R_i]$
with ratio between the values of $[R_0]$ and the others is always $2$ to
$-1$. 
Constants in these expressions are
normalized using the instanton amplitude ${\cal F}_n$ 
\begin{equation}
\frac{\gamma_{n_1}}{d_{\beta=[R_0]}}
=\frac{2\cdot 4^{n_1}{\cal F}_{n_1}}{\Gamma(4n_1-2)}.   
\end{equation}
\begin{table}[htop]
\begin{equation*}
\begin{array}{|c|l|}\hline
N_f&({\bf P}^1_b;R_0,R_1,\cdots,R_{N_f})\\\hline
0&(D_3;D_2)\\
1&(D_4;D_5, D_2)\\
2&(D_5;D_6, D_2+D_3, D_4)\\
3&(D_5;D_6+D_7, D_3+D_4, D_4, D_6)\\\hline
\end{array}
\end{equation*}
\caption{ Basis of $H_2(V;{\bf Z})$.
$D_i$ is the divisor in ${\bf P}_{base}$ 
corresponding to the 1-cone generated by $\nu_i$ in Table \ref{onecones}.}
\label{newbasis}
\end{table}

\section{Conclusion}
\hspace{5mm}
In this article, we presented the calculation 
for the instanton expansions in the $N=2$ $SU(2)$ gauge theories 
with $1 \leq N_f \leq 3$ massive hypermultiplets 
through the geometric engineering.
We checked the equivalence of the Yukawa coupling 
at the gauge theory limit and $\partial_a^3 {\cal F}_{gauge}$, 
and conjectured the pattern of the
distribution of the world-sheet instanton numbers. 
This proposal matches with general expectation 
in the geometric engineering of $N=2$ gauge theories
that the asymptotic growth of 
the world-sheet instanton numbers
is controlled by the instanton amplitude of the gauge theory.
Further, it might be the universal phenomenon for the mirror pair 
of Calabi-Yau manifolds that the 
distribution of the world-sheet instanton numbers
is governed by a singularity of the moduli space
where the discriminant of the B-model manifold becomes zero.    
We also analyzed the asymptotic form of the instanton amplitude 
of the gauge theory with massless hypermultiplets,
making use of the singularity of the moduli space at $u=\Lambda^2$,
and observed the characteristic factor. 
In principle, it should be possible to clarify 
how the factor originates from the analysis 
on the instanton background in the gauge theory
\cite{DoKhMa, AoHaSaWa, ItSa}.
There has been some developments in the direct evaluation 
of the instanton amplitude of Seiberg-Witten theories 
\cite{Ho, Ne, FlPo, BrFuMoTa} and so on.
It would be very interesting to relate this with the 
localization technique of the world-sheet instanton numbers 
\cite{ChKlYaZa}.

The geometric engineering of the $N=2$ $SU(2)$ gauge theory 
with $N_f=4$ hypermultiplets remains to be done.
It would be very interesting to understand how 
physics of the $N_f=4$ theory are geometrically realized 
in the local mirror model.
Then, we mention an extension of the $N=2$ $SU(2)$ gauge theories 
into five-dimensional gauge theories.
The five-dimensional theories receive no instanton corrections
and it is known how to construct geometrically 
the gauge theories with $N_f \leq 4$ hypermultiplets \cite{EgKa}. 
It would be also interesting to tailor the local mirror models
to the five-dimensional gauge theories exactly.
Meanwhile, there has been a surge of developments
in our understanding of supersymmetric gauge theories \cite{DiVa}.
It would be interesting 
to develop a tool for our instanton expansions in this context.

\section*{Acknowledgement}
\hspace{5mm}
We thank T. Eguchi, S. Hosono and M. Jinzenji for helpful discussions.
Research of M.N. is supported by Japanese Society for the Promotion of
Science under Post-doctorial Research Program (No. 0206911).

\section*{Appendix A: GKZ-hypergeometric System}
\hspace{5mm}
In this appendix we give the definition of the GKZ-hypergeometric 
differential system
(GKZ-system for short). 
For details, see \cite{Sturmfels, GeKaZe, SaStTa, HoLiYa}.

Let $A$ be an $n\times (n+k)$ matrix of integers, which has the
following properties: (a) the columns $A_1,\cdots,A_{n+k}$ of
$A$ generate the lattice ${\bf Z}^n$, (b)
there exist integers $(c_1,\cdots,c_{n+k})$ such that 
$\sum_{i=1}^n c_i \, (i\mbox{-th row of }A)=\overbrace{(1,\cdots,1)}^{n+k}$.
Let $\beta\in{\bf C}^n$. Then  the GKZ hypergeometric system is the 
left ideal in the Weyl algebra of dimensions $n+k$. 
We denote the variables by 
$a_1,\cdots,a_{n+k}$. Its generators are
\begin{align}
{\cal Z}_i&=\sum_{i=1}^{n+k}A_{ij}\theta_{a_j}-\beta_j, 
\hskip 0.2 in (1\leq i\leq n),
\\
I_A&=\Bigl\{
\prod_{i;l_i>0}\Big({{\partial}\over{\partial a_i}}\Big)^{l_i}-
\prod_{i;l_i<0}\Big({{\partial}\over{\partial a_i}}\Big)^{-l_i}
\Big |\,\, l\in L
\Bigr\},
\\
L&=\bigl\{
l\in{\bf Z}^{n+k}\big | A\cdot l=\overbrace{(0,\cdots,0)}^{n}
\bigr\}.
\end{align}
$I_A$ is called the toric ideal.
By the property (b), the system is regular holonomic.

The finite set of generators of the toric ideal can be obtained as follows 
(Algorithm 4.5 of \cite{Sturmfels}). 
We write the column vectors $A_i \, (1\leq i\leq n+k)$ 
of $A$ as $A_i^+-A_i^-$, where $j$-th component of 
$A_i^+$ (resp. $A_i^-$) is the $j$-th component of 
$A_i$ if it is positive (resp. negative), otherwise zero.
And consider the ideal $I_0$ in ${\bf Q} \, \langle
x_1, \cdots, x_{n+k}, t_0, t_1, \cdots, t_n
\rangle$ whose generators are
$t_0t_1\cdots t_n-1$ and 
$ x_i(\prod_{j;(A_i)_j>0}t_j^{(A_i)_j})-
(\prod_{j;(A_i)_j<0}t_j^{-(A_i)_j})\, (1\leq i\leq n+k)$. 
Then the generators of $I_0\cap {\bf Q}\,[x_1, \cdots, x_{n+k}]$
are the generators of $I_A$ with the identification of $x_i$ and 
$\partial_{a_i}$.

Consider the case where the first row of $A$ is 
$(1, \cdots, 1)$ and write $A$ in the following form
\begin{eqnarray}
A=\left(
\begin{array}{ccc}
1&\cdots&1\\
\nu_1&\cdots&\nu_{n+k}
\end{array}
\right)
,\hskip 0.3 in
\nu_i\in {\bf Z}^{n-1}, 
\hskip 0.2 in (1\leq i\leq n+k).
\end{eqnarray}
Then there is the following formal solution to the GKZ-system 
in the integral form
\begin{equation}
\int \, P(X)^{\beta_1} \, {{dX_1}\over{X_1^{\beta_2+1}}}
\wedge\cdots\wedge{{dX_{n-1}}\over{X_{n-1}^{\beta_n+1}}},
\end{equation}
where
$
P(X)=\sum_{i=1}^{n+k}a_i\prod_{j=1}^{n-1}X_j^{(\nu_i)_j}
$.
The proof is found in \cite{GeKaZe}.
This statement for $\beta_1=0$
is trivial. 
However,
the analysis of the local B-model 
suggests that 
$\int\,\log P(X) \prod_{j=1}^{n-1}\frac{dX_j}{X_j^{\beta_{j+1}+1}}$
might be the formal solution to the GKZ-system with  $\beta_1=0$.

\section*{Appendix B: World-sheet Instanton Numbers}
\hspace{5mm}
Following tables are the world-sheet instanton numbers
of low degree for Model $0$ and $1$ \cite{ChKlYaZa}.

\noindent Model $0$ : $d_{n_1,n_2}$.
The degree $n_1$ (resp. $n_2$) grows down (resp. right). 
\begin{center}
$\left[
\begin{array}{ccccccc}
d_{0,0}&-2&0&0&0&0&0\\
d_{1,0}&-2&-4&-6&-8&-10&-12\\
d_{2,0}&0&0&-6&-32&-110&-288\\
d_{3,0}&0&0&0&-8&-110&-756\\
d_{4,0}&0&0&0&0&-10&-288\\
d_{5,0}&0&0&0&0&0&-12\\
\end{array}
\right]$
\end{center}

\noindent Model $1$ : $d_{n_1,n_2,n_3}$. 
The degree $n_2$ (resp. $n_3$) grows down (resp. right).
\begin{center}
\begin{tabular}{|c|c|c|c|}\hline
$n_1$&
$d_{n_1,n_2,n_3}$
&$n_1$&
$d_{n_1,n_2,n_3}$
\\\hline
0&$\left[\begin{array}{cccccc}
d_{0,0,0}&2&0&0&0&0\\
2&-4&0&0&0&0\\
0&0& 0&0&0&0\\
0&0& 0&0&0&0\\
0&0&0 &0&0&0\\
0&0&0 &0&0&0\\
\end{array}\right]$
&1&$\left[\begin{array}{cccccc}
d_{1,0,0}&0&0&0&0&0\\
2&-4&0&0&0&0\\
0&6&-8&0&0&0\\
0&0&10&-12&0&0\\
0&0&0&14&-16&0\\
0&0&0&0&18&-20\\
\end{array}\right]$
\\\hline
2&$\left[\begin{array}{cccccc}
d_{2,0,0}&0&0&0&0&0\\
0&0&0&0&0&0\\
0&0&0&0&0&0\\
0&0&10&-12&0&0\\
0&0&-12&70&-64&0\\
0&0&0&-64&270&-220\\
\end{array}\right]$
&3&$\left[\begin{array}{cccccc}
d_{3,0,0}&0&0&0&0&0\\
0&0&0&0&0&0\\
0&0&0&0&0&0\\
0&0&0&0&0&0\\
0&0&0&14&-16&0\\
0&0&0&-64&270&-220\\
\end{array}\right]$\\\hline
4&$\left[\begin{array}{cccccc}
d_{4,0,0}&0&0&0&0&0\\
0&0&0&0&0&0\\
0&0&0&0&0&0\\
0&0&0&0&0&0\\
0&0&0&0&0&0\\
0&0&0&0&18&-20\\
\end{array}\right]$
&5&$\left[\begin{array}{cccccc}
d_{5,0,0}&0&0&0&0&0\\
0&0&0&0&0&0\\
0&0&0&0&0&0\\
0&0&0&0&0&0\\
0&0&0&0&0&0\\
0&0&0&0&0&0\\
\end{array}\right]$\\\hline
\end{tabular}
\end{center}

\section*{Appendix C: Yukawa Coupling at the Gauge Theory Limit}
\hspace{5mm}
We present the form of
the Yukawa coupling $F_{uuu}$ (or $F_{vvv}, F_{www}$
for $N_f=2, 3$) in the leading order 
at $\epsilon\to 0$. 
The parameters are chosen as $m=m_1$
for $N_f=1$, and as in Table \ref{ztoep} for $N_f=2,3$.
$\triangle_2, \triangle_3$ 
in the denominator of the form for $N_f=2, 3$ are also given below.

\begin{center}
\begin{tabular}{|c|l|}\hline
$N_f$ & \\\hline
0&$F_{uuu}={{\epsilon^2}\over{8(u^2-\Lambda^4)}}$
\\\hline
1&
$F_{uuu}=
{{\epsilon^2}}\frac{
64(-3u+4m^2)}{
-256u^2(u-m^2)+32\Lambda^3(9um-8m^3)-27\Lambda^6}$
\\\hline
2&
$F_{vvv}=\frac{\epsilon^2}{\triangle_2}
[-8m^4+16m^2n^2-8n^4+12m^2v+12n^2v-4v^2+{\Lambda}^2 (m^2 +2n^2-v)]$
\\\hline
3&
$F_{www}={{\epsilon^2}\over{\triangle_3}}
[-32m^4p^2+64m^2n^2p^2-32n^4p^2+24m^4w-48m^2n^2w+24n^4w$\\
&~~~~~~~~~~~~~~~~~~$
+48m^2p^2w+48n^2p^2w-32m^2w^2-32n^2w^2-16p^2w^2+8w^3$\\
&~~~~~~~~~~~~~~~~~$
+\Lambda(6m^5-12m^3n^2+6mn^4-9m^4p+10m^2n^2p-n^4p-4m^3p^2$\\
&~~~~~~~~~~~~~~~~~~~$
+28mn^2p^2+4m^2p^3+8n^2p^3-4m^3w-28mn^2w+10m^2pw$\\
&~~~~~~~~~~~~~~~~~~~$-2n^2pw+4mp^2w-4p^3w-2mw^2-pw^2)$\\
&~~~~~~~~~~~~~~~~~$+\Lambda^2(m^4-2m^2n^2+n^4-2m^3p+2mn^2p$\\
&~~~~~~~~~~~~~~~~~~~$+m^2p^2+3n^2p^2-m^2w-3n^2w+2mpw-p^2w)]$
\\\hline
\end{tabular}
\end{center}
\begin{eqnarray}
&&\triangle_2=
-48 v^2(m^4 +{( n^2 - v )}^2-2m^2(n^2+v))+4m^2p^3+8n^2p^3-4m^3w\nonumber\\
&&~~~~~~~~~~-28mn^2w+10m^2pw-2n^2pw+4mp^2w-4p^3w-2mw^2-pw^2
\nonumber\\
&&~~~~~~~
+24{\Lambda}^2(2m^6-2n^6+4n^4v-n^2v^2-v^3-m^4(6n^2+5v)+m^2(6n^4+n^2v+4v^2))
\nonumber\\
&&~~~~~~~
-3{\Lambda}^4(m^4-8n^4+8n^2v+v^2-2m^2(10n^2+v))+3{\Lambda}^6n^2,
\nonumber 
\end{eqnarray}
\begin{eqnarray}
&&\triangle_3=
\frac{-5}{8}[
256m^4p^2w^2-512m^2n^2p^2w^2
+256n^4p^2w^2-256m^4w^3+512m^2n^2w^3
\nonumber\\
&&~~~~~~~~~~~~~~~
-256n^4w^3-512m^2p^2w^3-512n^2p^2w^3+512m^2w^4+512n^2w^4\nonumber\\
&&~~~~~~~~~~~~~~~
+256p^2w^4-256w^5\nonumber\\
&&~~~~~~~~~~~~~
+\Lambda(-256m^6p^3+768m^4n^2p^3-768m^2n^4p^3+256n^6p^3+288m^6pw\nonumber\\
&&~~~~~~~~~~~~~~~~~
-864m^4n^2pw+864m^2n^4pw-288n^6pw+128m^5p^2w-256m^3n^2p^2w\nonumber\\
&&~~~~~~~~~~~~~~~~~
+128mn^4p^2w+640m^4p^3w-128m^2n^2p^3w-512n^4p^3w-192m^5w^2\nonumber\\
&&~~~~~~~~~~~~~~~~~
+384m^3n^2w^2-192mn^4w^2-672m^4pw^2+64m^2n^2pw^2+608n^4pw^2\nonumber\\
&&~~~~~~~~~~~~~~~~~~~
-256m^3p^2w^2-512mn^2p^2w^2-512m^2p^3w^2+128n^2p^3w^2+384m^3w^3\nonumber\\
&&~~~~~~~~~~~~~~~~~~~
+640mn^2w^3+480m^2pw^3-224n^2pw^3+128mp^2w^3+128p^3w^3\nonumber\\
&&~~~~~~~~~~~~~~~~~~~
-192mw^4-96pw^4)\nonumber\\
&&~~~~~~~~~~~~~~~
+{\Lambda}^2(-27m^8+108m^6n^2-162m^4n^4+108m^2n^6
-27n^8+72m^7p-216m^5n^2p\nonumber\\
&&~~~~~~~~~~~~~~~~~~~
+216m^3n^4p-72mn^6p-56m^6p^2+160m^4n^2p^2-152m^2n^4p^2+48n^6p^2\nonumber\\
&&~~~~~~~~~~~~~~~~~~~
-32m^5p^3+352m^3n^2p^3-320mn^4p^3+16m^4p^4-320m^2n^2p^4-128n^4p^4\nonumber\\
&&~~~~~~~~~~~~~~~~~~~
+60m^6w-84m^4n^2w-12m^2n^4w+36n^6w-120m^5pw-400m^3n^2pw\nonumber\\
&&~~~~~~~~~~~~~~~~~~~
+520mn^4pw +136m^4p^2w+192m^2n^2p^2w+56n^4p^2w+64m^3p^3w\nonumber\\
&&~~~~~~~~~~~~~~~~~~~
+32mn^2p^3w
-32m^2p^4w+128n^2p^4w-66m^4w^2+324m^2n^2w^2-2n^4w^2\nonumber\\
&&~~~~~~~~~~~~~~~~~~~
+24m^3pw^2
-152mn^2pw^2-104m^2p^2w^2-160n^2p^2w^2-32mp^3w^2+16p^4w^2\nonumber\\
&&~~~~~~~~~~~~~~~~~~~
+60m^2w^3+36n^2w^3+24mpw^3+24p^2w^3-27w^4)\nonumber
\end{eqnarray}
\begin{eqnarray}
&&~~~~~~~~~~~~~~~
+{\Lambda}^3(-4m^7+44m^5n^2-76m^3n^4
+36mn^6+12m^6p-140m^4n^2p+116m^2n^4p\nonumber\\
&&~~~~~~~~~~~~~~~~~~~
+12n^6p-12m^5p^2+168m^3n^2p^2-28mn^4p^2+4m^4p^3-40m^2n^2p^3-60n^4p^3
\nonumber\\ 
&&~~~~~~~~~~~~~~~~~~~
-48mn^2p^4+16n^2p^5+8m^5w-16m^3n^2w+8mn^4w-24m^4pw-16m^2n^2pw\nonumber\\
&&~~~~~~~~~~~~~~~~~~~
+40n^4pw+24m^3p^2w+8mn^2p^2w-8m^2p^3w+24n^2p^3w-4m^3w^2+36mn^2w^2\nonumber\\
&&~~~~~~~~~~~~~~~~~~~
+12m^2pw^2-36n^2pw^2-12mp^2w^2+4p^3w^2)\nonumber\\
&&~~~~~~~~~~~~~~~
+{\Lambda}^4(4m^4n^2-8m^2n^4+4n^6
-16m^3n^2p+16mn^4p+24m^2n^2p^2-8n^4p^2\nonumber\\
&&~~~~~~~~~~~~~~~~~~~
-16mn^2p^3+4n^2p^4)].\nonumber
\end{eqnarray}

\section*{Appendix D: Differential Operators for the Periods}
\hspace{5mm}
We show the differential operators (\ref{equationpqr})  
for $N_f=0,1$ and $2$.

\begin{center}
\begin{tabular}{|c|l|}\hline
$N_f$&$
{\cal P}=P(u,m_i)
\partial_u^3+Q(u,m_i)\partial_u^2+R(u,m_i)\partial_u
$\\\hline
0&$\begin{array}{l}
P=u^2-\Lambda^4\\
Q=2u\\
R=\frac{1}{4}
\end{array}$
\\\hline
1&$
\begin{array}{l}
P=(4m^2-3u)(-256m^2u^2+256u^3+\Lambda^3m(256m^2-288u)+27\Lambda^6)\\
Q=-2048m^4u+3840m^2u^2-1536u^3-384\Lambda^3m^3+81\Lambda^6\\
R=-8(32m^4-72m^2u+24u^2+9\Lambda^3m) 
\end{array}
$\\\hline
2&$
\begin{array}{l}
P=[-4(2m^4+2n^4-3n^2v+v^2-m^2(4n^2+3v))+\Lambda^2(m^2+2n^2-v)]  
\\
\hskip 0.3 in 
\times 
[16v^2(m^4+(n^2-v)^2-2m^2(n^2+v))
\\
\hskip 0.5 in 
-8\Lambda^2(2m^6-2n^6+4n^4v-n^2v^2-v^3-m^4(6n^2+5v)
\\
\hskip 0.7 in
+m^2(6n^4+n^2v+4v^2))
\\
\hskip 0.5 in
+\Lambda^4(m^4-8n^4+8n^2v+v^2-2m^2(10n^2+v))+\Lambda^6n^2]
\\
Q=
-64v(4m^8 - m^6 ( 16 n^2 + 15 v )+{( n^2 - v ) }^2(4n^4-7n^2 v + 2 v^2 )   
\\\hskip 0.6 in
+m^4(24n^4+15n^2 v+20 v^2)+m^2 ( -16 n^6 +15n^4v+8n^2v^2-11v^3))   
\\\hskip 0.35 in
-16{\Lambda}^2 (8m^8-4n^8-4n^6v+27n^4v^2-24n^2v^3+5v^4-2m^6(10n^2+13v)  
\\\hskip 0.6 in
+ 3m^4(4n^4+16n^2v+11v^2)+2m^2(2n^6-9n^4v+6n^2v^2-10v^3)) 
\\
\hskip 0.35 in
+4{\Lambda}^4(7m^6-4n^6+9m^4(7n^2-2v)
-12n^4v+21n^2v^2-4v^3\\
\hskip 0.6 in 
+3m^2(32n^4-20n^2v+5 v^2))  
\\\hskip 0.35 in 
+{\Lambda}^6(-m^4-28m^2n^2-4n^4+2m^2v+12n^2v-v^2)+\Lambda^8 n^2   
\\
R=
-16(2m^8-m^6( 8 n^2 + 9v)+{( n^2 - v ) }^2(2n^4-5n^2v+v^2)   
\\\hskip 0.6 in
+m^4(12n^4+9n^2v+13v^2)-m^2(8n^6-9n^4v+2n^2v^2+7v^3)) 
\\\hskip 0.35 in
+4{\Lambda}^2(m^6+8n^6-22n^4v+16n^2v^2-2v^3-2m^4(15n^2+2v)
\\\hskip 0.6 in
+m^2(21n^4+2n^2v+5v^2 ))   
\\\hskip 0.35 in  
+{\Lambda}^4(-m^4-13m^2n^2-10n^4+2m^2v+13n^2v-v^2)+\Lambda^6n^2 
\end{array}$
\\\hline
\end{tabular}
\end{center}



\begin{thebibliography}{99}
\bibitem{KaKlVa}
S. Katz, A. Klemm and C. Vafa,
``{\it Geometric Engineering of Quantum Field Theories}'',
Nucl. Phys. ${\bf B497}$ (1997) 173, hep-th/9609239.
\bibitem{KaMaVa}
S. Katz, P. Mayr and C. Vafa,
``{\it Mirror Symmetry and Exact Solution of $4D$ $N=2$ Gauge Theories: I}'',
Adv. Theor. Math. Phys. ${\bf 1}$ (1998) 53, hep-th/9706110.
\bibitem{AgGr}
M. Aganagic and M. Gremm,
``{\it Exact Solutions for Some $N=2$ Supersymmetric $SO(N)$ Gauge 
Theories with Vectors and Spinors}'', Nucl. Phys. ${\bf B524}$ (1998)
207, hep-th/9712011.
\bibitem{HaTe}
J. Hashiba and S. Terashima, 
``{\it Geometry and $N=2$ Exceptional Gauge Theories}'',
JHEP ${\bf 9909}$ (1999) 020, hep-th/9909032.
\bibitem{SeWi}
N. Seiberg and E. Witten,
``{\it Electric-Magnetic Duality,
Monopole Condensation, and Confinement
in $N=2$ Supersymmetric Yang-Mills Theory}'',
Nucl. Phys. ${\bf B426}$ (1994) 19, hep-th/9407087;
``{\it Monopoles, Duality and Chiral Symmetry Breaking in 
$N=2$ Supersymmetric QCD}'', Nucl. Phys. ${\bf B431}$ (1994) 484,
hep-th/9408099.
\bibitem{KaVa}
S. Katz and C. Vafa,
``{\it Matter from Geometry}'', Nucl. Phys. ${\bf B497}$ (1997) 146,
hep-th/9606086.
\bibitem{ChKlYaZa} 
T.-M. Chiang, A. Klemm, S.-T. Yau and E. Zaslow,
``{\it Local Mirror Symmetry: Calculations and Interpretations}'',
Adv. Theor. Math. Phys. ${\bf 3}$ (1999) 495, hep-th/9903053.
\bibitem{Ohta}
Y. Ohta,
``{\it Prepotential of $N=2$ $SU(2)$ Yang-Mills Gauge Theory 
Coupled with a Massive Matter Multiplet}'',
J. Math. Phys. ${\bf 37}$ (1996) 6074, hep-th/9604051;
``{\it Prepotential of $N=2$ $SU(2)$ Yang-Mills Theories 
Coupled with  Massive Matter Multiplets}'',
J. Math. Phys. ${\bf 38}$ (1997) 682, hep-th/9604059.
\bibitem{MaSu}
T. Masuda and H. Suzuki, 
``{\it Prepotential of $N=2$ Supersymmetric Yang-Mills Theories
in the Weak Coupling Region}'',
Int. J. Mod. Phys. ${\bf A13}$ (1998) 1495, hep-th/9609065;
``{\it Periods and Prepotential of $N=2$ $SU(2)$  Supersymmetric 
Yang-Mills Theory with Massive Hypermultiplets}'', 
Int. J. Mod. Phys. ${\bf A12}$ (1997) 3413, hep-th/9609066;
``{\it On explicit Evaluations Around the Conformal Point in
 $N=2$ Supersymmetric Yang-Mills Theories}'',
Nucl. Phys. ${\bf B495}$ (1997) 149, hep-th/9612240.
\bibitem{IsMuNuSh}
J.M. Isidro, A. Mukherjee, J.P. Nunes and H.J. Shnitzer,
``{\it A New Derivation of the Picard-Fuchs equations for Effective
$N=2$ Super Yang-Mills Theories}'',
Nucl. Phys. ${\bf B492}$ (1997) 647, hep-th/9606116;
``{\it On the Picard-Fuchs Equations for 
Massive $N=2$ Seiberg-Witten theories}'',
Nucl. Phys. ${\bf B502}$ (1997) 363, hep-th/9704174.
\bibitem{Eguchi}
T. Eguchi, private communication.
\bibitem{CadeGrPa}
P. Candelas, X. de la Ossa, P.S. Green and L. Parkes,
``{\it A Pair of Calabi-Yau Manifolds as an Exactly Soluble 
Superconformal Theory}'',
Nucl. Phys. ${\bf B359}$ (1991) 21.
\bibitem{HaOz}
A. Hanany and Y. Oz,
``{\it On the Quantum Moduli Space of Vacua of $N=2$ 
Supersymmetric $SU(N_c)$ Gauge Theories}'',
Nucl. Phys. ${\bf B452}$ (1995) 283, hep-th/9505075.
\bibitem{ItYa}
K. Ito and S.-K. Yang,
``{\it Prepotentials in $N=2$ $SU(2)$ Supersymmetric Yang-Mills Theory with 
Massless Hypermultiplets}'',
Phys. Lett. ${\bf B366}$ (1996) 165, hep-th/9507144;
``{\it Picard-Fuchs Equations and Prepotentials in $N=2$
Supersymmetric QCD}'', hep-th/9603073.
\bibitem{KlLeTh}
A. Klemm, W. Lerche and S. Theisen,
``{\it Nonperturbative Effective Actions of 
$N=2$ Supersymmetric Gauge Theories}'',
Int. J. Mod. Phys. ${\bf A11}$ (1996) 1929, hep-th/9505150.
\bibitem{KlLeThYa}
A. Klemm, W. Lerche, S. Theisen and S. Yankielowicz, 
``{\it Simple Singularities and $N=2$ Supersymmetric Yang-Mills Theory}'',
Phys. Lett. ${\bf B344}$ (1995) 169, hep-th/9411048.
\bibitem{Batyrev}
V.V. Batyrev,
``{\it Dual Polyhedra and Mirror Symmetry for 
Calabi-Yau Hypersurfaces in Toric Varieties}'',
J. Algebraic Geom. ${\bf 3}$ (1994) 493, math.AG/9310003.
\bibitem{HoKlThYa}
S. Hosono, A. Klemm, S. Theisen and S.-T. Yau,
``{\it Mirror Symmetry, Mirror Map and Applications to 
Calabi-Yau Hypersurfaces}'',
Commun. Math. Phys. 167 (1995) 301, hep-th/9308122;
``{\it Mirror Symmetry, Mirror Map and Applications to 
Complete Intersection Calabi-Yau spaces}'',
Nucl. Phys. ${\bf B433}$ (1995) 501-554, hep-th/9406055.
\bibitem{CoKa}
D.A. Cox and S. Katz,
``{\it Mirror Symmetry and Algebraic Geometry}'',
AMS, Providence, Rhode Island, (1999).
\bibitem{GrRy}
I.S. Gradshteyn and I.M. Ryzhik,
``{\it Table of Integrals, Series, and Products}'',
Sixth Edition, Academic  Press, (2000).
\bibitem{DoKhMa}
N. Dorey, V.V. Khoze and M.P. Mattis,
``{\it Multi-Instanton Calculus in $N=2$ Supersymmetric Gauge Theory}'',
Phys. Rev. ${\bf D54}$ (1996) 2921, hep-th/9603136;
``{\it Multi-Instanton Check of the Relation
Between the Prepotential ${\cal F}$ 
and the Modulus $u$ in $N=2$ SUSY Yang-Mills Theory}'',
Phys. Lett. ${\bf B390}$ (1997) 205, hep-th/9606199;
``{\it Multi-Instanton Calculus in $N=2$ Supersymmetric
Gauge Theory II. Coupling to Matter}'',
Phys. Rev. ${\bf D54}$ (1996) 7832, hep-th/9607202.
\bibitem{AoHaSaWa}
H. Aoyama, T. Harano, M. Sato and S. Wada,
``{\it Multi-instanton Calculus in $N=2$ Supersymmetric QCD}'',
Phys. Lett. ${\bf B388}$ (1996) 331, hep-th/9607076;
T. Harano and M. Sato,
``{\it Multi-instanton Calculus versus Exact Results in $N=2$
Supersymmetric QCD}'', Nucl. Phys. ${\bf B484}$ (1997) 167,
	hep-th/9608060.
\bibitem{ItSa}
K. Ito and N. Sasakura,
``{\it One-instanton Calculations in $N=2$ Supersymmetric $SU(N_c)$
Yang-Mills Theory}'', Phys. Lett. ${\bf B382}$ (1996) 95, hep-th/9602073; 
``{\it Exact and Microscopic One-instanton Calculations in $N=2$
Supersymmetric Yang-Mills Theories}'', Nucl. Phys. ${\bf B484}$ (1997)
141, hep-th/9608054;
``{\it One-instanton Calculations in $N=2$ $SU(N_c)$ Supersymmetric
	QCD}'', Mod. Phys. Lett. ${\bf A12}$ (1997) 205, hep-th/9609104.
\bibitem{Ho}
T.J. Hollowood,
``{\it Calculating the 
Prepotential by Localization on the Moduli Space of Instantons
}'', JHEP {\bf 0203} (2002) 038,
hep-th/0201075.
\bibitem{Ne}
N.A. Nekrasov, 
``{\it Seiberg-Witten Prepotential From Instanton Counting }'',
hep-th/0206161.
\bibitem{FlPo}
R. Flume and R. Poghossian, 
``{\it An Algorithm for the Microscopic Evaluation of 
the Coefficients of the
Seiberg-Witten Prepotential}'', hep-th/0208176.
\bibitem{BrFuMoTa}
U. Bruzzo, F. Fucito, J.F. Morales and A. Tanzini,
``{\it Multi-instanton Calculus and Equivariant Cohomology}'',
hep-th/0211108.
\bibitem{EgKa}
T. Eguchi and H. Kanno, 
``{\it Five-Dimensional Gauge Theories and Local Mirror Symmetry}'',
Nucl. Phys. ${\bf B586}$ (2000) 331, hep-th/0005008.
\bibitem{DiVa}
R. Dijkgraaf and C. Vafa,
``{\it Matrix Models, Topological Strings, and
Supersymmetric Gauge Theories}'', Nucl. Phys. ${\bf B644}$ (2002)
3, hep-th/0206255;
``{\it On Geometry and Matrix Models}'', Nucl. Phys. ${\bf B644}$ 21,
hep-th/0207106;
``{\it A Perturbative Window into Non-perturbative Physics}'',
hep-th/0208048;
``{\it ${\cal N}=1$ Supersymmetry, Deconstruction 
and Bosonic Gauge Theories}'', hep-th/0302011.
\bibitem{Sturmfels}
B. Sturmfels,
``{\it Gr\"obner Bases and Convex Polytopes}'',
University Lecture Notes Vol. 8,
AMS, Providence, Rhode Island, (1996).
\bibitem{GeKaZe}
I.M. Gel'fand, M.M. Kapranov and A.V. Zelevinskii, 
``{\it Hypergeometric Functions and Toral Manifolds}'', 
Funct. Anal. Appl. ${\bf 23}$ (1989) 94;
``{\it Generalized Euler Integrals and $A$-Hypergeometric Functions}'', 
Adv. Math. ${\bf 84}$ (1990) 255.
\bibitem{SaStTa}
M. Saito, B. Sturmfels and N. Takayama, 
``{\it Gr\"obner Deformations of Hypergeometric Differential Equations}'',
Algorithms and Computation in Mathematics Vol. 6, Springer, (2000).
\bibitem{HoLiYa} 
S. Hosono, B.H. Lian  and S.-T. Yau,
``{\it GKZ-Generalized Hypergeometric Systems in Mirror Symmetry of 
 Calabi-Yau Hypersurfaces}'',
Commun. Math. Phys. ${\bf 182}$ (1996) 535, alg-geom/9511001.
\end{thebibliography}
\end{document}